\documentclass[
	reprint,
	superscriptaddress,
	showkeys,
	aps,
	pra,
	longbibliography,
	nofootinbib
]{revtex4-2}

\usepackage[utf8]{inputenc}

\usepackage[hyperref]{xcolor}
	\definecolor{goethe-blau}{cmyk}{1.0,0.2,0.0,0.4}
	\definecolor{hellgrau}{cmyk}{0.04,0.04,0.05,0.02}
	\definecolor{sandgrau}{cmyk}{0.12,0.09,0.13,0.0}
	\definecolor{dunkelgrau}{cmyk}{0.25,0.25,0.30,0.75}
	\definecolor{emo-rot}{cmyk}{0.04,1.0,0.8,0.07}
	\definecolor{purple}{cmyk}{0.08,1.0,0.3,0.36}
	\definecolor{senfgelb}{cmyk}{0.01,0.25,1.0,0.05}
	\definecolor{gruen}{cmyk}{0.62,0.4,0.87,0.09}
	\definecolor{magenta}{cmyk}{0.08,0.86,0.12,0.12}
	\definecolor{orange}{cmyk}{0.0,0.7,1.0,0.04}
	\definecolor{sonnengelb}{cmyk}{0.0,0.12,0.95,0.0}
	\definecolor{helles-gruen}{cmyk}{0.4,0.17,0.81,0.07}
	\definecolor{lichtblau}{cmyk}{0.8,0.0,0.06,0.04}

\usepackage[
	colorlinks,
	pdfpagelabels,
	breaklinks,
	pdfstartview=FitH,
	bookmarksopen=true,
	bookmarksnumbered=true,
	bookmarksopenlevel=2,
	plainpages=false,
	hypertexnames=false,
	citecolor=emo-rot,
	linkcolor=goethe-blau,
	urlcolor=purple,
	pdftitle={Inhomogeneous condensation in the Gross-Neveu model in noninteger spatial dimensions. II.},
	pdfauthor={Laurin Pannullo, Adrian Koenigstein},
	pdfkeywords={GN model, inhomogeneous phases, moat regime, stability analysis}
]{hyperref}

\usepackage{mathtools}	
\usepackage{amsmath} 
\usepackage{amssymb} 
\usepackage{bm} 
\usepackage{slashed} 
\usepackage{tensor} 
\usepackage{upgreek} 

\usepackage{bookmark} 
\usepackage{graphicx} 
\usepackage{placeins} 

\usepackage{booktabs} 
\usepackage{dcolumn} 
\usepackage{multirow}

\usepackage[capitalise,sort,compress]{cleveref}

\allowdisplaybreaks 

\newcommand{\sdim}{d}

\newcommand{\dd}{\mathrm{d}}

\renewcommand{\S}{\mathcal{S}}

\newcommand{\Seff}{\S_\mathrm{eff}}

\newcommand{\Ueff}{\bar{U}}

\newcommand{\Sd}{S_\sdim}

\newcommand{\gtwo}{\Gamma^{(2)}}

\newcommand{\barpsi}{\bar{\psi}}

\newcommand{\pFq}[2]{{}_{#1}F_{#2}}

\newcommand{\vdistance}{\vphantom{\bigg(\bigg)}}
\newcommand{\Vdistance}{\vphantom{\Bigg(\Bigg)}}

\renewcommand{\Re}{\mathrm{Re}}

\newcommand{\ii}{\mathrm{i}}
\newcommand{\ee}{\mathrm{e}}

\newcommand{\nf}{n_\mathrm{f}}

\newcommand{\const}{\mathrm{const.}}

\newcommand{\sigmaminvac}{\bar{\sigma}_0}

\newcommand{\gammachiral}{{\gamma_{\mathrm{ch}}}}

\newcommand{\barsigma}{\bar{\sigma}}

\newcommand{\barmu}{\bar{\mu}}

\newcommand{\dimDirac}{{d_\gamma}}

\newcommand{\ffcoupling}{\lambda}

\DeclareMathOperator{\Li}{\mathrm{Li}}

\DeclareMathOperator{\arcoth}{\mathrm{arcoth}}
\DeclareMathOperator{\artanh}{\mathrm{artanh}}

\newcommand{\Det}{\mathrm{Det}}

\newcommand{\Def}{Def.~}
\newcommand{\Defs}{Defs.~}

\newcounter{numrefs}
\makeatletter
\newcommand{\Rcite}[1]{%
	\setcounter{numrefs}{0}
	\@for\@temp:=#1\do{\stepcounter{numrefs}}
	\ifnum\value{numrefs}>1%
	Refs.~\cite{#1}%
	\else%
	Ref.~\cite{#1}%
	\fi%
}
\makeatother
\newcommand{\RciteSingle}[2][]{%
	Ref.~\cite[#1]{#2}%
}

\newcommand{\eg}{e.g.}

\usepackage[acronym]{glossaries-extra}
\glssetcategoryattribute{acronym}{nohyperfirst}{true}
\setabbreviationstyle[acronym]{long-short}

\newacronym{uv}{UV}{ultra violet}
\newacronym{ir}{IR}{infra red}
\newacronym{wrt}{w.r.t.}{with respect to}
\newacronym{gn}{GN}{Gross-Neveu}
\newacronym{ip}{IP}{inhomogeneous phase}
\newacronym{sp}{SP}{symmetric phase}
\newacronym{hbp}{HBP}{homogeneously broken phase}
\newacronym{pt}{PT}{phase transition}
\newacronym{qft}{QFT}{quantum field theory}
\newacronym{qcd}{QCD}{Quantum Chromodynamics}
\newacronym{frg}{FRG}{functional renormalization group}
\DeclareMathOperator{\arsinh}{arsinh}

\begin{document}



\title{
	Inhomogeneous condensation in the Gross-Neveu model in noninteger spatial dimensions \texorpdfstring{$1 \leq \sdim < 3$. II.\\}{1 leq d less 3. II.}
	Nonzero temperature and chemical potential
}

\author{Adrian Koenigstein}
	\email{adrian.koenigstein@uni-jena.de}
	\affiliation{
		Theoretisch-Physikalisches Institut,
		Friedrich-Schiller Universität Jena,
		D-07743 Jena,
		Germany.
	}

\author{Laurin Pannullo}
	\email{lpannullo@physik.uni-bielefeld.de}		
	\affiliation{
		Fakultät für Physik,
		Universität Bielefeld,
		D-33615 Bielefeld,
		Germany.
	}

\date{\today}

\begin{abstract}
	We continue previous investigations of the (inhomogeneous) phase structure of the Gross-Neveu model in a noninteger number of spatial dimensions ($1 \leq \sdim < 3$) in the limit of an infinite number of fermion species ($N \to \infty$) at (non)zero chemical potential $\mu$ \cite{Pannullo:2023cat}.
	In this work, we extend the analysis from zero to nonzero temperature $T$.
	
	The phase diagram of the Gross-Neveu model in $1\leq \sdim < 3$ spatial dimensions is well known under the assumption of spatially homogeneous condensation with both a symmetry broken and a symmetric phase present for all spatial dimensions.
	In $\sdim = 1$ one additionally finds an inhomogeneous phase, where the order parameter, the condensate, is varying in space.
	Similarly, phases of spatially varying condensates are also found in the Gross-Neveu model in $\sdim = 2$ and $\sdim = 3$, as long as the theory is not fully renormalized, i.e., in the presence of a regulator.
	For $\sdim = 2$, one observes that the inhomogeneous phase vanishes, when the regulator is properly removed (which is not possible for $\sdim = 3$ without introducing additional parameters).

	In the present work, we use the stability analysis of the symmetric phase to study the presence (for $1 \leq \sdim < 2$) and absence (for $2 \leq \sdim < 3$) of these inhomogeneous phases and the related moat regimes in the fully renormalized Gross-Neveu model in the $\mu, T$-plane.
	We also discuss the relation between ``the number of spatial dimensions'' and ``studying the model with a finite regulator'' as well as the possible consequences for the limit $\sdim \to 3$.
\end{abstract}


\maketitle

\tableofcontents

\section{Introduction}

	The \gls{gn} model is arguably one of the most simple theories that describe (self-)interacting fermions.
	Despite this fact and having been formulated 50 years ago \cite{Gross:1974jv}, its chiral phase diagram in the $\mu,T$-plane in various number of spatial dimensions $\sdim$ is still under investigation today.
	Within this work, we aim to contribute to this research by focusing on so-called \glspl{ip} of spatially oscillating condensates in noninteger spatial dimensions. (We refer to \Rcite{Buballa:2014tba} for a review on \glspl{ip}.)

\subsection{General contextualization}

	Within the $N\to\infty$ limit, one finds that bosonic quantum fluctuations are suppressed \cite{Gross:1974jv}, which immensely simplifies calculations and enables mostly analytic approaches.
	Thus, it is not surprising that the most complete picture of the thermodynamics of the \gls{gn} model is within the special $N\to\infty$ limit, see, e.g., \Rcite{Dashen:1974xz,Jacobs:1974ys,Harrington:1974tf,Harrington:1974te,Wolff:1985av,Schnetz:2004vr,Schnetz:2005ih,Inagaki:1994ec,Ahmed:2018tcs,Buballa:2020nsi,Narayanan:2020uqt,Soler:1970xp,Klimenko:1986uq,Klimenko:1987gi,Rosenstein:1988dj,Rosenstein:1988pt,Rosenstein:1990nm}.
	Still, even with these simplifications the \gls{gn} model can be seen as a prototype \gls{qft} that shares a lot of features with more realistic \glspl{qft}.
	It is asymptotically free and undergoes dimensional transmutation, leading to a condensation of fermion-anti-fermion pairs in the \gls{ir} in vacuum, which is similar to \gls{qcd} and \gls{qcd}-like theories.
	However, there are also important relations between the \gls{gn} model and various models from solid state physics as well as numerous extensions of the model that are used as toy model \glspl{qft}, such that studying the model within different setups remains an interesting task on its own but is also of relevance as reference work.
	For further reading, we refer to \Rcite{Thies:2006ti,Koenigstein:2023wso}.

\subsection{Recap of central results}

	In $1+1$ dimensions the \gls{gn} model exhibits three distinct chiral phases \cite{Thies:2003kk,Schnetz:2004vr}.
	At low temperatures and chemical potential the discrete chiral $\mathbb{Z}_2$ symmetry is spontaneously broken and one finds the so-called \gls{hbp} that is characterized by a nonzero chiral condensate, which is constant in space.
	At moderate and high temperatures one finds the \gls{sp} --  a gas-like phase, which is characterized by a vanishing chiral condensate.
	Especially relevant for our work is the \gls{ip} at low temperatures and moderate and large $\mu$, where the chiral condensate is non-vanishing and exhibits a spatial dependence.
	Thus, in addition to chiral symmetry also translational invariance is spontaneously broken.
	This phase is associated with negative values of the bosonic two-point function for some range of spatial bosonic momenta \cite{Nakano:2004cd,Koenigstein:2021llr,Abuki:2011pf,deForcrand:2006zz,Wagner:2007he,Tripolt:2017zgc,Nickel:2009ke,Winstel:2019zfn,Carignano:2019ivp,Buballa:2018hux,Buballa:2020nsi,Buballa:2020xaa,Winstel:2021yok,Pannullo:2021edr,Winstel:2022jkk,Braun:2014fga}.
	Assuming a second order \gls{pt} to the \gls{sp}, a necessary condition for the \gls{ip} is a negative wave-function renormalization even though this is not a sufficient criterion \cite{Braun:2014fga,Roscher:2015xha,Pisarski:2020gkx}.
	Nevertheless, regions of negative wave-function renormalization in the $\mu,T$-plane are interesting on their own, since they feature a nontrivial momentum structure of the two-point function and the dispersion relation.
	These regions that can be larger than the actual \gls{ip} were labeled as moat regimes recently and could play an important role in the hadronization process in heavy-ion collisions \cite{Pisarski:2020gkx,Pisarski:2021qof,Rennecke:2021ovl,Rennecke:2023xhc}.
	As discussed in \Rcite{Koenigstein:2021llr} the $1 + 1$ \gls{gn} model has a moat regime, which extends over large parts of the $\mu,T$-plane and thus serves as a toy model for this phenomenon.

	It was found that the number of spatial dimensions has a profound impact on the phase structure.
	In $2+1$ dimensions, one finds that the \gls{ip} and the moat regime are only present  at finite regulator and vanish in the renormalized limit \cite{Buballa:2020nsi,Narayanan:2020uqt,Pannullo:2023one}.
	Still, one observes an \gls{hbp} for small $\mu$ and $T$ and an \gls{sp} for large $\mu$ and $T$ \cite{Klimenko:1987gi,Rosenstein:1988dj}.
	
	The situation is less clear in $3+1$ dimensions, where the model is non-renormalizable (without introducing additional parameters) and the value of the regulator and the choice of the regularization scheme has a drastic impact on the phase structure \cite{Pannullo:2022eqh,pannulloInhomogeneousPhasesMoat2023,Pasqualotto:2023hho,Partyka:2008sv,Kohyama:2015hix}.
	Most certainly one also finds an \gls{hbp} and an \gls{sp}, while the existence of an \gls{ip} can be regarded as disputed as it heavily depends on the regularization.

	In an effort to understand why the \gls{ip} is absent in $2+1$ dimensions and to what extent the existence of the phase is a regulator artifact in $3+1$ dimensions, \Rcite{Pannullo:2023cat} investigated the \gls{gn} model in noninteger $\sdim$ spatial dimensions in order to interpolate between the known integer dimensional results and to additional mimic the effect of dimensional regularization.\footnote{Note, that we will still use the same terminology \gls{hbp}, \gls{sp}, and \gls{ip} as well as symmetry breaking etc.\ for noninteger dimensions, even though chiral symmetry and the concept of spatial oscillation might not be well-defined in a noninteger number of spatial dimensions.
	It might be better to talk about instabilities of spatially constant condensates etc..
	However, this leads to a needless complication of the discussion.}
	This study was conducted at $T=0$, which sufficed to illuminate that (a) the \gls{ip} is present for $1 \leq \sdim<2$ and vanishes exactly in $2 = \sdim$, and (b) one does not find the phase in $2 < \sdim < 3$ in a renormalized setup, which implies that its existence in $\sdim=3$ is caused by a finite regulator.

\subsection{Research objective}

	In the present work, we extend this investigation of spatially inhomogeneous condensation in (non)integer number of spatial dimensions from zero to nonzero temperature in order to map out the $\sdim$-dependence of the \gls{ip} and the moat regime.
	This study therefore complements the previously mentioned study in \Rcite{Pannullo:2023cat} as well as \Rcite{Inagaki:1994ec}.
	The latter study already investigated the phase diagram of the \gls{gn} model in continuous dimensions $1 \leq \sdim < 3$ under the assumption of spatially homogeneous condensation.
	Therefore, we aim at closing a gap in the literature about the \gls{gn} model.

	Furthermore, we hope that this work contributes to the general discussion about necessary conditions for the presence/absence of \glspl{ip} in arbitrary models and theories.
	Here, especially our findings about the role of the spatial dimensionality may be essential to understand the general criteria for the formation of inhomogeneous condensates.

\subsection{Structure}

	This work is structured as follows:
	In \cref{sec:gn_model_in_medium} we recapitulate some basic mathematical aspects of the \gls{gn} model.
	We present the four-fermion action, the bosonized version of the model in the $N \to \infty$ limit and we discuss the quantities that are relevant to map out the phase diagram.
	Here, we also explain our regularization and renormalization prescription as well as the stability analysis -- the method to detect inhomogeneous condensation.

	Afterwards, in \cref{sec:results}, we turn to the results.
	We present the evaluation of the above expressions for some points in the $\mu,T$-plane and different dimensions $\sdim$.
	We show sample plots for the two-point function and wave-function renormalization.
	Most importantly, we present the dependence of the phase diagram and especially of the \gls{ip} and moat regime on the number of spatial dimensions $\sdim$.

	We conclude and comment on our results in \cref{sec:conclusions_and_outlook} and provide a brief outlook to possible consequences of our findings and followup questions.

	Our work is accompanied by a large number of appendices, where we present all relevant details of this work.
	We hope that the amount of technical details might help the interested reader to easily reproduce and/or build on our work.
	We also consider these appendices as a compilation of the most relevant formulae for the \gls{gn} model in $\sdim < 3$ within the $N\to\infty$ limit.

\section{The Gross-Neveu model in \texorpdfstring{$1 \leq \sdim < 3$}{1 <= d < 3} spatial dimensions in medium}
\label{sec:gn_model_in_medium}

	In this chapter, we introduce the \gls{gn} model in $\sdim + 1$ dimensions, where $\sdim$ is the number of spatial dimensions.
	We work at (non)zero temperature $T$ and (non)zero chemical potential $\mu$ and briefly recapitulate the derivation of the grand canonical/effective potential, the bosonic two-point function as well as the bosonic wave-function renormalization.
	All calculations and results are in the limit $N \to \infty$, where $N$ is the number of fermion species.
	The quantities that are presented in this chapter are required for the computation of the phase diagram, the stability analysis, and the detection of a possible \gls{ip} and/or moat regime, see also \Rcite{Koenigstein:2021llr,Pannullo:2023one,Buballa:2020nsi} for similar analyses and results that arise as the limiting cases for integer $\sdim$.

\subsection{The action and the potential}

	The microscopic action of the \gls{gn} model is given by \cite{Gross:1974jv}
		\begin{align}
			& \S [ \barpsi, \psi ] =	\Vdistance
			\\
			={}& \int_{0}^{\frac{1}{T}} \dd \tau \int \dd^\sdim x \,\big[ \barpsi \, ( \slashed{\partial} + \gamma^0 \mu ) \, \psi - \tfrac{\ffcoupling}{2 N} \, ( \barpsi \, \psi )^2 \big] \, .	\Vdistance	\nonumber
		\end{align}
	Here, $\barpsi = \bar{\psi} ( \tau, \vec{x} \, )$ and $\psi = \psi ( \tau, \vec{x} \, )$ are the fermion fields, where $x^i \in ( - \infty, \infty )$, $i \in \{ 1, \ldots, \sdim \}$, are the spatial coordinates of a $\sdim$-dimensional Euclidean space and $\tau \in [ 0, \frac{1}{T} )$ denotes the coordinate of the compactified temporal direction that mimics the (inverse) temperature $T$.
	The fermions have antiperiodic boundary conditions in the compact direction, come in $N$ different species\footnote{Occasionally, these species are referred to as different colors or flavors.} and transform as spinors.
	We use a $\dimDirac$-dimensional representation of the gamma matrices of the corresponding Clifford algebra,
		\begin{align}
			&	\{ \gamma^\mu, \gamma^\nu \}_+ = 2 \delta^{\mu \nu} \openone_{\dimDirac}  \, ,	&&	\mu, \nu \in \{ 0, 1, \ldots, \sdim \} \, .
		\end{align}	
	This generalizes to noninteger $d$ dimensions \cite{tHooft:1972tcz} and we use the Kronecker delta as the components of the Euclidean metric.
	Furthermore, we use $\ffcoupling$ for the four-fermion coupling and introduce the fermion chemical potential $\mu$ in the standard way.
	
	For a detailed discussion of the symmetries of this model in different integer dimensions, we refer for example to \Rcite{Koenigstein:2023wso,Urlichs:2007zz}.

	In order to study four-fermion models (especially in the $N \to \infty$ limit) one convenient approach is to bosonize the theory in the \gls{uv} via a Hubbard-Stratonovich transformation \cite{Stratonovich:1957,Hubbard:1959ub}.
	Here, the four-fermion interaction is replaced by an auxiliary real scalar bosonic field $\phi$.
	On the level of the partition function the equivalent action is \cite{Gross:1974jv,Dashen:1974xz,Wolff:1985av} 
		\begin{align}
			& \S [ \barpsi, \psi, \phi ] =	\Vdistance	\label{eq:gross-neveu_action_bosonized}
			\\
			={}& \int_{0}^{\frac{1}{T}} \dd \tau \int \dd^\sdim x \, \big[ \tfrac{1}{2 \ffcoupling} \, ( h \phi )^2 + \barpsi \, \big( \slashed{\partial} + \gamma^0 \mu + \tfrac{h}{\sqrt{N}} \, \phi \big) \, \psi \big] \, ,	\Vdistance	\nonumber
		\end{align}
	where we also introduced the Yukawa coupling $h$ to obtain canonical energy dimensions for $\phi$.
	Now, in addition to the functional integration over fermion fields, one also has to integrate over the bosonic field $\phi$.
	Since loop corrections for the Yukawa coupling are suppressed for $N \to \infty$, see, \eg, \Rcite{Braun:2011pp,Koenigstein:2023wso}, it is convenient to absorb the Yukawa coupling in the field $\phi$ which is afterwards of dimension energy in any number of spatial dimensions.
	In addition, to correctly take the limit $N \to \infty$ one rescales the boson field with $\sqrt{N}$.
	Hence, we introduce
		\begin{align}
			\sigma = \tfrac{h}{\sqrt{N}} \, \phi
		\end{align}
	as the new bosonic degree of freedom.

	It is simple to show, see, e.g., \Rcite{Pannullo:2019}, that 
		\begin{align}
			\langle \sigma \rangle \propto \langle \barpsi \, \psi \rangle \, .
		\end{align}
	In an even number of spacetime dimensions $\sdim + 1$ the formation of a nonzero expectation value of $\sigma$ therefore signals the breaking of the discrete $\mathbb{Z}_2$ chiral symmetry
		\begin{align}
			&	\psi \mapsto \gammachiral \psi \, ,	&&	\barpsi \mapsto - \barpsi \, \gammachiral \, ,	&&	\sigma \mapsto - \sigma
		\end{align}
	of the \gls{gn} action and the formation of a condensate.
	In an odd number of spacetime dimensions as well as for noninteger dimensions the matrix $\gammachiral$, which fulfills
		\begin{align}
			&	\{ \gamma^\mu, \gammachiral \}_+ = 0  \, ,	&&	\mu \in \{ 0, 1, \ldots, \sdim \} \, ,
		\end{align}
	is either nonexisting or its definition is ambiguous (depending on $\dimDirac$).
	For  detailed discussions of these situations, we refer to \Rcite{tHooft:1972tcz,Gies:2009da,Urlichs:2007zz}.
	Still, it is possible to study the action \labelcref{eq:gross-neveu_action_bosonized} and analyze for which $\mu$ and $T$ one finds (non)vanishing expectation values of the scalar field $\langle \sigma \rangle$.
	This even holds true if one allows for spatial modulations of this condensate.
	Regions in the $\mu, T$-plane with $\langle \sigma \rangle \neq 0$ are denoted as phases of spatially (in)homogeneous condensation/symmetry breaking, while regions with $\langle \sigma \rangle = 0$ are called gas like/symmetric phases.

	The standard way to proceed from \cref{eq:gross-neveu_action_bosonized} is to perform the functional integration over the fermion fields and absorb the resulting fermion determinant in an effective action for the boson field $\sigma$.
	The resulting action $\Seff$ in the probability distribution in the thermal partition function is
		\begin{align}
			&	\tfrac{1}{N} \, \Seff [ \sigma ] =	\Vdistance	\label{eq:gross-neveu_action_effective_large_n}
			\\
			={}& \int_{0}^{\frac{1}{T}} \dd \tau \int \dd^\sdim x \, \tfrac{\sigma^2}{2 \ffcoupling} - \ln \Det \big[ \beta ( \slashed{\partial} + \gamma^0 \mu + \sigma ) \big] \, ,	\Vdistance	\nonumber
		\end{align}
	where we already divided by the number of species $N$ and $\Det$ denotes a functional determinant.

	Considering the limit $N \to \infty$, this implies that the only field configurations that contribute in the partition function and the expectation values are those that minimize the effective action $\Seff$.
	This is equivalent to studying the full quantum effective action (the generating functional for one-particle-irreducible vertex functions) and only taking into account the contribution of the fermionic quantum fluctuations \cite{Peskin:1995ev,Goldstone:1961eq,Wetterich:2001kra}.
	In any case, for arbitrary modulations in spatial and temporal directions, the minimization of \cref{eq:gross-neveu_action_effective_large_n} is a highly challenging task.
	It might not even be a well-posed problem for noninteger dimensions.
	However, assuming that the field configuration with least (effective) action is constant in space and time, \eg, $\sigma ( \tau, \vec{x} \, ) = \barsigma = \const$ the problem simplifies drastically.
	We define the (homogeneous) effective potential
		\begin{align}
			\Ueff ( \barsigma, \mu, T, \sdim ) = \tfrac{1}{N} \, \tfrac{1}{\beta V} \, \Seff [ \barsigma ],
		\end{align}
	which is the effective action for homogeneous fields per species and spacetime volume.
	The eigenvalues of the Dirac operator for homogeneous fields $\bar{\sigma}$ are those of free fermions with mass $m=\bar{\sigma}$.
	Thus, the evaluation of $\Ueff$	yields,
		\begin{align}
			& \Ueff ( \barsigma, \mu, T, \sdim ) =	\Vdistance	\label{eq:effective_potential_main}
			\\
			={}& \tfrac{\barsigma^2}{2 \ffcoupling} - \tfrac{1}{\beta V} \ln \Det \big[ \beta ( \slashed{\partial} + \gamma^0 \mu + \barsigma ) \big] =	\Vdistance	\nonumber
			\\
			={}& \tfrac{\barsigma^2}{2 \ffcoupling} - \tfrac{\dimDirac}{2} \, l_0 ( \barsigma, \mu, T, \sdim ) \, ,	\Vdistance	\nonumber
		\end{align}
	where $l_0$ is the Matsubara sum and momentum integral over the $\log$ of the eigenvalues.
	Its evaluation is presented in \cref{app:ellzero}.
	
	By determining the global and local minima of \cref{eq:effective_potential_main}, we obtain the homogeneous phase diagram including the spinodal lines.
	We denote the homogeneous field configuration that corresponds to the global minimum of the homogeneous effective potential for a given $\mu$ and $T$ by $\bar{\Sigma}(\mu,T)$.
	
	The derivative of the homogeneous effective potential \gls{wrt} the homogeneous field $\barsigma$
	\begin{align}
		& \tfrac{\dd}{\dd \bar{\sigma}} \, \Ueff ( \barsigma, \mu, T, \sdim ) =	\Vdistance	\label{eq:gap_equation_main} 	\\
		={}& \bar{\sigma} \, \big( \tfrac{1}{\ffcoupling} - \dimDirac \, l_1 ( \barsigma, \mu, T, \sdim ) \big)	\Vdistance	\nonumber
	\end{align}
	is used to express the gap equation, which is of central importance in the renormalization of this model as discussed in \cref{sec:renormalization}.
	The quantity $l_1$ is again a Matsubara sum and spatial momentum integration.
	Its evaluation is discussed in \cref{app:ellone}.

\subsection{The bosonic two-point function}

	The bosonic two-point function at bosonic spatial momentum $q = | \vec{q} \, |$ for a homogeneous bosonic field $\barsigma$ is given by
		\begin{align}
			& \Gamma^{(2)} ( \barsigma, \mu, T, q, \sdim ) =	\vdistance	\label{eq:gamma2_main}
			\\
			={}& \tfrac{1}{\ffcoupling} - \dimDirac \big[ l_1 ( \barsigma, \mu, T, \sdim ) - \tfrac{1}{2} \, ( q^2 + 4 \, \barsigma^2 ) \, l_2 ( \barsigma, \mu, T, q, \sdim ) \big] \, ,	\vdistance	\nonumber
		\end{align}
	where the quantities $l_1$ and $l_2$ are fermionic Matsubara sums and loop momentum integrals that are discussed in \cref{app:ellone,app:elltwo}.
	The derivation of this quantity in the \gls{gn} model is discussed in great detail in \Rcite{Buballa:2020nsi,Koenigstein:2021llr,Pannullo:2023cat,Koenigstein:2023wso} and shall not be repeated here. 
	Simply speaking, one obtains the bosonic two-point function, by (1.) taking two  functional derivatives \gls{wrt} to $\sigma$ of \cref{eq:gross-neveu_action_effective_large_n}, (2.) evaluating the result at vanishing external Matsubara frequencies and for $\sigma ( \tau, \vec{x} \, ) = \barsigma = \const$.
	In our analysis, we evaluate the two-point function at the global minimum of the effective potential $\bar{\Sigma} ( \mu, T )$ for the specific $\mu$ and $T$.

\subsection{The bosonic wave-function renormalization}

	Another quantity of interest is the so-called bosonic wave-function renormalization given by
		\begin{align}
			z ( \barsigma, \mu, T, \sdim ) = \tfrac{1}{2} \, \tfrac{\dd^2}{\dd q^2} \, \Gamma^{(2)} ( \barsigma, \mu, T, q, \sdim ) \Big|_{q = 0} \, , \label{eq:zdef}
		\end{align}
	which is the coefficient of the kinetic contribution $\frac{1}{2} \, ( \partial_\mu \sigma )^2$ to the quantum effective action \cite{Dashen:1974xz}.
	After a short calculation, which is summarized in \cref{app:wave_function_renormalization} one finds 
		\begin{align}
			& z ( \barsigma, \mu, T, \sdim ) =	\vdistance	\label{eq:wave-function_renormalization_main}
			\\
			={}& \tfrac{\dimDirac}{2} \, \big[ l_2 ( \barsigma, \mu, T, 0, \sdim ) - \tfrac{8}{6} \, \barsigma^2 \, l_3 ( \barsigma, \mu, T, \sdim ) \big] \, ,	\vdistance	\nonumber
		\end{align}
	where we introduced $l_3$ as another Matsubara sum and integral that is evaluated in \cref{app:ellthree}.
	If the wave-function renormalization is evaluated at the global minimum of the effective potential $\bar{\Sigma}$, we denote it by $Z$, i.e., $Z(\mu,T,d)\equiv z(\bar{\Sigma}(\mu,T),\mu,T,d)$.

\subsection{Regularization of vacuum contributions}

	Some of the previously listed quantities contain contributions with \gls{uv}-divergent integrals.
	For $d<3$, these are the vacuum parts of $l_0$ and $l_1$, which require a \gls{uv} regularization to render calculations tractable.
	We regularize the theory with a spatial momentum cutoff that confines the integration of spatial fermionic loop momenta to a $d$-dimensional sphere of radius $\Lambda$.
	This scheme certainly has its drawbacks in the investigation of inhomogeneous condensation as we explicitly break translational invariance.
	However, as is discussed in  \cref{sec:renormalization}, one can renormalize the theory for $d < 3$.
	Thus, it is possible to eventually remove the regulator completely by sending $\Lambda \to \infty$, allowing us to make the assumptions that $\Lambda$ is large.

	Regularizing the vacuum part of $l_0$ and $l_1$ with the spatial momentum cutoff and expanding the result for $\Lambda/\barsigma\gg 1$ results in
			\begin{align}
				&	l_0^\Lambda ( \barsigma, 0, 0, \sdim ) =	\Vdistance	\label{eq:expansion_l_0}
				\\
				={}& \tfrac{\Sd}{( 2 \uppi )^\sdim} \, \tfrac{1}{2} \, \bigg( - \tfrac{\Gamma ( \frac{\sdim + 2}{2} ) \, \Gamma ( - \frac{\sdim + 1}{2} )}{\sdim \sqrt{\uppi}} \, \barsigma^{\sdim + 1} + \tfrac{2}{\sdim + 1} \, \Lambda^{\sdim + 1} +	\Vdistance	\nonumber
				\\
				& \quad + \Lambda^\sdim \, \Big[ \tfrac{\barsigma^2}{\sdim - 1} \, \tfrac{1}{\Lambda} + \tfrac{\barsigma^4}{4 ( 3 - \sdim )} \, \big( \tfrac{1}{\Lambda} \big)^3 + \mathcal{O} \big( \tfrac{1}{\Lambda^5} \big) \Big] \bigg) \Vdistance	\nonumber
			\end{align}
		and
			\begin{align}
				&	l_1^\Lambda ( \barsigma, 0, 0, \sdim ) =	\Vdistance	\label{eq:expansion_l_1}
				\\
				={}& \tfrac{\Sd}{( 2 \uppi )^\sdim} \, \tfrac{1}{2} \, \bigg( \tfrac{\Gamma ( \frac{\sdim + 2}{2} ) \, \Gamma ( - \frac{\sdim + 1}{2} )}{\sdim \, \sqrt{\uppi}} \, \big( - \tfrac{\sdim + 1}{2} \big) \, \barsigma^{\sdim - 1} +	\Vdistance	\nonumber
				\\
				& \quad + \Lambda^{\sdim} \, \Big[ \tfrac{1}{\sdim - 1} \, \tfrac{1}{\Lambda} + \tfrac{\barsigma^2}{2 ( 3 - \sdim )} \, \tfrac{1}{\Lambda^3} + \mathcal{O} \big( \tfrac{1}{\Lambda^5} \big) \Big] \bigg) \, ,	\Vdistance	\nonumber
			\end{align}
		where the $l_x^\Lambda$ are the regularized versions \labelcref{eq:l_0_sigma_mu_0_d_Lambda,eq:l_1_sigma_mu_0_d_Lambda} of the respective quantities, and $\Sd$ is given by \cref{eq:sd}.
		These expansions are obtained by applying the formula \cref{eq:expansions_2f1}, whose origin is discussed in \cref{app:expansion}.
		In the last step we also used \cref{eq:gamma_function_recurrence_relation} in order to have identical prefactors of the first terms in \cref{eq:expansion_l_0,eq:expansion_l_1}.

\subsection{Renormalization}
\label{sec:renormalization}

	The \gls{gn} model naturally experiences spontaneous symmetry breaking in the vacuum for all $\sdim$ for $N \to \infty$, which is exploited in the renormalization procedure.
	We impose as renormalization condition that the auxiliary field $\sigma$ assumes the homogeneous nonzero value $\sigmaminvac$ in the vacuum.
	The coupling $\ffcoupling$ is tuned such that this condition is fulfilled and the divergences are absorbed.
	Within the limit of $N \to \infty$, one finds that this is successful for spatial dimensions $d<3$, see \Rcite{Rosenstein:1988pt,Hands:1992be,Inagaki:1994ec}.
	This can easily be understood by looking at \cref{eq:expansion_l_0}, where a divergence $\propto \barsigma^4$ arises for $d\geq3$.
	There is no coupling in the action that can compensate this divergence and removing the cutoff is only possible by introducing more parameters.

\subsubsection{The gap equation}

	Our renormalization condition can be expressed by the gap equation in vacuum
		\begin{align}
			\tfrac{\dd}{\dd \barsigma} \, \Ueff ( \barsigma, \mu=0, T=0, \sdim ) \Big|_{\barsigma=\sigmaminvac}=0. \label{eq:gapequation}
		\end{align}

	Inserting \cref{eq:gap_equation_main} and rearranging the equation for nonzero $\sigmaminvac$ fixes the value of the coupling (its cutoff dependence) as
		\begin{align}
			\tfrac{1}{\ffcoupling} ={}& \dimDirac \, l_1^\Lambda ( \sigmaminvac, 0, 0, \sdim ) =	\vdistance	\label{eq:uv_scaling_quartic_coupling}
			\\
			={}& \dimDirac \, \tfrac{\Sd}{( 2 \uppi )^\sdim} \, \tfrac{1}{2} \, \bigg( \tfrac{\Gamma ( \frac{\sdim + 2}{2} ) \, \Gamma ( - \frac{\sdim + 1}{2} )}{\sdim \, \sqrt{\uppi}} \, \big( - \tfrac{\sdim + 1}{2} \big) \, | \sigmaminvac |^{\sdim - 1} +	\Vdistance	\nonumber
			\\
			& \quad + \Lambda^{\sdim} \, \Big[ \tfrac{1}{\sdim - 1} \, \tfrac{1}{\Lambda} + \tfrac{\sigmaminvac^2}{2 ( 3 - \sdim )} \, \tfrac{1}{\Lambda^3} + \mathcal{O} \big( \tfrac{1}{\Lambda^5} \big) \Big] \bigg) \, ,	\Vdistance	\nonumber
		\end{align}
	where we again assumed that $\Lambda$ is large, i.e., $\Lambda/\barsigma_0 \gg 1$ and therefore simply applied \cref{eq:expansion_l_1} for $\barsigma \to \sigmaminvac$.

\subsubsection{Renormalization of the effective potential}

	Hence, for the effective potential 	in the presence of the UV cutoff we find
		\begin{align}
			& \Ueff^\Lambda ( \barsigma, \mu, T, \sdim ) =	\vdistance	\label{eq:effective_potential_regularized_main}
			\\
			={}& \tfrac{\dimDirac}{2} \, \big[ \barsigma^2 \, l_1^\Lambda ( \sigmaminvac, 0, 0, \sdim ) - l_0^\Lambda ( \barsigma, \mu, T, \sdim ) \big] \, ,	\vdistance	\nonumber
		\end{align}
	which in vacuum evaluates to
		\begin{align}
			& \Ueff ( \barsigma, 0, 0, \sdim ) =	\vdistance
			\\
			={}& \tfrac{\dimDirac}{2} \, \tfrac{\Sd}{( 2 \uppi )^\sdim} \Big[ \tfrac{( \sdim + 1) \, \Gamma ( \frac{\sdim + 2}{2}) \, \Gamma ( - \frac{\sdim + 1}{2})}{2 \sdim \sqrt{\uppi}} \Big( - \tfrac{\sigmaminvac^{\sdim - 1} \, \barsigma^2}{2} + \tfrac{\barsigma^{\sdim + 1}}{\sdim + 1} \Big) +	\vdistance	\nonumber
			\\
			& + \Lambda^\sdim \Big[ \tfrac{\barsigma^4}{8 ( 3 - \sdim )} \, \tfrac{1}{\Lambda^3} + \mathcal{O}\big( \tfrac{1}{\Lambda^5} \big) \Big] \Big]	\vdistance	\nonumber.
		\end{align}
	The last line vanishes for $\sdim < 3$ in the limit of $\Lambda \to \infty$, see also \Rcite{Pannullo:2023cat}.
	We then obtain the renormalized homogeneous effective potential
		\begin{align}
			& \Ueff ( \barsigma, \mu, T, \sdim ) =	\Vdistance	\label{eq:potential_main_text}
			\\
			={}& \tfrac{\dimDirac}{2} \, \tfrac{\Sd}{( 2 \uppi )^\sdim} \, \bigg[ \tfrac{\Gamma ( \frac{\sdim}{2} ) \Gamma ( - \frac{\sdim + 1}{2} ) ( \sdim + 1)}{4 \sqrt{\uppi}} \, \big( \tfrac{1}{\sdim + 1} \, | \barsigma |^{\sdim + 1} - \tfrac{1}{2} \, \sigmaminvac^{\sdim - 1} \, \barsigma^2 \big)  +	\Vdistance	\nonumber
			\\
			& \quad - T \int_{0}^{\infty} \dd p \, p^{\sdim - 1} \ln \bigl[ 1 + \exp \big( - \tfrac{E + \mu}{T} \big) \bigr] +	\Vdistance	\nonumber
			\\
			& \quad + ( \mu \to - \mu ) \bigg] \, ,	\Vdistance	\nonumber
		\end{align}
	whose derivation and special limits of the parameters $\barsigma, \mu, T, \sdim $ are presented in \cref{app:effective_potential}.
	This result is equivalent to the result presented already in \Rcite{Inagaki:1994ec,Wipf:2021mns}.
	
\subsubsection{Renormalization of the two-point function}

	The unrenormalized two-point function \cref{eq:gamma2_main} contains the divergent contribution $l_1$.
	By inserting the regularized value for the coupling $\ffcoupling$ and inserting the regularized expression for $l_1$, one obtains
		\begin{align}
			& \Gamma^{(2) \Lambda} ( \barsigma, \mu, T, q, \sdim ) =	\vdistance	\label{eq:gamma2_regularized_main}
			\\
			={}& \dimDirac \big[ l_1^\Lambda ( \sigmaminvac, 0, 0, \sdim ) - l_1^\Lambda ( \barsigma, \mu, T, \sdim ) +	\vdistance	\nonumber
			\\
			&\hphantom{\dimDirac \big[}+ \tfrac{1}{2} \, ( q^2 + 4 \, \barsigma^2 ) \, l_2 ( \barsigma, \mu, T, q, \sdim ) \big] \, .	\vdistance	\nonumber
		\end{align}
	In the limit $\Lambda \to \infty$, one finds that the divergent parts are absorbed by the coupling and we find for the renormalized two-point function the expression
	\begin{align}
		& \Gamma^{(2)} ( \barsigma, \mu, T, q, \sdim ) =	\Vdistance	\label{eq:gamma2_main_text}
		\\
		={}& \tfrac{\dimDirac}{2} \, \tfrac{\Sd}{( 2 \uppi )^\sdim} \, \bigg[ \big( | \sigmaminvac |^{\sdim - 1} - | \barsigma |^{\sdim - 1} \big) \, \tfrac{\Gamma ( \frac{1 - \sdim}{2} ) \, \Gamma ( \frac{\sdim}{2} )}{2 \sqrt{\uppi}} +	\Vdistance	\nonumber
		\\
		& \quad + \int_{0}^{\infty} \dd p \, p^{\sdim - 1} \, \tfrac{1}{E} \big[ \nf \big( \tfrac{E + \mu}{T} \big) + \nf \big( \tfrac{E - \mu}{T} \big) \big] +	\Vdistance	\nonumber
		\\
		& \quad + \big( \tfrac{q^2}{4} + \barsigma^2 ) \int_{0}^{1} \dd x \int_{0}^{\infty} \dd p \, p^{\sdim - 1} \, \tfrac{1}{\tilde{E}^3} \times	\Vdistance	\nonumber
		\\
		& \qquad \times \big[ 1 - \nf \big( \tfrac{\tilde{E} + \mu}{T} \big) + \tfrac{\tilde{E}}{T} \, \big[ \nf^2 \big( \tfrac{\tilde{E} + \mu}{T} \big) - \nf \big( \tfrac{\tilde{E} + \mu}{T} \big) \big] +	\Vdistance	\nonumber
		\\
		& \qquad \quad + ( \mu \to - \mu ) \big] \bigg] \, .	\Vdistance	\nonumber
	\end{align}
	The definition of $\tilde{E}$ is given in \cref{eq:tildeenergy}, and detailed steps of the renormalization as well as special limits in the parameters $ \barsigma, \mu, T, q, \sdim$ are presented in \cref{app:bosonic_two-point_function}.

\subsection{The stability analysis}

	While the investigation of the homogeneous phase structure can be conducted via the one-dimensional minimization of the homogeneous effective potential $\Ueff$ \gls{wrt} the variable $\bar{\sigma}$, one cannot minimize the effective action for an arbitrary inhomogeneous field configuration.
	Typically, one either resorts to an ansatz for the field modulation, as, e.g., in \Rcite{Thies:2003kk,Schnetz:2004vr,Urlichs:2007zz,Basar:2009fg,Nickel:2009wj,Narayanan:2020uqt}, or conduct a stability analysis, which is the approach that we consider.
	Here, we only summarize the strategy of this technique and the final quantities.
	We refer to \Rcite{Koenigstein:2021llr} for a detailed derivation and discussion at the example of the $(1+1)$-dimensional \gls{gn} model and to \Rcite{Pannullo:2023cat,pannulloInhomogeneousPhasesMoat2023} for some further details of the stability analysis in noninteger dimensions.
	Other works that relied on this or related techniques are for example \Rcite{Pannullo:2023one,Motta:2023pks,Pannullo:2021edr,Koenigstein:2021llr,Buballa:2020nsi,Buballa:2020xaa,Fu:2019hdw,Buballa:2018hux,Tripolt:2017zgc,Braun:2014fga,Nakano:2004cd,Roscher:2015xha,Braun:2014fga,Braun:2015fva}.

	The strategy of this technique is to apply inhomogeneous perturbations to a homogeneous field configuration and expand the effective action in powers of this perturbation.
	When applying this expansion at the global homogeneous minimum, one finds that the first nonzero correction is the second order term quadratic in the perturbations.
	A negative sign of its coefficient signals that an inhomogeneous field configuration is energetically favored over the homogeneous expansion point.
	This coefficient is given by the bosonic two-point function $\gtwo$ with the external, spatial bosonic momentum corresponding to the momentum of the inhomogeneous perturbation.
	On a strictly formal level, we are therefore simply searching for spatial momenta $q$, where the two-point function \labelcref{eq:gamma2_main_text} takes negative values, if it is evaluated at the global minimum of the potential \labelcref{eq:potential_main_text}.

	Closely related to the \gls{ip} is the so-called moat regime \cite{Pisarski:2021qof,Rennecke:2023xhc}, which is characterized by a two-point function featuring a minimum at a finite momentum.
	This is also realized in an \gls{ip}, where, however, the minimum of the two-point function is necessarily negative.
	A simple criterion for the detection of the moat regime is that the wave-function renormalization \labelcref{eq:zdef} evaluated at the homogeneous minimum of the potential \labelcref{eq:potential_main_text} assumes a negative value.\footnote{This criterion assumes that there are no minima in the two-point function at finite momentum that are separated from the origin by a local maximum.
	This situation would also correspond to a moat regime, but blind to our criterion.
	Such a situation can indeed occur in non-renormalizable (pseudo)scalar four-fermion models like the $3+1$-dimensional Nambu-Jona-Lasinio model, when considering chemical potentials which are larger than the regulator \cite{pannulloInhomogeneousPhasesMoat2023}.
	However, we are not aware that these models also exhibit such properties in the fully renormalized limit and, thus, the wave-function renormalization appears as an appropriate criterion.}

\section{Results}
\label{sec:results}

	The results as discussed in the following are an extension of the results presented in \Rcite{Pannullo:2023cat}, which conducted the stability analysis of the $(d+1)$-dimensional \gls{gn} model at $T=0$.
	Therefore, we concentrate our presentation on the effects by nonzero $T$ and on the phase diagram as a function of $\sdim$ as a whole.
	We refer to \Rcite{Pannullo:2023cat} for further results, which analyze the $d$-dependence of the stability analysis.
	Furthermore, we refer to \Rcite{Koenigstein:2021llr} for a detailed discussion of the results for $d=1$ and to \Rcite{Buballa:2020nsi,Pannullo:2023one} for $d=2$.

\subsection{The two-point function}

\begin{figure}
			\centering
			\includegraphics{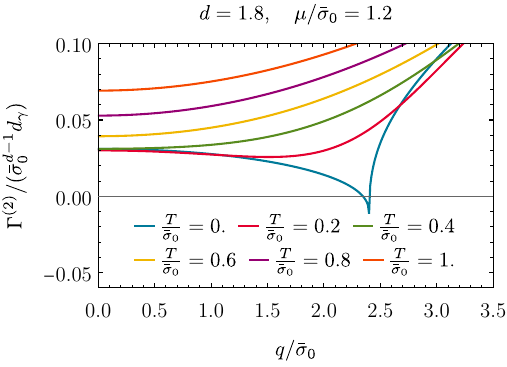}
			\caption{The two-point function $\gtwo$ evaluated at the global homogeneous minimum $\bar{\Sigma} ( \mu, T )$ at $\mu/\sigmaminvac = 1.2$ and $d = 1.8$ for various temperatures $T/\sigmaminvac$ as a function of the bosonic momentum $q/\sigmaminvac$.}
			\label{fig:twp_TScan_d18}
		\end{figure}

\begin{figure}
			\centering
			\includegraphics{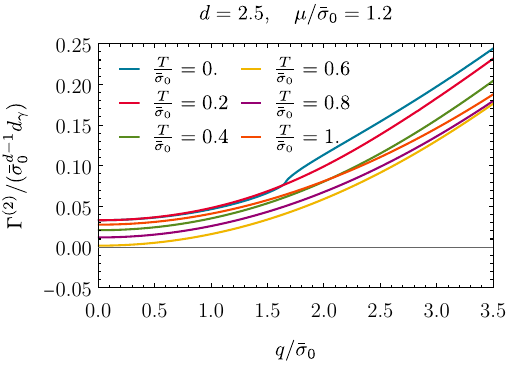}
			\caption{The two-point function $\gtwo$ evaluated at the global homogeneous minimum $\bar{\Sigma} ( \mu, T )$ at $\mu/\sigmaminvac = 1.2$ and $d = 2.5$ for various temperatures $T/\sigmaminvac$ as a function of the bosonic momentum $q/\sigmaminvac$.}
			\label{fig:twp_TScan_d25}
		\end{figure}

	We start the discussion by providing two example plots of the bosonic two-point function as a function of the bosonic momentum $q$.
	We chose $\mu/\sigmaminvac=1.2$ and various temperatures for $d=1.8$ in \cref{fig:twp_TScan_d18} and for $d=2.5$ in \cref{fig:twp_TScan_d25}, because most of the relevant effects are visible for these choices of $\mu$ and $T$ and the two values of $d$ are representative for the behavior for $1 \leq \sdim <2$ and $2 < \sdim$ respectively.

	We find that the non-analytic points at $q=2\mu$, which are present at $T=0$, see \Rcite{Pannullo:2023cat}, are completely smoothed out already at rather low temperatures.
	For $\sdim < 2$, the former nonanalytic point turns into a nontrivial global minimum for some temperature range, while at $\sdim > 2$ no structure reminiscent of the cusp at $T=0$ remains.
	In this way for $\sdim < 2$ the instability towards an \gls{ip} signaled by a negative $\gtwo$ at $T=0$ (blue curve in \cref{fig:twp_TScan_d18}) vanishes at some $T > 0$.
	A moat regime remains signaled by the negative curvature of $\gtwo$ at $q=0$ (red curve in \cref{fig:twp_TScan_d18}), which results in a nontrivial global positive minimum.
	For large temperatures we universally find a convex shape with no remains of the \gls{ip} or a moat regime.
	Before we close the discussion, let us remark that the $T$-order of the curves in \cref{fig:twp_TScan_d25} is correct:
	The offset of the two-point function is the bosonic curvature mass, which vanishes at the second order \gls{pt}.
	At $\mu/\sigmaminvac = 1.4$ one starts in the \gls{hbp} at $T/\sigmaminvac = 0$ and crosses the \gls{pt} to the \gls{sp} at approximately $T/\sigmaminvac = 0.6$ by increasing $T$, see also \cref{fig:fullPD_1_8}.
	Of course, we could have plotted the two-point function for $\sdim = 2.5$ in a region, where the global minimum of the potential is trivial, hence for some $\mu/\sigmaminvac \gtrsim 1.5$ similar to $\sdim = 1.8$.
	This, however, would have been even less interesting and we therefore opted for a comparison of $\sdim = 1.8$ and $\sdim = 2.5$ at the same $\mu/\sigmaminvac$.

\subsection{The wave-function renormalization}

\begin{figure}
			\centering
			\includegraphics{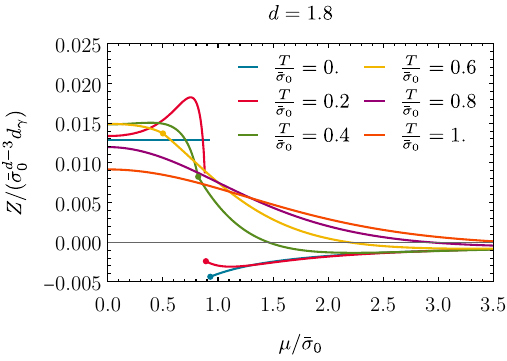}
			\caption{The wave-function renormalization $Z$ evaluated at the minimum of the potential for $d = 1.8$ for various temperatures $T$ as a function of the chemical potential $\mu$.
			The dots mark the chemical potential, where the spatially homogeneous global minimum of the potential turns (non)trivial, see \cref{fig:fullPD_1_8}.}
			\label{fig:z_MuScan_1_8}
		\end{figure}

\begin{figure}
			\centering
			\includegraphics{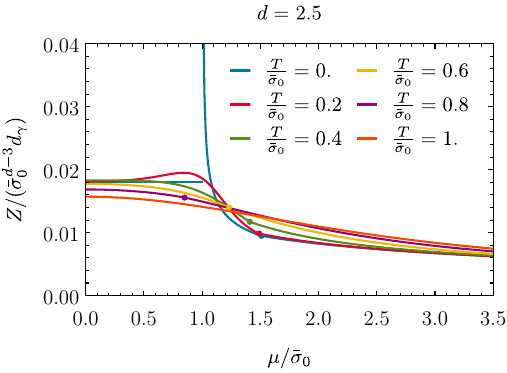}
			\caption{The wave-function renormalization $Z$ evaluated at the minimum of the potential for $d = 2.5$ for various temperatures $T$ as a function of the chemical potential $\mu$.
			The $(T = 0)$-curve only stays finite due to finite computational $\mu$-resolution.
			The dots mark the chemical potential, where the spatially homogeneous global minimum of the potential turns (non)trivial, see \cref{fig:fullPD_1_8}.}
			\label{fig:z_TScan_2_5}
		\end{figure}

	The second quantity of interest is the wave-function renormalization $Z$, which serves as the indicator for the moat regime.
	It is shown as a function of the chemical potential evaluated at the global homogeneous minimum $\bar{\Sigma} ( \mu, T )$ at various temperatures for $d=1.8$ in \cref{fig:z_MuScan_1_8} and for $d=2.5$ in \cref{fig:z_TScan_2_5}.
	The two values of $\sdim$ are again representative for the behavior for $1 \leq \sdim < 2$ and $2 < \sdim$ respectively.
	The dots on the curves indicate the chemical potential that corresponds to the homogeneous \gls{pt} at the given temperature.
	

	For $\sdim < 2$, one finds at $T = 0$ that $Z$ is constant for small $\mu/\sigmaminvac$ and jumps to a negative value at the homogeneous \gls{pt}.
	The constant behavior is a consequence of the Silver blaze property as \cite{Cohen:2003kd,Marko:2014hea,Braun:2017srn,Braun:2020bhy}, which is no longer fulfilled for nonzero $T$.
	The jump vanishes exactly at temperatures, where one does no longer find a first order \gls{pt} in $\mu$-direction under the assumption of homogeneous condensation, i.e., at the temperature of the critical endpoint, see also \cref{fig:fullPD_1_8}.
	Above this temperature, one finds that the chemical potential beyond which the wave-function renormalization is negative moves to higher values.
	This is the behavior that we would expect based on the results for $\sdim=1$ as presented in \Rcite{Koenigstein:2021llr}.
	However, one finds a moat regime for all temperatures for sufficiently large $\mu/\sigmaminvac$.

	For $\sdim > 2$, one finds at $T=0$ that $Z$ is constant for small $\mu$ and diverges at $\mu=\sigmaminvac$, which is again a manifestation of Silver Blaze.
	This nonanalytic behavior is smoothened already for small temperatures and one finds that $Z$ smoothly changes with $\mu$.
	However, most importantly for the present investigation is, that the wave-function renormalization is always positive for arbitrary values of $T$ and $\mu$.
	This implies the total absence of an \gls{ip} and also the total absence of a moat regime for $\sdim > 2$.

\begin{figure}
		\centering
		\includegraphics[width=1\linewidth]{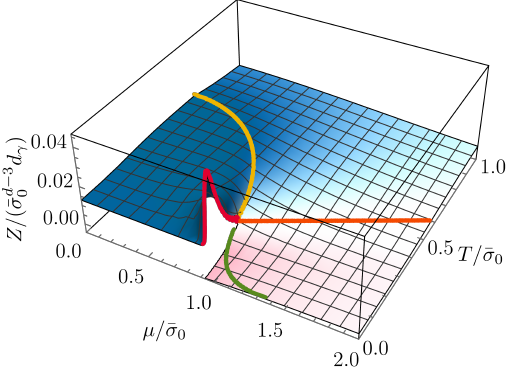}
		\caption{The wave-function renormalization $Z$ at $d=1.8$ in the $(\mu,T)$-plane.
			The yellow curve corresponds to the second order \gls{hbp}-\gls{sp} \gls{pt}, the red curve to the first order \gls{hbp}-\gls{sp} \gls{pt}, the green curve to the second order \gls{ip}-\gls{sp} \gls{pt}, and the orange curve separates regions of positive and negative wave-function renormalization.}
		\label{fig:z_TMuPlane_1_8}
	\end{figure}

\begin{figure}
		\centering
		\includegraphics[width=1\linewidth]{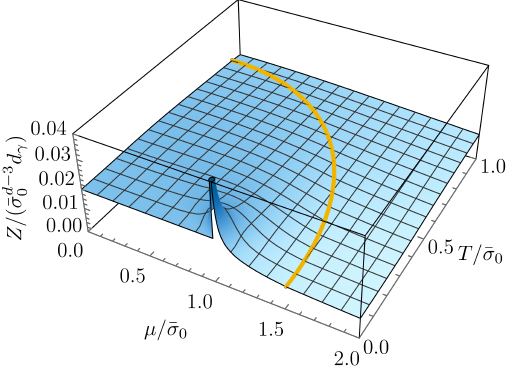}
		\caption{The wave-function renormalization $Z$ at $d=2.5$ in the $(\mu,T)$-plane.
			The yellow curve marks the second order \gls{hbp}-\gls{sp} \gls{pt}}
		\label{fig:z_TMuPlane_2_5}
	\end{figure}

	For the sake of clearness, we also provide two density plots of the wave-function renormalization in the $\mu,T$-plane for $\sdim = 1.8$ and $\sdim = 2.5$ in \cref{fig:z_TMuPlane_1_8,fig:z_TMuPlane_2_5}.
	(The \cref{fig:z_MuScan_1_8,fig:z_TScan_2_5} are just sections of these density plots at constant $T$.)
	Here, it is again clearly visible that there is no moat regime and \gls{ip} for $\sdim>2$, where $Z$ is always positive.
	For $\sdim = 1.8$, however, we find a similar structure as for $\sdim = 1$ (see \Rcite{Koenigstein:2021llr}).
	The moat regime is present in the region of the \gls{ip} and \gls{sp} below the straight line that originates from $\mu = T = 0$ and passes through the critical point.
	This straight line is associated with a vanishing  quartic coefficient of the effective potential.
	Hence, the coefficient in front of $\barsigma^4$ changes its sign along this line as a function of $\mu$ and $T$, if one expands \cref{eq:potential_main_text} in $\barsigma$, see \Rcite{Actor:1986zf,Ahmed:2018tcs,Stoll:2021ori}.
	It can be shown \cite{Actor:1986zf} that this coefficient is proportional to the bosonic wave-function renormalization \labelcref{eq:wave-function_renormalization_main}, if the wave-function renormalization $z$ is evaluated at the trivial point $\barsigma/\sigmaminvac = 0$.
	This is the correct evaluation point only in the \gls{sp}, which is however the important region for the moat regime.
	Hence, the straight line separates the regime of negative and positive $Z$ in the \gls{sp}.


\subsection{The phase diagram}
	\label{sec:result_pd}

\begin{figure}
	\centering
	\includegraphics{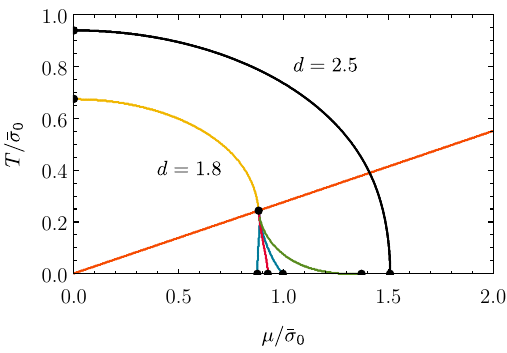}
	\caption{The phase diagram at $d=1.8$ (colored lines) and $\sdim = 2.5$ (black line) in the $(\mu,T)$-plane.
	The plot shows for $\sdim = 1.8$: the second order \gls{pt} (\gls{hbp} $\leftrightarrow$ \gls{sp}) in yellow, first order phase \gls{pt} (\gls{hbp} $\leftrightarrow$ \gls{sp}) red, the second order \gls{pt} (\gls{ip} $\leftrightarrow$ \gls{sp}) in green, the spinodal lines in blue, the line of vanishing quartic coefficient of the potential ($=$ vanishing $z(\bar{\sigma}=0)$) in orange.
	The black line is the second order \gls{pt} (\gls{hbp} $\leftrightarrow$ \gls{sp}) for $\sdim = 2.5$.
	Black dots mark the endpoints of the lines.
	}
	\label{fig:fullPD_1_8}
\end{figure}

	Next, we turn to the full phase diagram of the \gls{gn} model as a function of the spatial dimension $\sdim$.
	Indeed, already in \cref{fig:z_TMuPlane_1_8,fig:z_TMuPlane_2_5} we basically plotted the phase structure for $\sdim = 1.8$ and $\sdim = 2.5$, which served as examples for $1 \leq \sdim < 2$ and $\sdim > 2$.
	For $1 \leq \sdim < 2$ one always finds a first order \gls{pt} between \gls{hbp} and \gls{sp} at low temperatures and a second order \gls{pt} for large temperatures, if one assumes spatially homogeneous condensation in the entire $\mu,T$-plane.
	On the other hand, for $\sdim \geq 2$ the \gls{pt} between the \gls{hbp} and \gls{sp} is always of second order (except for $\sdim=2$ and $T = 0$ \cite{Klimenko:1987gi,Rosenstein:1988dj}).
	This result was already found in \Rcite{Inagaki:1994ec}, where it was also observed that the critical point moves down to $T = 0$ and that the \gls{hbp} enlarges, while going continuously from $\sdim = 1$ to $\sdim = 2$.
	However, when allowing for the formation of spatially inhomogeneous condensates and searching for these via the stability analysis, this picture is modified for $1 \leq \sdim < 2$, while nothing happens at $\sdim \geq 2$.
	Already at $T = 0$ it was found in \Rcite{Pannullo:2023cat} that the stability analysis reveals an \gls{ip} for $1 \leq \sdim < 2$.
	This phase extends to $\mu/\sigmaminvac = \infty$ for $\sdim = 1$ at $T = 0$ but shrinks and has a second order \gls{pt} to the \gls{sp} at some finite critical $\mu$ when $1 < \sdim < 2$.
	The \gls{pt} between the \gls{hbp} and the \gls{ip} cannot be resolved correctly with this method as is discussed in detail in \Rcite{Braun:2014fga,Koenigstein:2021llr}. 
	Still, for $\sdim = 1$ the analytic solution is well-known \cite{Thies:2003br,Schnetz:2004vr} and served as a test field for the stability analysis in \Rcite{Koenigstein:2021llr}.
	For a direct comparison, of the situation in $1 \leq \sdim < 2$ and $\sdim > 2$, we again used $\sdim = 1.8$ and $\sdim = 2.5$ and plotted both phase diagrams together in \cref{fig:fullPD_1_8}.
	For reference, we also included the spinodal lines (the lines that engulf the region in the phase diagram, where the effective potential has a global and local minimum/minima), plotted in blue.

\begin{figure}
	\centering
	\includegraphics{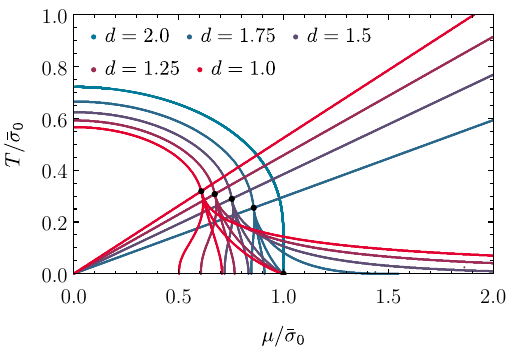}
	\caption{The boundaries of the \gls{hbp} (assuming homogeneous condensation), the boundary \gls{ip} to the \gls{sp}, and the moat regime, and the spinodal lines for spatial dimensions $d \in \{1.0,1.25,1.5,1.75,2.0\}$ in the $(\mu,T)$-plane.}
	\label{fig:fullPD}
\end{figure}

\begin{figure}
	\centering
	\includegraphics{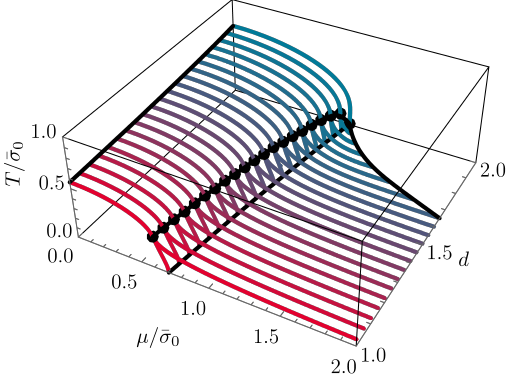}
	\caption{The first and second order phase boundary of the \gls{hbp}-\gls{sp} \gls{pt} for a translational invariant bosonic field and the phase boundary between the \gls{ip} and \gls{sp} in the $(\mu,T,d)$-space.
	Different colored lines correspond to different values of $\sdim$.
	(The \gls{hbp}-\gls{ip} \gls{pt} is not plotted and not detectable within our approach.)}
	\label{fig:fullPD_3D}
\end{figure}

	However, to really observe the effect of dimensionality on the \gls{ip} and the moat regime, we prepared \cref{fig:fullPD,fig:fullPD_3D}, which clearly show that the phase diagram for $1 < \sdim < 2$ is similar to the phase diagram at $\sdim = 1$ (except for the finite extent of the \gls{ip} at $T = 0$) and that $\sdim = 2$ is the strict upper bound for the existence of an \gls{ip}.
	Still, it is remarkable that the \gls{ip} and the moat regime vanish rather slowly as a function of $\sdim$ and an \gls{ip} is still clearly visible for our last plotted curve at $\sdim = 1.95$.
	Nonetheless, this behavior was actually expected, because the same slow convergence was already observed for the critical point in \Rcite{Inagaki:1994ec}, which turns into the Lifshitz point, if one allows for spatially inhomogeneous condensation.
	Let us remark at this point that we did not prepare extra plots for the situation at $2 < \sdim < 3$, since we do not find any signal of spatially inhomogeneous condensation and/or a moat regime for any $T$ and $\mu$ (this was already observed at $T=0$ in \Rcite{Pannullo:2023cat}).
	Therefore, the situation of exclusively spatially homogeneous condensation is already fully covered by \Rcite{Inagaki:1994ec} with which our results agree.
	We believe that the example of $\sdim = 2.5$ is absolutely sufficient within this work to understand the situation for $2 < \sdim < 3$.

	We close this section by summarizing that we find moat regimes and phases, where the condensate varies in space, in the \gls{gn} model for $N \to \infty$ for $1 \leq \sdim < 2$, while they are absent for $\sdim \geq 2$.
	
\section{Conclusions and outlook}
\label{sec:conclusions_and_outlook}

	Finally, we want to summarize our results, draw several conclusions, and provide an outlook to possible follow-up projects.

\subsection{Summary}

	In the present work, we investigated the phase diagram of the \gls{gn} model at (non)zero fermion chemical potential $\mu$ and (non)zero temperature $T$ in the limit of an infinite number of fermion species, $N \to \infty$, as a function of the spatial dimension $\sdim$.
	We focused on $1 \leq \sdim < 3$, where the model is renormalizable and solely depends on the fixation of a single dimensionful parameter.
	We used the vacuum fermion mass to fix the scales and worked in the renormalized limit.
	The focus of this work is the detection and the dependence on the number of spatial dimensions of the \glspl{ip} and a possible moat regime.
	We found that the well-known result for $\sdim = 1$ \cite{Thies:2003br,Schnetz:2004vr,Schnetz:2005ih,Schnetz:2005vh} generalizes to $1 < \sdim < 2$ and one detects an instability of the \gls{sp} that signals the presence of an \gls{ip}.
	Furthermore, one always finds an even larger moat regime at large chemical potential.
	Both, the \gls{ip} and the moat regime vanish, when $\sdim$ approaches $\sdim = 2$.
	For $\sdim \geq 2$ we do not find any indication of an \gls{ip} in terms of an instability.
	The same applies to the moat regime.
	While, we cannot exclude any kind of inhomogeneous condensation, which is not detectable via a stability analysis, such a situation is, however, highly unlikely to be present in this model.
	The reason is that the \gls{pt} between the \gls{sp} and an \gls{ip} is generally expected to be of second order, which enables the use of this method (see \Rcite{Koenigstein:2021llr} for a detailed discussion about the range of validity of this method).

	Apart from these novel results, we implicitly and explicitly confirmed various existing literature results for the \gls{gn} model for integer $\sdim = 1$ \cite{Dashen:1974xz,Wolff:1985av,Thies:2003kk,Schnetz:2004vr} and $\sdim = 2$ \cite{Klimenko:1987gi,Rosenstein:1988dj,Urlichs:2007zz,Narayanan:2020uqt,Buballa:2020nsi} as well as some of the results from \Rcite{Inagaki:1994ec} for continuous $\sdim$. 

	We also provide several appendices with detailed material that may be of general use for follow-up or related projects.

\subsection{Conclusion}

	Already in \Rcite{Pannullo:2023cat} we speculated about the relevance of the number of spatial dimensions $\sdim$ on the formation of spatially inhomogeneous ground states in the \gls{gn} and other models.
	Furthermore, \Rcite{Narayanan:2020uqt,Buballa:2020nsi,Pannullo:2022eqh} showed that the effects of the presence of a finite regulator or an effective \gls{uv}/\gls{ir} cutoff in terms of a spatial lattice or a finite spatial box play an important role for the presence/absence of spatially inhomogeneous condensation. 
	Here, we want to continue this discussion and believe that the present investigation sheds light on the previous findings.
	Let us therefore briefly summarize the findings for the \gls{gn} model at $N \to \infty$ up to this point:
	In $\sdim = 1$ there is an exact solution for the phase diagram and one finds spatially inhomogeneous condensation \cite{Thies:2003br,Schnetz:2004vr}.
	This feature seems to be robust at $N \to \infty$ even in the presence of a spatial lattice etc. \cite{Wagner:2007he,Heinz:2015lua}.
	For $\sdim = 2$ and for $N \to \infty$ there is also an exact solution for the phase diagram, but there are no \glspl{ip} in the renormalized limit \cite{Buballa:2020nsi,Narayanan:2020uqt}.
	However, in the presence of some \gls{uv} or \gls{ir} regulator/cutoff one recovers an \gls{ip} and a phase diagram that has some similarities with the situation in $\sdim = 1$ \cite{Buballa:2020nsi,Narayanan:2020uqt}.
	The size of the \gls{ip} and the shape of the \gls{hbp} thereby strongly depends on the value of the regulator/cutoff and one finds results that are closer to $\sdim = 1$ or $\sdim = 2$ depending on the strength of the regularization.
	In $\sdim = 3$ there are several models that are similar to the \gls{gn} model at $N \to \infty$ which support the presence of spatially inhomogeneous condensation, while the extent of the \gls{ip} usually depends on the cutoff/model parameters (renormalization like in this work is not possible and the phase diagram usually depends on at least two parameters).
	If we compare these results to our findings, we come to the conclusion that spatially inhomogeneous condensation seems to be a dimensional effect.
	For the studies discussed previously and for the present study the situation is always the same:
	As soon as the (effective) dimensionality of the model is reduced to $1 + \Delta \sdim$ dimensions, where $\Delta \sdim \in [ 0, 1 )$, we find an \gls{ip}.
	However, as soon as there are two or more full-featured spatial dimensions available, the \gls{ip} vanishes.
	In fact, it does not play a role if the number of spatial dimensions is directly reduced via dimensional regularization or as worked out in the present study using $\sdim$ as a continuous parameter or even via an \gls{uv}/\gls{ir} cutoff as, e.g., in \Rcite{Nakano:2004cd,Nickel:2009wj,Pannullo:2022eqh,pannulloInhomogeneousPhasesMoat2023,Buballa:2020nsi,Narayanan:2020uqt,Carignano:2014jla}. 
	In particular, the latter case can be viewed as a reduction of a full dimension to a fractional/part of a dimension by restricting the system to a finite spatial box (\gls{ir} cutoff) or coarse lattice (\gls{uv} cutoff).
	A similar effect is observed in a recent work \Rcite{Lenz:2023gsq}, which analyzed the homogeneous phase structure of the $2+1$-dimensional \gls{gn} model in a finite volume.
	The finite volume causes the critical endpoint to be located at a finite temperature, which is a feature that is limited to $\sdim<2$ in the infinite volume.
	
	These results also seem to be in line with the observation that one-dimensional ansatz functions for inhomogeneous condensates usually appear to be the most promising and energetically favored solutions, if an \gls{ip} is present at all \cite{Urlichs:2007zz,Nickel:2009wj,Carignano:2012sx,Pannullo:2021edr}.

	Of course, at this point the immediate question that arises is:
	What is the underlying nature and physical principle behind this strong relation to a single spatial dimension?
	So far, we were not able to come up with a conclusive answer, but we hope that this work might be an important step to start the search in the right direction.
	The Peirls instability, which is the origin of the \gls{ip} in $\sdim=1$ \cite{Schnetz:2004vr,Thies:2006ti} and a one-dimensional effect that cannot be directly generalized to higher dimensional systems, might be a good starting point in terms of a physics understanding and correct interpretation.

\subsection{Outlook}

	Now, that we have mostly settled the situation for the \gls{gn} model at $N \to \infty$ there are basically four main directions to proceed.
	
\subsubsection{One-dimensional ansatz functions}

	It was found in $\sdim=1$, that the stability analysis is not able to detect the portion of the \gls{ip}, where it is energetically favored, but the homogeneous expansion point $\bar{\Sigma}$ is finite \cite{Koenigstein:2021llr}.
	As mentioned in \cref{sec:result_pd}, we expect the same situation to occur for $1<\sdim<2$ in our calculations with some part of the \gls{ip} in the vicinity of the first order \gls{pt} between \gls{hbp} and \gls{sp} to be missing.
	A way to improve on this would be to consider a one-dimensional ansatz function embedded in the $\sdim$-dimensional space.
	Such a procedure was considered in $(3+1)$-dimensional models in \Rcite{Nickel:2009wj}, which treats the perpendicular space in such a general way that it can be generalized from $d_\perp=2$ to noninteger $d_\perp=d-1$ dimensions.
	It is expected that the additional portion of the \gls{ip} is not particularly large, because it quite limited in size in $\sdim=1$ and likely shrinks even further with increasing $\sdim$.
	Nevertheless, this step would yield the \textit{complete} phase diagram of this model.\footnote{This is based on the assumption that also in noninteger dimensions $1$-dimensional kink-antikink modulations motivated from the solution of the \gls{gn} model in $d=1$ are the preferred shape of the inhomogeneous condensate.}
	
\subsubsection{Finite regulator or finite volume }
	
	To solidify the concept of effective dimensionality, it would be fascinating to carry out the present investigation not in the renormalized limit, but at a finite  \gls{uv} regulator. 
	In this way, one could (a) connect smoothly to the $3+1$ model results by extending our analysis to $\sdim=3$ and (b) one could investigate the interplay between the explicit number of spatial dimension $\sdim$ and the effective dimensional reduction introduced by the \gls{uv} regulator.
	The latter could also be investigated by considering a finite volume, which introduces an \gls{ir} regularization that should also lead to an effective dimensional reduction.

\subsubsection{Finite \texorpdfstring{$N$}{N}}

	While the $N\to\infty$ limit was essential to investigate the analytic structure of the \gls{gn} model for noninteger $\sdim$, this semi-classical limit is fairly different from the behavior that we would expect of a \gls{qft}.
	The general observation is that bosonic quantum fluctuations as they would occur for finite $N$ weaken ordered phases in such models (see, e.g.~\Rcite{Stoll:2021ori,Lenz:2020bxk,Tripolt:2017zgc,Scherer:2013pda,Lakaschus2021}) and therefore likely do not enable the emergence of an \gls{ip} for $\sdim\geq2$.	
	
	For $\sdim<2$, it is highly likely that the \gls{ip} vanishes altogether.
	The most pathological aspect of the $N\to\infty$ limit is the fact that it circumvents the Coleman-Mermin-Wagner-Hohenberg-Berezinskii theorem \cite{Mermin:1966,Wagner:2007he,Berezinsky:1970fr,Coleman:1973ci,Hohenberg:1967} (or related arguments for discrete symmetries \cite{Dashen:1974xz,Landau:1980mil,Landau:1937obd,Koenigstein:2023wso,Witten:1978qu}) in $\sdim=1$, which would normally forbid any type of condensation at a finite temperature.
	Accordingly, it was found in \Rcite{Stoll:2021ori,Koenigstein:2023wso}, that there is no symmetry breaking at finite $T$ and $N$ in $\sdim=1$.
	These effects likely suppress any \gls{ip} for $\sdim<2$, which all in all suggests that these models do not exhibit an \gls{ip} in any dimensions for finite $N$.
	
	Nevertheless, it might be interesting to consider this model for finite $N$.
	While there is no condensation in $\sdim=1$ at finite $T$, one finds an \gls{hbp} in $\sdim=2$ \cite{Scherer:2013pda}.
	Thus, considering the \gls{gn} model for noninteger $\sdim$ might be instructive to observe how the theory evolves from a system without any symmetry breaking to the system with a broken symmetry.
	A functional method that admits the formulation of the theory for an arbitrary $\sdim$ such as the Functional Renormalization Group \cite{Wetterich:1992yh,Kopietz:2010zz,Dupuis:2020fhh} might be the optimal framework for extending our analysis to finite $N$.
	
\subsubsection{Consequences for higher dimensional models and QCD}
	
	Our analysis shows that four-fermion models with scalar interaction channels in the limit of $N \to \infty$ only exhibit an \gls{ip} for $d<2$ or via an effective reduction of the dimensionality of the system.
	This suggests that an \gls{ip} might exist in \gls{qcd} only when the low-energy behavior of \gls{qcd} is not only described by scalar-pseudoscalar four-fermion interactions, but by interactions that would admit an \gls{ip} for higher dimensions in other than the scalar and pseudoscalar channel.
	This is most likely the case at finite chemical potential, where it was found that the relevant interaction channels are diquark interactions \cite{Ruester:2005jc,Braun:2019aow}, which have not been systematically studied with respect to the \gls{ip}.
	Moreover, in this regime vector interactions become relevant, which were found to mix with scalar modes at finite densities and to play an important role in the homogeneous phase transition near the critical endpoint in \gls{qcd} \cite{Haensch:2023sig}.
	Such a mixing might even induce an instability that results in a spatial modulation of the condensates \cite{Schindler:2019ugo,Schindler:2021otf}.	
	Another mechanism that could cause the existence of an \gls{ip} in \gls{qcd} is that it has in fact a lower effective dimension, \eg, caused by additional strong magnetic fields.
	While these are aspects that are yet to be understood, they certainly imply important questions that should be answered in an effort to investigate \glspl{ip} and the moat regime in \gls{qcd}.
	However, there are also indications that a possible \gls{ip} would be completely destabilized by the Goldstone bosons from chiral symmetry breaking \cite{Pisarski:2020dnx}.

\begin{acknowledgments}
	A.~K.\ and L.~P.\ thank J.~Braun, H.~Gies, G.~Markó, R.~D.~Pisarski, D.~H.~Rischke, M.~J.~Steil, J.~Stoll, M.~Wagner, M.~Winstel, A.~Wipf, N.~Zorbach for fruitful discussions about this work.
	A.~K.\ and L.~P.\ thank R.~Pisarski, A.~Wipf,  M.~Winstel and N.~Zorbach for useful comments on the manuscript.
	A.~K.\ and L.~P.\ especially thank S.~Floerchinger and G.~Endrődi for valuable discussions and for their general support at the TPI in Jena and the faculty of physics at the University of Bielefeld, respectively.

	A.~K.\ and L.~P.\ also like to thank M.~J.~Steil, because the adaptive mesh-refinement algorithm that was used to generate the data of the phase diagrams is based on his work and code.
			
	A.~K.\ and L.~P.\ acknowledge support from  the \textit{Helmholtz Graduate School for Hadron and Ion Research}, the \textit{Giersch Foundation} and the \textit{Deutsche Forschungsgemeinschaft} (DFG, German Research Foundation) through the Collaborative Research Center TransRegio CRC-TR 211 “Strong-interaction matter under extreme conditions” -- project number 315477589 -- TRR 211.
	
	All numeric results as well as the figures in this work were obtained and designed using \texttt{Mathematica} \cite{Mathematica:13.0}.
\end{acknowledgments}

\appendix

\section{Conventions}

\subsubsection{Fourier transformations}

	In this work, we use the following conventions for Fourier transformations.
	For the bosonic field we have
		\begin{align}
			& \varphi ( \tau, \vec{x} \, ) =	\Vdistance
			\\
			={}& T \sum_{n = - \infty}^{\infty} \int_{- \infty}^{\infty} \frac{\dd^\sdim p}{( 2 \uppi )^\sdim} \, \tilde{\varphi} ( \omega_n, \vec{p} \, ) \, \ee^{+ \ii ( \omega_n \tau + \vec{p} \cdot \vec{x} \, )} \, ,	\Vdistance	\nonumber
			\\
			& \tilde{\varphi} ( \omega_n, \vec{p} \, ) =	\Vdistance
			\\
			={}& \int_{0}^{\frac{1}{T}} \dd \tau \int_{- \infty}^{\infty} \frac{\dd^\sdim x}{( 2 \uppi )^\sdim} \, \varphi ( \tau, \vec{x} \, ) \, \ee^{- \ii ( \omega_n \tau + \vec{p} \cdot \vec{x} \, )} \, ,	\Vdistance	\nonumber
		\end{align}
	while the fermion fields are Fourier-transformed according to
		\begin{align}
			& \psi ( \tau, \vec{x} \, ) =	\Vdistance
			\\
			={}& T \sum_{n = - \infty}^{\infty} \int_{- \infty}^{\infty} \frac{\dd^\sdim p}{( 2 \uppi )^\sdim} \, \tilde{\psi} ( \nu_n, \vec{p} \, ) \, \ee^{+ \ii ( \nu_n \tau + \vec{p} \cdot \vec{x} \, )} \, ,	\Vdistance	\nonumber
			\\
			& \tilde{\psi} ( \nu_n, \vec{p} \, ) =	\Vdistance
			\\
			={}& \int_{0}^{\frac{1}{T}} \dd \tau \int_{- \infty}^{\infty} \frac{\dd^\sdim x}{( 2 \uppi )^\sdim} \, \psi ( \tau, \vec{x} \, ) \, \ee^{- \ii ( \nu_n \tau + \vec{p} \cdot \vec{x} \, )} \, ,	\Vdistance	\nonumber
			\\
			& \barpsi ( \tau, \vec{x} \, ) =	\Vdistance
			\\
			={}& T \sum_{n = - \infty}^{\infty} \int_{- \infty}^{\infty} \frac{\dd^\sdim p}{( 2 \uppi )^\sdim} \, \tilde{\barpsi} ( \nu_n, \vec{p} \, ) \, \ee^{- \ii ( \nu_n \tau + \vec{p} \cdot \vec{x} \, )} \, ,	\Vdistance	\nonumber
			\\
			& \tilde{\barpsi} ( \nu_n, \vec{p} \, ) =	\Vdistance
			\\
			={}& \int_{0}^{\frac{1}{T}} \dd \tau \int_{- \infty}^{\infty} \frac{\dd^\sdim x}{( 2 \uppi )^\sdim} \, \barpsi ( \tau, \vec{x} \, ) \, \ee^{+ \ii ( \nu_n \tau + \vec{p} \cdot \vec{x} \, )} \, .	\Vdistance	\nonumber
		\end{align}
	Hereby, the corresponding Matsubara frequencies for the discretized energies are
		\begin{align}
			&	\omega_n = 2 \uppi T n \, ,	&&	\nu_n = 2 \uppi T \big( n + \tfrac{1}{2} \big) \, ,	\label{eq:matsubara_frequency}
		\end{align}
	which stem from the (anti-)periodic boundary conditions in $\tau$-direction at $\tau = \frac{1}{T}$ for (fermions) bosons.

\subsubsection{Fermi-Dirac distribution function}

	We define the Fermi-Dirac distribution function as follows  \cite{Dirac:1926jz,Fermi:1926}
		\begin{align}
			\nf ( x ) = \tfrac{1}{\ee^x + 1} = \tfrac{1}{2} \, \big[ 1 - \tanh \big( \tfrac{x}{2} \big) \big] \, .	\label{eq:fermi-dirac-distribution}
		\end{align}
	Especially for the numeric implementation we exclusively use the representation in terms of $\tanh$.
	In addition, we present two useful identities, which are also part of the derivation of the explicit analytic expressions in this work as well as the numeric implementation,
		\begin{align}
			\nf^\prime ( x ) ={}&  \nf^2 ( x ) - \nf ( x ) = - \tfrac{1}{4 \cosh^2 ( \frac{x}{2} )} \, ,	\vdistance	\label{eq:fermi-dirac-distribution_d}
			\\
			\nf^{\prime \prime} ( x ) ={}& 2 \, \nf^3 ( x ) - 3 \, \nf^2 ( x ) + \nf ( x ) = \tfrac{\sinh ( \frac{x}{2} )}{4 \cosh^3 ( \frac{x}{2} )} \, .	\vdistance	\label{eq:fermi-dirac-distribution_dd}
		\end{align}
	``Primes'' denote derivatives \gls{wrt} $x$.

	For the derivation of the zero-temperature limits of some formulae of this work, we repeatedly need the following limits.
		\begin{align}
			\lim_{T \to 0} \nf \big( \tfrac{E \pm | \mu |}{T} \big) =
			\begin{cases}
				0 \, ,	\vphantom{\Big(\Big)}
				\\
				\Theta \Big( \tfrac{| \mu |}{E} - 1 \Big) \, .
			\end{cases} 
		\end{align}
	Here, $E$ is the energy, $\mu$ the chemical potential, and $\Theta$ is the Heaviside function.
	In addition,
		\begin{align}
			& \lim_{T \to 0} \tfrac{E}{T} \, \big[ \nf^2 \big( \tfrac{E \pm \mu}{T} \big) - \nf \big( \tfrac{E \pm \mu}{T} \big) \big] \overset{\labelcref{eq:fermi-dirac-distribution_d}}{=}	\vdistance
			\\
			={}& \lim_{T \to 0} - \tfrac{E}{4 T \cosh^2 ( \frac{E \pm \mu}{2 T} )} =	\vdistance	\nonumber
			\\
			={}& - \tfrac{E}{| \mu |} \, \delta \Big( \tfrac{E}{| \mu |} \pm \mathrm{sgn} ( \mu ) \Big) \, ,	\vdistance	\nonumber
		\end{align}
	where $\delta$ is the Dirac-delta distribution.

\subsubsection{Abbreviations and definitions}

	For the sake of a compact notation and better readability, we define several quantities.
	First, we introduce the fermion energy/dispersion relation
		\begin{align}
			E \equiv \sqrt{p^2 + \barsigma^2} \, ,	\label{eq:fermion_energy}
		\end{align}
	where $\barsigma$ is the background field and fermion mass.
	For calculations at nonzero chemical potential, in particular at $T = 0$, it is useful to define the reduced chemical potential,
		\begin{align}
			\barmu \equiv \sqrt{\mu^2 - \barsigma^2} \, ,	\label{eq:barmu}
		\end{align}
	which reduces the chemical potential by the fermion mass.
	In the presence of an external momentum with absolute value $q$ the shifted fermion energy/dispersion relation is defined by
		\begin{align}
			\tilde{E} \equiv \sqrt{p^2 + \tilde{\Delta}^2} \, .	\label{eq:tildeenergy}
		\end{align}
	Here,
		\begin{align}
			\tilde{\Delta} \equiv \sqrt{\barsigma^2 + q^2 \, x \, ( 1 - x )}	\label{eq:tildedelta}
		\end{align}
	is the shifted squared fermion mass and $x \in [ 0, 1 ]$ the Feynman parameter.
	For $\barsigma=0$, this reduces to a shifted momentum
	\begin{align}
		\tilde{p}=\sqrt{p^2+q^2 x (1-x)}.
	\end{align}
	Finally, we also need the reduced and shifted chemical potential,
		\begin{align}
			\tilde{\mu} \equiv \sqrt{\mu^2 - \tilde{\Delta}^2} \, .	\label{eq:tildemu}
		\end{align}

\section{Formulary}

	In this appendix we present a collection of useful formulae, integral evaluations, and expansions that are repeatedly used in our calculations.

\subsection{Spherical symmetric integration}

	Most of the momentum integrals in this work are of the hyperspherical type.
	The integrand is usually only a function of the absolute value of the momentum such that the angular integration can be performed,
		\begin{align}
			\int \frac{\dd^\sdim p}{( 2 \uppi )^\sdim} f ( | \vec{p} \, | ) = \tfrac{\Sd}{( 2 \uppi )^\sdim} \int_{0}^{\infty} \dd p \, p^{\sdim - 1} \, f ( p ) \, .	\label{eq:spherically_symmetric_p_integral}
		\end{align}
	Here, we introduced
		\begin{align}
			\Sd = \tfrac{2 \uppi^{\frac{\sdim}{2}}}{\Gamma ( \frac{\sdim}{2} )} \, , \label{eq:sd}
		\end{align}
	which is the surface of the $\sdim$-dimensional sphere.

\subsection{Transcendental functions}

	A lot of the explicit formulae for the effective potential, bosonic wave-function renormalization, and the bosonic two-point function can be expressed in terms of known functions.
	For the sake of a self-contained presentation, we provide these functions with links to references for further reading in this appendix.
	We hope that this reduces unnecessary look ups and literature searches to a minimum for the reader.

\subsubsection{Gamma functions}

	The gamma function is given in terms of its integral representation by the following expression \cite[Eq.~6.1.1]{abramowitz+stegun},
		\begin{align}
			\Gamma ( z ) = \int_{0}^{\infty} \dd t \, t^{z - 1} \, \ee^{- t} \, .	\label{eq:gamma_function}
		\end{align}
	It fulfills the defining relation,
		\begin{align}
			\Gamma ( z + 1 ) = z \, \Gamma ( z ) \, .		\label{eq:gamma_function_recurrence_relation}
		\end{align}
	For this work, we make use of the Laurent series representation \cite[Eq.~5.7.1]{NIST:DLMF}
		\begin{align}
			\Gamma ( z ) = \tfrac{1}{z} - \upgamma + \mathcal{O} ( z ) \, ,	\label{eq:gamma_function_laurent_series}
		\end{align}
	which we use for an expansion about $z = 0$.

	The polygamma function with integer index is defined in terms of derivatives of the conventional gamma function.
	However, there is also an integral representation \cite[Eq.~6.4.1]{abramowitz+stegun}.
		\begin{align}
			\psi^{(n)} ( s ) ={}& \tfrac{\dd^n}{\dd s^n} \, \psi ( s ) = \tfrac{\dd^{n + 1}}{\dd s^{n + 1}} \, \ln \Gamma ( s ) =	\Vdistance	\label{eq:polygamma_function}
			\\
			={}& ( - 1 )^{n + 1} \int_{0}^{\infty} \dd t \, \frac{t^n \, \ee^{- s t}}{1 - \ee^{- t}}	\Vdistance	\nonumber
		\end{align}

\subsubsection{Riemann zeta function}

	Some formulae of this work can be expressed in terms of the Riemann zeta function \cite[Eq.~23.2.7]{abramowitz+stegun},
		\begin{align}
			\zeta ( s ) = \frac{1}{\Gamma ( s )} \int_{0}^{\infty} \dd t \, \frac{t^{s - 1}}{\ee^t - 1} \, .	&&	\Re ( s ) > 1 \, .
		\end{align}

\subsubsection{Dirichlet eta function}

	We also define the Dirichlet eta function via the Riemann zeta function \cite[Eq.~23.2.19]{abramowitz+stegun} and in terms of an integral,
		\begin{align}
			\eta ( s ) = \big( 1 - 2^{1 - s} \big) \, \zeta ( s ) = \frac{1}{\Gamma ( s )} \int_{0}^{\infty} \dd t \, \frac{t^{s-1}}{\ee^t + 1} \, .	\label{eq:dirichlet_eta_function}
		\end{align}

\subsubsection{Polylogarithm}

	It is well-known that some integrals over Bose-Einstein or Fermi-Dirac distribution functions can be expressed in terms of (incomplete) polylogarithms.
	The polylogarithm is defined via the following integral \cite[Eq.~25.12.10]{NIST:DLMF},
		\begin{align}
			\Li_s ( z ) = \frac{1}{\Gamma ( s )} \int_{0}^{\infty} \dd t \, \frac{t^{s - 1}}{\ee^t/z - 1} \, ,
		\end{align}
	which reduces to the Dirichlet eta function \labelcref{eq:dirichlet_eta_function} for $z = - 1$,
		\begin{align}
			\Li_s ( - 1 ) = - \eta ( s ) \, .
		\end{align}
	A special definition, which is used in this work is the symmetrized derivative of the polylogarithm \gls{wrt} its index $s$,
		\begin{align}
			& \mathrm{DLi}_{2n} ( y ) =	\vdistance \label{eq:DLi_definition}
			\\
			={}& \big[ \tfrac{\partial}{\partial s} \, \Li_s \big( - \ee^y \big) + \tfrac{\partial}{\partial s} \, \Li_s \big( - \ee^{- y} \big) \big]_{s = 2n} =	\vdistance	\nonumber
			\\
			={}& - \delta_{0, n} \, \big( \log ( 2 \uppi ) + \upgamma \big) +	\vdistance	\nonumber
			\\
			& + ( - 1 )^{1 - n} \, ( 2 \uppi )^{2 n} \, \Re \Big( \psi^{( - 2 n )} \big( \tfrac{1}{2} + \tfrac{\ii}{2 \uppi} \, y \big) \Big) \, .	\vdistance	\nonumber
		\end{align}
	Here, $\upgamma$ is the Euler-Mascheroni constant, $\psi^{(n)} ( s )$ is the polygamma function \labelcref{eq:polygamma_function}, and the last equality holds for $2 n \leq 0$.
	It turned out that using the last relation is more stable and accurate when it comes to numeric evaluation \cite{Stoll:2021ori,Koenigstein:2021llr}.

\subsubsection{Hypergeometric Function}

	The hypergeometric function is defined by \cite[Eq.~15.1.1]{abramowitz+stegun}
		\begin{align}
			\pFq{2}{1} ( \alpha, \beta; \gamma; z ) = \tfrac{\Gamma ( \gamma )}{\Gamma ( \alpha ) \, \Gamma ( \beta )} \, \sum_{n = 0}^{\infty} \tfrac{\Gamma ( \alpha + n ) \, \Gamma ( \beta + n )}{\Gamma ( \gamma + n )} \, \tfrac{z^n}{n!} \, ,	\label{eq:hypergeometric_f_series_definition}
		\end{align}
	where $| z | < 1$.
	The integral representation (and analytic continuation of \cref{eq:hypergeometric_f_series_definition}) reads \cite[Eq.~15.3.1]{abramowitz+stegun},
		\begin{align}
			&	\pFq{2}{1} ( \alpha, \beta; \gamma; z ) =	\Vdistance	\label{eq:hypergeometric_f_integral_definition}
			\\
			={}& \tfrac{\Gamma ( \gamma )}{\Gamma ( \beta ) \, \Gamma ( \gamma - \beta )} \int_{0}^{1} \dd t \, t^{\beta - 1} \, ( 1 - t )^{\gamma - \beta - 1} \, ( 1 - t z )^{- \alpha} \, .	\Vdistance	\nonumber
		\end{align}
	This formula is valid as long as $\Re ( \gamma ) > \Re ( \beta ) > 0$.

	A particular useful (linear) transformation formula is
	\cite[Eq.~15.3.7]{abramowitz+stegun}
		\begin{align}
			&	\pFq{2}{1} ( \alpha, \beta; \gamma; z ) =	\vdistance	\label{eq:hypergeometric_f_transformatio_formula_inverse_z}
			\\
			={}& \tfrac{\Gamma ( \gamma ) \, \Gamma ( \beta - \alpha )}{\Gamma ( \beta ) \, \Gamma ( \gamma - \alpha )} \, ( - z )^{- \alpha} \, \pFq{2}{1} \big( \alpha, 1 - \gamma + \alpha; 1 - \beta + \alpha; \tfrac{1}{z} \big) +	\vdistance	\nonumber
			\\
			& + \tfrac{\Gamma ( \gamma ) \, \Gamma ( \alpha - \beta )}{\Gamma ( \alpha ) \, \Gamma ( \gamma - \beta )} \, ( - z )^{- \beta} \, \pFq{2}{1} \big( \beta, 1 - \gamma + \beta; 1 - \alpha + \beta; \tfrac{1}{z} \big)	\vdistance	\nonumber
		\end{align}
	and is valid for $| \mathrm{arg} ( - z ) | < \uppi $.
	It can be used to expand the hypergeometric function for large $| z |$.

\subsection{Integrals and an expansion}

	Next, we present two important integrals for this work as well as an expansion that is used several times.

\subsubsection{First special integral}

	Repeatedly, we are confronted with integrals that are of the type
		\begin{align}
			& \int_{0}^{\Uplambda} \dd p \, p^{\sdim - 1} \, \tfrac{1}{E^n} =	\Vdistance	\label{eq:vacuum_integration_2F1}
			\\
			={}& \int_{0}^{\Uplambda} \dd p \, p^{\sdim - 1} \, ( p^2 + \Updelta^2 )^{- \frac{n}{2}} \overset{\labelcref{eq:substitution_p_t}}{=}	\Vdistance	\nonumber
			\\
			={}& \tfrac{\Uplambda^\sdim}{| \Updelta |^n} \, \tfrac{1}{2} \int_{0}^{1} \dd t \, t^{\frac{\sdim - 2}{2}} \, \big( 1 + t \, \tfrac{\Uplambda^2}{\Updelta^2} \big)^{- \frac{n}{2}} =	\Vdistance	\nonumber
			\\
			={}& \tfrac{\Uplambda^\sdim}{| \Updelta |^n} \, \tfrac{1}{2} \int_{0}^{1} \dd t \, t^{\frac{\sdim}{2} - 1} \, \big( 1 - t \big)^{\frac{\sdim + 2}{2} - \frac{\sdim}{2} - 1} \, \big( 1 + t \, \tfrac{\Uplambda^2}{\Updelta^2} \big)^{- \frac{n}{2}} \overset{\labelcref{eq:hypergeometric_f_integral_definition}}{=}	\Vdistance	\nonumber
			\\
			={}& \tfrac{\Uplambda^\sdim}{| \Updelta |^n} \, \tfrac{1}{2} \, \tfrac{\Gamma ( \frac{\sdim}{2} ) \, \Gamma ( \frac{\sdim + 2}{2} - \frac{\sdim}{2} )}{\Gamma ( \frac{\sdim + 2}{2} )} \, \pFq{2}{1} \big( \tfrac{n}{2}, \tfrac{\sdim}{2}; \tfrac{\sdim + 2}{2}; - \tfrac{\Uplambda^2}{\Updelta^2} \big) \overset{\labelcref{eq:gamma_function_recurrence_relation}}{=}	\Vdistance	\nonumber
			\\
			={}& \tfrac{\Uplambda^\sdim}{| \Updelta |^n} \, \tfrac{1}{\sdim} \, \pFq{2}{1} \big( \tfrac{n}{2}, \tfrac{\sdim}{2}; \tfrac{\sdim + 2}{2}; - \tfrac{\Uplambda^2}{\Updelta^2} \big) \, .	\Vdistance	\nonumber
		\end{align}
	We used the substitution
		\begin{align}
			&	t \, \Uplambda^2 = p^2 \, ,	&&	\dd t \, \Uplambda^2 = 2 p \, \dd p \, ,	\label{eq:substitution_p_t}
		\end{align}
		as well as \cref{eq:gamma_function_recurrence_relation,eq:hypergeometric_f_integral_definition}.

\subsubsection{Second special integral}
	
	Another integral that appears several times during our calculations is
		\begin{align}
			& \int_{0}^{\Uplambda} \dd p \, p^{a} \, \tfrac{1}{E^n} \, \delta \Big( \tfrac{E}{| \mu |} - 1 \Big) =	\Vdistance	\label{eq:vacuum_integration_delta}
			\\
			={}& \int_{0}^{\Uplambda} \dd p \, p^{a} \, \tfrac{1}{E^n} \, \tfrac{1}{| f^\prime ( p_0 ) |} \, \delta ( p - p_0 ) =	\Vdistance	\nonumber
			\\
			={}& \upmu^{a - 1} \, | \mu |^{2 - n} \, .	\Vdistance	\nonumber
		\end{align}
	We used that $E^2 = p^2 + \Updelta^2$ and defined
		\begin{align}
			\upmu^2 = \mu^2 - \Updelta^2 \, .
		\end{align}
	The integral was evaluated using
		\begin{align}
			\delta \big( f ( x ) \big) = \sum_i \tfrac{1}{| f^\prime ( x_i ) |} \, \delta ( x - x_i ) \, ,
		\end{align}
	where $x_i$ are the roots of $f ( x )$.
	For $\mu^2 > \Updelta^2$, thus $\upmu > 0$,
		\begin{align}
			p_0 = \sqrt{\mu^2 - \Updelta^2} = \upmu
		\end{align}
	is the zero of
		\begin{align}
			f = \tfrac{E}{| \mu |} - 1 \, ,
		\end{align}
	while
		\begin{align}
			f^\prime = \tfrac{p}{| \mu | E} \, .
		\end{align}
	Evaluating $p^a$, $\frac{1}{E^n}$, and $f^\prime$ at $p_0$ one obtains the above results.
	
\subsection{Expansion}
\label{app:expansion}

	At several points in this work, \eg, for sending the \gls{uv} cutoff $\Lambda$ to infinity or to evaluate certain expressions at the symmetric point $\barsigma \to 0$, we need an asymptotic expansion formula for the hypergeometric function $\pFq{2}{1} ( \alpha, \beta; \gamma; z )$ for large $| z |$.
	This formula is found by inserting the series representation \labelcref{eq:hypergeometric_f_series_definition} in the linear transformation formula \labelcref{eq:hypergeometric_f_transformatio_formula_inverse_z}.
	In particular, we find,
		\begin{align}
			& \lim_{\frac{\Uplambda^2}{\Updelta^2} \to \infty} \pFq{2}{1} \big( \tfrac{n}{2}, \tfrac{\sdim}{2}; \tfrac{\sdim + 2}{2}; - \tfrac{\Uplambda^2}{\Updelta^2} \big) =	\vdistance	\label{eq:expansions_2f1}
			\\
			={}& \sdim \, \Big[ \big| \tfrac{\Updelta}{\Uplambda} \big|^\sdim \, \tfrac{\Gamma ( \frac{\sdim + 2}{2} ) \, \Gamma ( \frac{n - \sdim}{2} )}{\sdim \,  \Gamma ( \frac{n}{2} )} + \big| \tfrac{\Updelta}{\Uplambda} \big|^n \, \Big( \tfrac{1}{\sdim - n}+ \tfrac{1}{2 - \sdim + n} \, \tfrac{n}{2} \, \tfrac{\Updelta^2}{\Uplambda^2} +	\vdistance	\nonumber
			\\
			& \quad  - \tfrac{1}{8 - 2 \sdim + 2 n} \, \tfrac{n}{2} \, \big( \tfrac{n}{2} + 1 \big) \, \tfrac{\Updelta^4}{\Uplambda^4} + \mathcal{O} \big( \tfrac{\Updelta^6}{\Uplambda^6} \big) \Big] \, .	\vdistance	\nonumber
		\end{align}

\section{Evaluation of \texorpdfstring{$l_0 ( \barsigma, \mu, T, \sdim )$}{l 0}}
\label{app:ellzero}

	In this appendix we provide details on the $l_0 ( \barsigma, \mu, T, \sdim )$-Matsubara sum and integral which occurs in the expression for the effective potential \labelcref{eq:effective_potential_main}.
	It is defined as follows
		\begin{align}
			&	l_0 ( \barsigma, \mu, T, \sdim ) \equiv	\Vdistance	\label{eq:l_0_sigma_mu_T_d}
			\\
			\equiv \, & \int \frac{\dd^\sdim p}{( 2 \uppi )^\sdim} \, \tfrac{1}{\beta} \sum_{n = - \infty}^{\infty} \ln \big( \beta^2 \big[ ( \nu_n - \ii \mu )^2 + E^2 \big] \big) =	\Vdistance	\nonumber
			\\
			={}& \int \frac{\dd^\sdim p}{( 2 \uppi )^\sdim} \, \big( E + T \ln \big[ 1 + \exp \big( - \tfrac{E + \mu}{T} \big) \big] +	\Vdistance	\nonumber
			\\
			& \quad + T \ln \big[ 1 + \exp \big( - \tfrac{E - \mu}{T} \big) \big] \big) + \const \, .	\Vdistance	\nonumber
		\end{align}
	Here, $\nu_n$ denotes the fermionic Matsubara frequencies \labelcref{eq:matsubara_frequency} and $E$ is the fermion energy \labelcref{eq:fermion_energy}.
	We used contour integration to evaluate the Matsubara sum.
	The infinite constant term can be ignored in what follows.
	It corresponds to an arbitrary normalization of the effective potential.

\subsection{For \texorpdfstring{$T = 0$}{T = 0}}

	The zero temperature limit of \cref{eq:l_0_sigma_mu_T_d}
		\begin{align}
			&	l_0 ( \barsigma, \mu, 0, \sdim ) =	\Vdistance
			\\
			={}& \tfrac{\Sd}{( 2 \uppi )^\sdim} \int_{0}^{\infty} \dd p \, p^{\sdim - 1} \, \Big[ E - ( E - | \mu | ) \, \Theta \Big( \tfrac{| \mu |}{E} - 1 \Big) \Big] \, ,	\Vdistance	\nonumber
		\end{align}
	where we used \cref{eq:spherically_symmetric_p_integral} to simplify the momentum integration with hyperspherical coordinates.
	Regularizing the \gls{uv} divergence with a sharp \gls{uv} cutoff and splitting the integral in $\mu$-(in)-dependent parts one obtains with \cref{eq:vacuum_integration_2F1},
		\begin{align}
			&	l_0^\Lambda ( \barsigma, \mu, 0, \sdim ) =	\Vdistance	\label{eq:l_0_sigma_mu_0_d_Lambda}
			\\
			={}& \tfrac{\Sd}{( 2 \uppi )^\sdim} \, \bigg( \int_{0}^{\Lambda} \dd p \, p^{\sdim - 1} \, E +	\Vdistance	\nonumber
			\\
			& \quad - \Theta \big( \tfrac{\barmu^2}{\barsigma^2} \big) \int_{0}^{\barmu} \dd p \, p^{\sdim - 1} \, ( E - | \mu | ) \bigg) =	\Vdistance	\nonumber
			\\
			={}& \tfrac{\Sd}{( 2 \uppi )^\sdim} \, \Big( \tfrac{| \barsigma |^{\sdim + 1}}{\sdim} \, \Big[ \big| \tfrac{\Lambda}{\barsigma} \big|^\sdim \, \pFq{2}{1} \big( - \tfrac{1}{2}, \tfrac{\sdim}{2}; \tfrac{\sdim + 2}{2}; - \tfrac{\Lambda^2}{\barsigma^2} \big) +	\Vdistance	\nonumber
			\\
			& \quad - \Theta \big( \tfrac{\barmu^2}{\barsigma^2} \big) \, \big| \tfrac{\barmu}{\barsigma} \big|^\sdim \, \big( \pFq{2}{1} \big( - \tfrac{1}{2}, \tfrac{\sdim}{2}; \tfrac{\sdim + 2}{2}; - \tfrac{\barmu^2}{\barsigma^2} \big) - \big| \tfrac{\mu}{\barsigma} \big| \big) \Big] \Big) \, .	\Vdistance	\nonumber
		\end{align}
	Here, we used the \Def\labelcref{eq:barmu} to facilitate a compact notation.

\subsection{For \texorpdfstring{$T \neq 0$}{T != 0}}

	In general, for nonzero $T$ we can still evaluate the vacuum contribution in \cref{eq:l_0_sigma_mu_T_d} and find with the $(\mu = 0)$-part of \cref{eq:l_0_sigma_mu_0_d_Lambda},
		\begin{align}
			&	l_0^\Lambda ( \barsigma, \mu, T, \sdim ) =	\Vdistance	\label{eq:l_0_sigma_mu_T_d_Lambda}
			\\
			={}& \tfrac{\Sd}{( 2 \uppi )^\sdim} \int_{0}^{\Lambda} \dd p \, p^{\sdim - 1} \, \big( E + T \ln \big[ 1 + \exp \big( - \tfrac{E + \mu}{T} \big) \big] +	\Vdistance	\nonumber
			\\
			& \quad + T \ln \big[ 1 + \exp \big( - \tfrac{E - \mu}{T} \big) \big] \big) =	\Vdistance	\nonumber
			\\
			={}& \tfrac{\Sd}{( 2 \uppi )^\sdim} \Big[ \tfrac{| \barsigma |^{\sdim + 1}}{\sdim} \, \big| \tfrac{\Lambda}{\barsigma} \big|^\sdim \, \pFq{2}{1} \big( - \tfrac{1}{2}, \tfrac{\sdim}{2}; \tfrac{\sdim + 2}{2}; - \tfrac{\Lambda^2}{\barsigma^2} \big) +	\Vdistance	\nonumber
			\\
			& \quad + \int_{0}^{\Lambda} \dd p \, p^{\sdim - 1} \, \big( T \ln \big[ 1 + \exp \big( - \tfrac{E + \mu}{T} \big) \big] +	\Vdistance	\nonumber
			\\
			& \qquad + T \ln \big[ 1 + \exp \big( - \tfrac{E - \mu}{T} \big) \big] \big) \Big] \, .	\Vdistance	\nonumber
		\end{align}
	Further evaluation of \cref{eq:l_0_sigma_mu_0_d_Lambda,eq:l_0_sigma_mu_T_d_Lambda} is performed after renormalization of the effective potential in \cref{app:effective_potential}.

\section{Evaluation of \texorpdfstring{$l_1 ( \barsigma, \mu, T, \sdim )$}{l 1}}
\label{app:ellone}
	In this appendix, we present explicit expressions for the $l_1 ( \barsigma, \mu, T, \sdim )$-Matsubara sum and integral which is part of the gap equation \labelcref{eq:gap_equation_main}, the regularized effective potential \labelcref{eq:effective_potential_regularized_main}, and the (regularized) bosonic two-point function \cref{eq:gamma2_main,eq:gamma2_regularized_main}.
	It is defined and evaluated as follows
		\begin{align}
			&	l_1 ( \barsigma, \mu, T, \sdim ) \equiv	\Vdistance	\label{eq:l_1_sigma_mu_T_d}
			\\
			\equiv \, & \int \frac{\dd^\sdim p}{( 2 \uppi )^\sdim} \, \tfrac{1}{\beta} \sum_{n = - \infty}^{\infty} \frac{1}{( \nu_n - \ii \mu )^2 + E^2} =	\Vdistance	\nonumber
			\\
			={}& \int \frac{\dd^\sdim p}{( 2 \uppi )^\sdim} \, \tfrac{1}{2 E} \, \big[ 1 - \nf \big( \tfrac{E + \mu}{T} \big) - \nf \big( \tfrac{E - \mu}{T} \big) \big] =	\Vdistance	\nonumber
			\\
			={}& \int \frac{\dd^\sdim p}{( 2 \uppi )^\sdim} \, \tfrac{1}{2 E} \, \big[ \tfrac{1}{2} \tanh \big( \tfrac{E + \mu}{2T} \big) + \tfrac{1}{2} \tanh \big( \tfrac{E - \mu}{2T} \big) \big] \, .	\Vdistance	\nonumber
		\end{align}
	Again, $\nu_n$ are the fermionic Matsubara frequencies \labelcref{eq:matsubara_frequency} and $E$ is the fermion energy \labelcref{eq:fermion_energy}.
	Additionally, we introduced the Fermi-Dirac distribution \labelcref{eq:fermi-dirac-distribution}, while it turned out that the $\tanh$-representation seems to be more stable and accurate for numeric computations.

\subsection{For \texorpdfstring{$T = 0$}{T = 0}}

	In the zero-temperature limit \cref{eq:l_1_sigma_mu_T_d} reduces to
		\begin{align}
			&	l_1 ( \barsigma, \mu, 0, \sdim ) =	\Vdistance
			\\
			={}& \tfrac{\Sd}{( 2 \uppi )^\sdim} \int_{0}^{\infty} \dd p \, p^{\sdim - 1} \, \tfrac{1}{2 E} \, \Big[ 1 - \Theta \Big( \tfrac{| \mu |}{E} - 1 \Big) \Big]	\Vdistance	\nonumber
		\end{align}
	where we made use of hyperspherical coordinates in momentum space, see \cref{eq:spherically_symmetric_p_integral}.
	Splitting $\mu$-(in)-dependent terms, using the abbreviation \labelcref{eq:barmu}, and introducing the \gls{uv} cutoff $\Lambda$ one can make use of \cref{eq:vacuum_integration_2F1} to arrive at
		\begin{align}
			&	l_1^\Lambda ( \barsigma, \mu, 0, \sdim ) =	\Vdistance	\label{eq:l_1_sigma_mu_0_d_Lambda}
			\\
			={}& \tfrac{\Sd}{( 2 \uppi )^\sdim} \, \tfrac{1}{2} \, \bigg( \int_{0}^{\Lambda} \dd p \, p^{\sdim - 1} \, \tfrac{1}{E} \, - \Theta \big( \tfrac{\barmu^2}{\barsigma^2} \big) \int_{0}^{| \barmu |} \dd p \, p^{\sdim - 1} \, \tfrac{1}{E} \bigg) =	\Vdistance	\nonumber
			\\
			={}& \tfrac{\Sd}{( 2 \uppi )^\sdim} \, \tfrac{1}{2} \, \Big( \tfrac{| \barsigma |^{\sdim - 1}}{\sdim} \, \Big[ \big| \tfrac{\Lambda}{\barsigma} \big|^\sdim \, \pFq{2}{1} \big( \tfrac{1}{2}, \tfrac{\sdim}{2}; \tfrac{\sdim + 2}{2}; - \tfrac{\Lambda^2}{\barsigma^2} \big) +	\Vdistance	\nonumber
			\\
			& \quad - \Theta \big( \tfrac{\barmu^2}{\barsigma^2} \big) \, \big| \tfrac{\barmu}{\barsigma} \big|^\sdim \, \pFq{2}{1} \big( \tfrac{1}{2}, \tfrac{\sdim}{2}; \tfrac{\sdim + 2}{2}; - \tfrac{\barmu^2}{\barsigma^2} \big) \Big] \Big) \, .	\Vdistance	\nonumber
		\end{align}

\subsection{For \texorpdfstring{$T \neq 0$}{T != 0}}

	Again, for $T \neq 0$ we solely evaluate the vacuum contribution of \cref{eq:l_1_sigma_mu_T_d} and again use the previous result \cref{eq:l_1_sigma_mu_0_d_Lambda} for $\mu = 0$.
	For the regularized expression one finds,
		\begin{align}
			&	l_1^\Lambda ( \barsigma, \mu, T, \sdim ) =	\Vdistance	\label{eq:l_1_sigma_mu_T_d_Lambda}
			\\
			={}& \tfrac{\Sd}{( 2 \uppi )^\sdim} \int_{0}^{\Lambda} \dd p \, p^{\sdim - 1} \, \tfrac{1}{2 E} \, \big[ 1 - \nf \big( \tfrac{E + \mu}{T} \big) - \nf \big( \tfrac{E - \mu}{T} \big) \big] =	\Vdistance	\nonumber
			\\
			={}& \tfrac{\Sd}{( 2 \uppi )^\sdim} \tfrac{1}{2} \bigg( \tfrac{1}{\sdim | \barsigma |} \, \Lambda^\sdim \, \pFq{2}{1} \big( \tfrac{1}{2}, \tfrac{\sdim}{2}; \tfrac{\sdim + 2}{2}; - \tfrac{\Lambda^2}{\barsigma^2} \big) +	\Vdistance	\nonumber
			\\
			& \quad - \int_{0}^{\Lambda} \dd p \, p^{\sdim - 1} \, \tfrac{1}{E} \big[ \nf \big( \tfrac{E + \mu}{T} \big) + \nf \big( \tfrac{E - \mu}{T} \big) \big] \bigg) \, .	\Vdistance	\nonumber
		\end{align}
	Further evaluation of \cref{eq:l_1_sigma_mu_0_d_Lambda,eq:l_1_sigma_mu_T_d_Lambda} is postponed to \cref{app:effective_potential,app:wave_function_renormalization,app:bosonic_two-point_function}.

\section{Evaluation of \texorpdfstring{$l_2 ( \barsigma, \mu, T, q, \sdim )$}{l 2}}
\label{app:elltwo}
	In the expression for the bosonic two-point function \labelcref{eq:gamma2_main} contains a $\vec{q}$-dependent part.
	Here, we show some simplifications for this contribution that reads
		\begin{align}
			& l_2 ( \barsigma, \mu, T, q, \sdim ) =	\Vdistance	\label{eq:l_2_sigma_mu_T_q_d}
			\\
			={}& \int \frac{\dd^\sdim p}{( 2 \uppi )^\sdim} \, \tfrac{1}{\beta} \sum_{n = - \infty}^{\infty} \frac{1}{( \nu_n - \ii \mu )^2 + \vec{p}^{\, 2} + \barsigma^2} \times	\Vdistance	\nonumber
			\\
			& \quad \times \frac{1}{( \nu_n - \ii \mu )^2 + ( \vec{p} + \vec{q} \, )^2 + \barsigma^2} \, .	\Vdistance	\nonumber
		\end{align}
	We already used $q = | \vec{q} \, |$ as an argument of $l_2$ instead of $\vec{q}$.
	This becomes clear in the following lines.
	In order to get rid of the nasty vectorial $\vec{q}$-shift in the second propagator we use the following Feynman parameter integral
		\begin{align}
			\frac{1}{a_1 \, a_2} = \int_{0}^{1} \dd x \, \frac{1}{[ a_1 \, x + a_2 \, ( 1 - x ) ]^2} \, .
		\end{align}
	Applying this to \cref{eq:l_2_sigma_mu_T_q_d} one obtains,
		\begin{align}
			& l_2 ( \barsigma, \mu, T, q, \sdim ) =	\Vdistance
			\\
			={}& \int \frac{\dd^\sdim p}{( 2 \uppi )^\sdim} \tfrac{1}{\beta} \sum_{n = - \infty}^{\infty} \int_{0}^{1} \dd x \times	\Vdistance	\nonumber
			\\
			& \quad \times \frac{1}{[ ( \nu_n - \ii \mu )^2 + ( \vec{p} + \vec{q} \, )^2 \, x + \vec{p}^{\, 2} \, ( 1 - x) + \barsigma^2 ]^2} =	\Vdistance	\nonumber
			\\
			={}& \int_{0}^{1} \dd x \int \frac{\dd^\sdim p}{( 2 \uppi )^\sdim} \tfrac{1}{\beta} \sum_{n = - \infty}^{\infty} \times	\Vdistance	\nonumber
			\\
			& \quad \times \frac{1}{[ ( \nu_n - \ii \mu )^2 + ( \vec{p} + \vec{q} \, x )^2 + \vec{q}^{\, 2} \, x \, ( 1 - x) + \barsigma^2 ]^2} =	\Vdistance	\nonumber
			\\
			={}& \int_{0}^{1} \dd x \int \frac{\dd^\sdim p}{( 2 \uppi )^\sdim} \, \tfrac{1}{\beta} \sum_{n = - \infty}^{\infty} \times	\Vdistance	\nonumber
			\\
			& \quad \times \frac{1}{[ ( \nu_n - \ii \mu )^2 + \vec{p}^{\, 2} + \vec{q}^{\, 2} \, x \,  ( 1 - x) + \barsigma^2 ]^2} \, .	\Vdistance	\nonumber
		\end{align}
	We exchanged the order of integration, substituted $\vec{p}^{\, \prime} = \vec{p} + \vec{q} \, x$, and immediately returned to the ``unprimed'' notation for $\vec{p}$.
	Using the \Defs\labelcref{eq:fermi-dirac-distribution,eq:fermi-dirac-distribution_d,eq:tildeenergy,eq:tildedelta,eq:tildemu}, 
		\begin{align}
			& l_2 ( \barsigma, \mu, T, q, \sdim ) =	\Vdistance
			\\
			={}& \int_{0}^{1} \dd x \int \frac{\dd^\sdim p}{( 2 \uppi )^\sdim} \, \tfrac{1}{4 \tilde{E}^3} \, \big[ 1 - \nf \big( \tfrac{\tilde{E} + \mu}{T} \big) +	\Vdistance	\nonumber
			\\
			& \quad + \tfrac{\tilde{E}}{T} \, \big[ \nf^2 \big( \tfrac{\tilde{E} + \mu}{T} \big) - \nf \big( \tfrac{\tilde{E} + \mu}{T} \big) \big] + ( \mu \leftrightarrow - \mu ) \big] =	\Vdistance	\nonumber
			\\
			={}& \int_{0}^{1} \dd x \int \frac{\dd^\sdim p}{( 2 \uppi )^\sdim} \, \tfrac{1}{4 \tilde{E}^3} \, \bigg( \tfrac{1}{2} \tanh \big( \tfrac{\tilde{E} + \mu}{2 T} \big) +	\Vdistance	\nonumber
			\\
			& \quad - \tfrac{\tilde{E}}{T} \, \tfrac{1}{4 \cosh^2 ( \frac{\tilde{E} + \mu}{2T} )} + ( \mu \leftrightarrow - \mu ) \bigg) \, .	\Vdistance	\nonumber
		\end{align}
	Thus, the dependence on $\vec{q}$ is actually a dependence on its absolute value $q$.

\subsection{For: \texorpdfstring{$T = 0$}{T = 0}}
	
	For $T = 0$ \cref{eq:l_2_sigma_mu_T_q_d} further reduces to
		\begin{align}
			& l_2 ( \barsigma, \mu, 0, q, \sdim ) =	\Vdistance
			\\
			={}& \tfrac{\Sd}{( 2 \uppi )^\sdim} \, \tfrac{1}{4} \int_{0}^{1} \dd x \int_{0}^{\infty} \dd p \, p^{\sdim - 1} \, \tfrac{1}{\tilde{E}^3} \, \Big[ 1 - \Theta \Big( \tfrac{| \mu |}{\tilde{E}} - 1 \Big) +	\Vdistance	\nonumber
			\\
			& \quad - \tfrac{\tilde{E}}{| \mu |} \, \Big[ \delta \Big( \tfrac{\tilde{E}}{| \mu |} + 1 \Big) + \delta \Big( \tfrac{\tilde{E}}{| \mu |} - 1 \Big) \Big] \Big] \, ,	\Vdistance	\nonumber
		\end{align}
	where we already used \cref{eq:spherically_symmetric_p_integral}, the hyperspherical coordinates for the momentum integration.
	\gls{uv} regularization of this expression is not needed for $\sdim < 3$.
	However, it can be useful to (1.) introduce an \gls{uv} cutoff $\Lambda$, (2.) use \cref{eq:vacuum_integration_2F1}, and (3.) study $\Lambda \to \infty$ with \cref{eq:expansions_2f1}.
	Splitting the integration into $\mu$-(in)-dependent parts leads to
		\begin{align}
			& l_2 ( \barsigma, \mu, 0, q, \sdim ) =	\Vdistance	\label{eq:l_2_sigma_mu_0_q_d}
			\\
			={}& \tfrac{\Sd}{( 2 \uppi )^\sdim} \, \tfrac{1}{4} \int_{0}^{1} \dd x \int_{0}^{\infty} \dd p \, p^{\sdim - 1} \, \tfrac{1}{\tilde{E}^3} \, +	\Vdistance	\nonumber
			\\
			& \quad - \Theta \big( \tfrac{\tilde{\mu}^2}{\tilde{\Delta}^2} \big) \, \int_{0}^{\tilde{\mu}} \dd p \, p^{\sdim - 1} \, \tfrac{1}{\tilde{E}^3} +	\Vdistance	\nonumber
			\\
			& \quad - \tfrac{1}{| \mu |} \int_{0}^{\infty} \dd p \, p^{\sdim - 1} \, \tfrac{1}{\tilde{E}^2} \, \delta \Big( \tfrac{\tilde{E}}{| \mu |} - 1 \Big) \Big] =	\Vdistance	\nonumber
			\\
			={}& \tfrac{\Sd}{( 2 \uppi )^\sdim} \, \tfrac{1}{4} \int_{0}^{1} \dd x \, \Big[ \tilde{\Delta}^{\sdim - 3} \, \tfrac{\Gamma ( \frac{3 - \sdim}{2} ) \, \Gamma ( \frac{\sdim}{2} )}{\sqrt{\uppi}} +	\Vdistance	\nonumber
			\\
			& \quad - \Theta \Big( \tfrac{\tilde{\mu}^2}{\tilde{\Delta}^2} \Big) \Big( \tfrac{\tilde{\mu}^\sdim}{\tilde{\Delta}^3} \, \tfrac{1}{\sdim} \, \pFq{2}{1} \big( \tfrac{3}{2}, \tfrac{\sdim}{2}; \tfrac{\sdim + 2}{2}; - \tfrac{\tilde{\mu}^2}{\tilde{\Delta}^2} \big) + \tfrac{\tilde{\mu}^{\sdim - 2}}{| \mu |} \Big) \Big] \, .	\Vdistance	\nonumber
		\end{align}
	For the $\mu$-dependent medium part we used \cref{eq:vacuum_integration_2F1,eq:vacuum_integration_delta}.
	
\subsection{For: \texorpdfstring{$T \neq 0$}{T != 0}}

	Of course, we also provide a simplification of \cref{eq:l_2_sigma_mu_T_q_d} for $T \neq 0$ by using the vacuum contribution of the previous result \labelcref{eq:l_2_sigma_mu_0_q_d},
		\begin{align}
			& l_2 ( \barsigma, \mu, T, q, \sdim ) =	\Vdistance	\label{eq:l_2_sigma_mu_T_q_d_final}
			\\
			={}& \tfrac{\Sd}{( 2 \uppi )^\sdim} \, \tfrac{1}{4} \int_{0}^{1} \dd x \bigg( \tilde{\Delta}^{\sdim - 3} \, \tfrac{\Gamma ( \frac{3 - \sdim}{2} ) \, \Gamma ( \frac{\sdim}{2} )}{\sqrt{\uppi}} +	\Vdistance	\nonumber
			\\
			& \quad - \int_{0}^{\infty} \dd p \, p^{\sdim - 1} \, \tfrac{1}{\tilde{E}^3} \, \big[ \nf \big( \tfrac{\tilde{E} + \mu}{T} \big) +	\Vdistance	\nonumber
			\\
			& \qquad - \tfrac{\tilde{E}}{T} \, \big[ \nf^2 \big( \tfrac{\tilde{E} + \mu}{T} \big) - \nf \big( \tfrac{\tilde{E} + \mu}{T} \big) \big] + ( \mu \leftrightarrow - \mu ) \big] \bigg) \, .	\Vdistance	\nonumber
		\end{align}

\section{Evaluation of \texorpdfstring{$l_3 ( \barsigma, \mu, T, \sdim )$}{l 3}}
\label{app:ellthree}

	In complete analogy to the previous appendices, we present an appendix for the partial evaluation of $l_3 ( \barsigma, \mu, T, \sdim )$ that is part of \cref{eq:wave-function_renormalization_main} for the bosonic wave-function renormalization.
	In terms of the (un)evaluated Matsubara sum and momentum integral it reads
		\begin{align}
			& l_3 ( \barsigma, \mu, T, \sdim ) =	\Vdistance	\label{eq:l_3_sigma_mu_T_d}
			\\
			={}& \int \frac{\dd^\sdim p}{( 2 \uppi )^\sdim} \, \tfrac{1}{\beta} \sum_{n = - \infty}^{\infty} \frac{1}{[ ( \nu_n - \ii \mu )^2 + E^2 ]^3} =	\Vdistance	\nonumber
			\\
			={}& \int \frac{\dd^\sdim p}{( 2 \uppi )^\sdim} \, \tfrac{3}{16 E^5} \, \big[ 1 - \nf \big( \tfrac{E + \mu}{T} \big) +	\Vdistance	\nonumber
			\\
			& \quad + \tfrac{E}{T} \, \big[ \nf^2 \big( \tfrac{E + \mu}{T} \big) - \nf \big( \tfrac{E + \mu}{T} \big) \big] +	\Vdistance	\nonumber
			\\
			& \quad - \big( \tfrac{E}{T} \big)^2 \, \big[ \tfrac{2}{3} \, \nf^3 \big( \tfrac{E + \mu}{T} \big) - \nf^2 \big( \tfrac{E + \mu}{T} \big) + \tfrac{1}{3} \, \nf \big( \tfrac{E + \mu}{T} \big) \big] +	\Vdistance	\nonumber
			\\
			& \quad + ( \mu \leftrightarrow - \mu ) \big] =	\Vdistance	\nonumber
			\\
			={}& \int \frac{\dd^\sdim p}{( 2 \uppi )^\sdim} \, \tfrac{3}{16 E^5} \, \bigg[ \tfrac{1}{2} \tanh \big( \tfrac{E + \mu}{2 T} \big) - \tfrac{E}{T} \, \tfrac{1}{4 \, \cosh^2 ( \frac{E + \mu}{2 T} )} +	\Vdistance	\nonumber
			\\
			& \quad - \big( \tfrac{E}{T} \big)^2 \, \tfrac{\sinh ( \frac{E + \mu}{2 T} )}{12 \, \cosh^3 ( \frac{E + \mu}{2 T} )} + ( \mu \leftrightarrow - \mu ) \bigg] \, .	\Vdistance	\nonumber
		\end{align}
	In this expression we used \cref{eq:matsubara_frequency,eq:fermi-dirac-distribution,eq:fermi-dirac-distribution_d,eq:fermi-dirac-distribution_dd,eq:fermion_energy}.
	
\subsection{For \texorpdfstring{$T = 0$}{T = 0}}

	In the zero-temperature limit \cref{eq:l_3_sigma_mu_T_d} turns into
		\begin{align}
			& l_3 ( \barsigma, \mu, 0, \sdim ) =	\Vdistance
			\\
			={}& \tfrac{\Sd}{( 2 \uppi )^\sdim} \int_{0}^{\infty} \dd p \, p^{\sdim - 1} \, \tfrac{3}{16 E^5} \, \Big[ 1 - \Theta \Big( \tfrac{| \mu |}{E} - 1 \Big) +	\Vdistance	\nonumber
			\\
			& \quad - \tfrac{E}{| \mu |} \, \Big[ \delta \Big( \tfrac{E}{| \mu |} + 1 \Big) + \delta \Big( \tfrac{E}{| \mu |} - 1 \Big) \Big] +	\Vdistance	\nonumber
			\\
			& \quad + \tfrac{1}{3} \, \big( \tfrac{E}{\mu} \big)^2 \, \Big[ \delta^\prime \Big( \tfrac{E}{| \mu |} + 1 \Big) + \delta^\prime \Big( \tfrac{E}{| \mu |} - 1 \Big) \Big] \Big] \, ,	\Vdistance	\nonumber
		\end{align}
	where we already made use of hyperspherical coordinates via \cref{eq:spherically_symmetric_p_integral}.
	Again, it is not necessary though still useful to introduce a \gls{uv} cutoff regulator $\Lambda$ to use \cref{eq:vacuum_integration_2F1} also for the vacuum contribution.
	Afterwards, the cutoff can be removed with the help of \cref{eq:expansions_2f1}.
	Splitting and evaluating the integrals in $\mu$-(in)-dependent contributions, one finds, using integration by parts and \cref{eq:vacuum_integration_2F1,eq:vacuum_integration_delta},
		\begin{align}
			& l_3 ( \barsigma, \mu, 0, \sdim ) =	\Vdistance	\label{eq:l_3_sigma_mu_0_d}
			\\
			={}& \tfrac{\Sd}{( 2 \uppi )^\sdim} \, \tfrac{3}{16} \, \bigg( \int_{0}^{\infty} \dd p \, p^{\sdim - 1} \, \tfrac{1}{E^5} - \Theta \big( \tfrac{\barmu^2}{\barsigma^2} \big) \, \int_{0}^{\barmu} \dd p \, p^{\sdim - 1} \, \tfrac{1}{E^5} +	\Vdistance	\nonumber
			\\
			& \quad - \tfrac{\sdim - 2}{3 | \mu |} \int_{0}^{\infty} \dd p \, p^{\sdim - 3} \, \tfrac{1}{E^2} \, \delta \Big( \tfrac{E}{| \mu |} - 1 \Big) +	\Vdistance	\nonumber
			\\
			& \quad - \tfrac{1}{3 | \mu |} \int_{0}^{\infty} \dd p \, p^{\sdim - 1} \, \tfrac{1}{E^4} \, \delta \Big( \tfrac{E}{| \mu |} - 1 \Big) \Big] =	\Vdistance	\nonumber
			\\
			={}& \tfrac{\Sd}{( 2 \uppi )^\sdim} \, \tfrac{3}{16} \, \Big[ \tfrac{2}{3} \, | \barsigma |^{\sdim - 5} \, \tfrac{\Gamma ( \frac{5 - \sdim}{2} ) \, \Gamma ( \frac{\sdim}{2})}{\sqrt{\uppi}} +	\Vdistance	\nonumber
			\\
			& \quad - \Theta \Big( \tfrac{\barmu^2}{\barsigma^2} \Big) \Big( \tfrac{\barmu^\sdim}{| \barsigma |^5} \, \tfrac{1}{\sdim} \, \pFq{2}{1} \big( \tfrac{5}{2}, \tfrac{\sdim}{2}; \tfrac{\sdim + 2}{2}; - \tfrac{\barmu^2}{\barsigma^2} \big) +	\Vdistance	\nonumber
			\\
			& \qquad + \tfrac{\sdim - 2}{3 | \mu |} \, \barmu^{\sdim - 4} + \tfrac{1}{3 | \mu |} \, \barmu^{\sdim - 2} \, \mu^{- 2} \Big) \Big] \, .	\Vdistance	\nonumber
		\end{align}
	Again, we used the compact notation \cref{eq:barmu}.

\subsection{For \texorpdfstring{$T \neq 0$}{T != 0}}

	At nonzero temperature we can use the vacuum part of \cref{eq:l_3_sigma_mu_0_d} to simplify \cref{eq:l_3_sigma_mu_T_d},
		\begin{align}
			& l_3 ( \barsigma, \mu, T, \sdim ) =	\Vdistance
			\\
			={}& \tfrac{\Sd}{( 2 \uppi )^\sdim} \, \tfrac{3}{16} \, \bigg( \tfrac{2}{3} \, | \barsigma |^{\sdim - 5} \, \tfrac{\Gamma ( \frac{5 - \sdim}{2} ) \, \Gamma ( \frac{\sdim}{2})}{\sqrt{\uppi}} +	\Vdistance	\nonumber
			\\
			& \quad - \int_{0}^{\infty} \dd p \, p^{\sdim - 1} \, \tfrac{1}{E^5} \, \big[ \nf \big( \tfrac{E + \mu}{T} \big) +	\Vdistance	\nonumber
			\\
			& \qquad - \tfrac{E}{T} \, \big[ \nf^2 \big( \tfrac{E + \mu}{T} \big) - \nf \big( \tfrac{E + \mu}{T} \big) \big] +	\Vdistance	\nonumber
			\\
			& \qquad + \big( \tfrac{E}{T} \big)^2 \, \big[ \tfrac{2}{3} \, \nf^3 \big( \tfrac{E + \mu}{T} \big) - \nf^2 \big( \tfrac{E + \mu}{T} \big) + \tfrac{1}{3} \, \nf \big( \tfrac{E + \mu}{T} \big) \big] +	\Vdistance	\nonumber
			\\
			& \qquad + ( \mu \leftrightarrow - \mu ) \big] \bigg) \, .	\Vdistance	\nonumber
		\end{align}

\section{The effective potential}
\label{app:effective_potential}

	In this appendix we turn to the detailed evaluation of the effective potential \cref{eq:effective_potential_main}.
	To this end, we calculate the renormalized limits of \cref{eq:effective_potential_regularized_main}, where we remove the \gls{uv} cutoff by sending $\Lambda \to \infty$.
	Explicit expressions for $\barsigma$, $\mu$, $T$ being zero or nonzero as well as the limiting cases of $\sdim = 1$ and $\sdim = 2$ are provided.
	For the sake of the conciseness, we collected links to every special case in \cref{tab:U_explicit_expressions} and subdivide the appendix according to the table.
		\setlength{\extrarowheight}{3pt}
		\begin{table}
			\caption{\label{tab:U_explicit_expressions}%
			Quick links to the equations for explicit evaluation of the effective potential $U ( \barsigma, \mu, T, \sdim )$.
			The formulae are simplified in terms of known functions as far as possible.
			}
			\begin{ruledtabular}
				\begin{tabular}{c c c c c c}
					$T$							&	$\barsigma$					&	$\mu$	&	$1 \leq \sdim < 3$				&	$\sdim = 1$	&	$\sdim =2$
					\\
					\hline
					\multirow{4}{*}{$\neq 0$}	&	\multirow{2}{*}{$\neq 0$}	&	$\neq 0$	&	\cref{eq:U_sigma_mu_T_d}	& 	\cref{eq:U_sigma_mu_T_1}	&	\cref{eq:U_sigma_mu_T_2}
					\\
					\cline{3-6}
												&								&	$= 0$		&	\cref{eq:U_sigma_0_T_d}		&	\cref{eq:U_sigma_0_T_1} 	&	\cref{eq:U_sigma_0_T_2}
					\\
					\cline{2-6}
												&	\multirow{2}{*}{$= 0$}		&	$\neq 0$	&	\cref{eq:U_0_mu_T_d}		&	\cref{eq:U_0_mu_T_1} 		&	\cref{eq:U_0_mu_T_2}
					\\
					\cline{3-6}
												&								&	$= 0$		&	\cref{eq:U_0_0_T_d}			&	\cref{eq:U_0_0_T_1} 		&	\cref{eq:U_0_0_T_2}
					\\
					\hline
					\multirow{4}{*}{$= 0$}		&	\multirow{2}{*}{$\neq 0$}	&	$\neq 0$	&	\cref{eq:U_sigma_mu_0_d}	& 	\cref{eq:U_sigma_mu_0_1}	&	\cref{eq:U_sigma_mu_0_2}
					\\
					\cline{3-6}
												&								&	$= 0$		&	\cref{eq:U_sigma_0_0_d}		& 	\cref{eq:U_sigma_0_0_1}		&	\cref{eq:U_sigma_0_0_2}
					\\
					\cline{2-6}
												&	\multirow{2}{*}{$= 0$}		&	$\neq 0$	&	\cref{eq:U_0_mu_0_d}		& 	\cref{eq:U_0_mu_0_1}		&	\cref{eq:U_0_mu_0_2}
					\\
					\cline{3-6}
												&								&	$= 0$		&	\cref{eq:U_0_0_0_d}			& 	\cref{eq:U_0_0_0_d}			&	\cref{eq:U_0_0_0_d}
				\end{tabular}
			\end{ruledtabular}
		\end{table}

\subsection{\texorpdfstring{$T \neq 0$}{T != 0}}

	We start with the $T \neq 0$ cases.

\subsubsection{\texorpdfstring{$T \neq 0, \barsigma \neq 0$}{T != 0, sigma != 0}}

	Considering $\barsigma \neq 0$ there are two cases to be distinguished.

\paragraph{\texorpdfstring{$T \neq 0, \barsigma \neq 0, \mu \neq 0$}{T != 0, sigma != 0, mu != 0}}

	Inserting \cref{eq:l_0_sigma_mu_T_d_Lambda,eq:l_1_sigma_mu_T_d_Lambda} in \cref{eq:effective_potential_regularized_main} one finds for the regularized potential
		\begin{align}
			& U^\Lambda ( \barsigma, \mu, T, \sdim ) =	\Vdistance	\label{eq:U_sigma_mu_T_d_Lambda}
			\\
			={}& \tfrac{\dimDirac}{2} \, \big[ \barsigma^2 \, l_1^\Lambda ( \sigmaminvac, 0, 0, \sdim ) - l_0^\Lambda ( \barsigma, \mu, T, \sdim ) \big] =	\Vdistance	\nonumber	
			\\
			={}& \tfrac{\dimDirac}{2} \, \tfrac{\Sd}{( 2 \uppi )^\sdim} \, \bigg[ \barsigma^2 \, \tfrac{1}{2} \, \tfrac{| \sigmaminvac |^{\sdim - 1}}{\sdim} \, \big| \tfrac{\Lambda}{\sigmaminvac} \big|^\sdim \, \pFq{2}{1} \big( \tfrac{1}{2}, \tfrac{\sdim}{2}; \tfrac{\sdim + 2}{2}; - \tfrac{\Lambda^2}{\sigmaminvac^2} \big) +	\Vdistance	\nonumber
			\\
			& \quad - \tfrac{| \barsigma |^{\sdim + 1}}{\sdim} \, \big| \tfrac{\Lambda}{\barsigma} \big|^\sdim \, \pFq{2}{1} \big( - \tfrac{1}{2}, \tfrac{\sdim}{2}; \tfrac{\sdim + 2}{2}; - \tfrac{\Lambda^2}{\barsigma^2} \big) +	\Vdistance	\nonumber
			\\
			& \quad - \int_{0}^{\Lambda} \dd p \, p^{\sdim - 1} \, \big( T \ln \big[ 1 + \exp \big( - \tfrac{E + \mu}{T} \big) \big] +	\Vdistance	\nonumber
			\\
			& \qquad + ( \mu \to - \mu ) \big) + \const \bigg] =	\Vdistance	\nonumber
			\\
			={}& \tfrac{\dimDirac}{2} \, \tfrac{\Sd}{( 2 \uppi )^\sdim} \, \bigg[ \tfrac{| \barsigma |}{\sdim} \, \Lambda^\sdim \, \Big( \tfrac{1}{2} \, \tfrac{| \barsigma |}{\sigmaminvac} \, \pFq{2}{1} \big( \tfrac{1}{2}, \tfrac{\sdim}{2}; \tfrac{\sdim + 2}{2}; - \tfrac{\Lambda^2}{\sigmaminvac^2} \big) +	\Vdistance	\nonumber
			\\
			& \quad - \pFq{2}{1} \big( - \tfrac{1}{2}, \tfrac{\sdim}{2}; \tfrac{\sdim + 2}{2}; - \tfrac{\Lambda^2}{\barsigma^2} \big) \Big) +	\Vdistance	\nonumber
			\\
			& \quad - \int_{0}^{\Lambda} \dd p \, p^{\sdim - 1} \, \big( T \ln \big[ 1 + \exp \big( - \tfrac{E + \mu}{T} \big) \big] +	\Vdistance	\nonumber
			\\
			& \qquad + ( \mu \to - \mu ) \big) \bigg] \, .	\Vdistance	\nonumber
		\end{align}
	Using the expansion of the hypergeometric function \labelcref{eq:expansions_2f1} we obtain the renormalized result by sending $\Lambda \to \infty$,
		\begin{align}
			& \Ueff ( \barsigma, \mu, T, \sdim ) =	\Vdistance	\label{eq:U_sigma_mu_T_d}
			\\
			={}& \tfrac{\dimDirac}{2} \, \tfrac{\Sd}{( 2 \uppi )^\sdim} \, \bigg[ \tfrac{\Gamma ( \frac{\sdim}{2} ) \Gamma ( - \frac{\sdim + 1}{2} ) ( \sdim + 1)}{4 \sqrt{\uppi}} \, \big( \tfrac{1}{\sdim + 1} \, | \barsigma |^{\sdim + 1} - \tfrac{1}{2} \, \sigmaminvac^{\sdim - 1} \, \barsigma^2 \big)  +	\Vdistance	\nonumber
			\\
			& \quad - T \int_{0}^{\infty} \dd p \, p^{\sdim - 1} \ln \big[ 1 + \exp \big( - \tfrac{E + \mu}{T} \big) \big] +	\Vdistance	\nonumber
			\\
			& \quad + ( \mu \to - \mu ) \bigg] \, .	\Vdistance	\nonumber
		\end{align}
	We remark that sending $\Lambda \to \infty$ is only possible for $\sdim < 3$, while for $\sdim \geq 3$ there are divergent $\barsigma$-dependent terms.

	Carefully taking the limit $\sdim \to 1$, we recover the result \cite{Stoll:2021ori}
		\begin{align}
			& \Ueff ( \barsigma, \mu, T, 1 ) =	\Vdistance	\label{eq:U_sigma_mu_T_1}
			\\
			={}& \tfrac{\dimDirac}{2 \uppi} \, \bigg[ \tfrac{\barsigma^2}{4} \, \Big( \ln \big( \tfrac{\barsigma^2}{\sigmaminvac^2} \big) - 1 \Big) +	\Vdistance	\nonumber
			\\
			& \quad - T \, \int_{0}^{\infty} \dd p \, \ln \big[ 1 + \exp \big( - \tfrac{E + \mu}{T} \big) \big]  + ( \mu \to - \mu ) \big) \bigg] \, .	\Vdistance	\nonumber
		\end{align}
	Also for $\sdim \to 2$ one recovers a well-known literature results \cite{Lenz:2018thesis}.
		\begin{align}
			& \Ueff ( \barsigma, \mu, T, 2 ) =	\vdistance	\label{eq:U_sigma_mu_T_2}
			\\
			={}& \tfrac{\dimDirac}{4 \uppi} \, \Big[ \barsigma^2 \, \Big( \tfrac{| \barsigma |}{3} - \tfrac{| \sigmaminvac |}{2} \Big) + T^2 \, | \barsigma | \Li_2 \big( - \exp \big( - \tfrac{| \barsigma | + \mu}{T} \big) \big) +	\vdistance	\nonumber
			\\
			& \quad + T^3 \Li_3 \big( - \exp \big( - \tfrac{| \barsigma | + \mu}{T} \big) \big) + ( \mu \to - \mu ) \Big] \, .	\vdistance	\nonumber
		\end{align}

\paragraph{\texorpdfstring{$T \neq 0, \barsigma \neq 0, \mu = 0$}{T != 0, sigma != 0, mu = 0}}

	Yet, it is straightforward to evaluate the previous expressions for $\mu = 0$.
	From \cref{eq:U_sigma_mu_T_d} we find
		\begin{align}
			& \Ueff ( \barsigma, 0, T, \sdim ) =	\Vdistance	\label{eq:U_sigma_0_T_d}
			\\
			={}& \tfrac{\dimDirac}{2} \, \tfrac{\Sd}{( 2 \uppi )^\sdim} \, \Big[ \tfrac{\Gamma ( \frac{\sdim}{2} ) \Gamma ( - \frac{\sdim + 1}{2} ) ( \sdim + 1)}{4 \sqrt{\uppi}} \, \big( \tfrac{1}{\sdim + 1} \, | \barsigma |^{\sdim + 1} - \tfrac{1}{2} \, \sigmaminvac^{\sdim - 1} \, \barsigma^2 \big) +	\Vdistance	\nonumber
			\\
			& \quad - 2 \, T \int_{0}^{\infty} \dd p \, p^{\sdim - 1} \ln \big[ 1 + \exp \big( - \tfrac{E}{T} \big) \big] \Big] \, ,	\Vdistance	\nonumber
		\end{align}
	while for $\sdim \to 1$ we can simply set $\mu = 0$ in \cref{eq:U_sigma_mu_T_1},
		\begin{align}
			\Ueff ( \barsigma, 0, T, 1 ) ={}& \tfrac{\dimDirac}{2 \uppi} \, \Big[ \tfrac{\barsigma^2}{4} \, \Big( \ln \big( \tfrac{\barsigma^2}{\sigmaminvac^2} \big) - 1 \Big) +	\Vdistance	\label{eq:U_sigma_0_T_1}
			\\
			& \quad - 2 \int_{0}^{\infty} \dd p \, T \ln \big[ 1 + \exp \big( - \tfrac{E}{T} \big) \big] \Big] \, .	\Vdistance	\nonumber
		\end{align}
	Similarly, for $\sdim \to 2$ and $\mu = 0$ we can use \cref{eq:U_sigma_mu_T_2} and find
		\begin{align}
			& \Ueff ( \barsigma, \mu, T, 2 ) =	\vdistance	\label{eq:U_sigma_0_T_2}
			\\
			={}& \tfrac{\dimDirac}{4 \uppi} \, \Big[ \barsigma^2 \, \Big( \tfrac{| \barsigma |}{3} - \tfrac{| \sigmaminvac |}{2} \Big) + 2 \, T^2 \, | \barsigma | \Li_2 \big( - \exp \big( - \tfrac{| \barsigma |}{T} \big) \big) +	\vdistance	\nonumber
			\\
			& + 2 \, T^3 \Li_3 \big( - \exp \big( - \tfrac{| \barsigma |}{T} \big) \big) \Big] \, .	\vdistance	\nonumber
		\end{align}

\subsubsection{\texorpdfstring{$T \neq 0, \barsigma = 0$}{T != 0, sigma = 0}}

	Next, we turn to the cases where $\barsigma = 0$, hence, the potential at the origin of field space.

\paragraph{\texorpdfstring{$T \neq 0, \barsigma = 0, \mu \neq 0$}{T != 0, sigma = 0, mu != 0}}

	From \cref{eq:U_sigma_mu_T_d} for $\barsigma \to 0$ we can directly infer
		\begin{align}
			& \Ueff ( 0, \mu, T, \sdim ) =	\Vdistance	\label{eq:U_0_mu_T_d}
			\\
			={}& - \tfrac{\dimDirac}{2} \, \tfrac{\Sd}{( 2 \uppi )^\sdim} \, T \int_{0}^{\infty} \dd p \, p^{\sdim - 1} \, \ln \big[ 1 + \exp \big( - \tfrac{p + \mu}{T} \big) \big] +	\Vdistance	\nonumber
			\\
			& \quad + ( \mu \to - \mu ) \, .	\Vdistance	\nonumber
		\end{align}
	Similarly, the limit $\barsigma \to 0$ of \cref{eq:U_sigma_mu_T_1} for $\sdim = 1$ is well defined and the remaining integral can be evaluated analytically \cite{Koenigstein:2023wso},
		\begin{align}
			& \Ueff ( 0, \mu, T, 1 ) =	\Vdistance	\label{eq:U_0_mu_T_1}
			\\
			={}& - \tfrac{\dimDirac}{2 \uppi} \, \int_{0}^{\infty} \dd p \, \big[ T \ln \big[ 1 + \exp \big( - \tfrac{p + \mu}{T} \big) \big] +	\Vdistance	\nonumber
			\\
			& \quad + ( \mu \to - \mu ) \big] =	\Vdistance	\nonumber
			\\
			={}& - \tfrac{\dimDirac}{2 \uppi} \, \big( \tfrac{\uppi^2}{6} \, T^2 + \tfrac{1}{2} \, \mu^2 \big) \, .	\Vdistance	\nonumber
		\end{align}
	For $\sdim = 2$ one arrives at
		\begin{align}
			& \Ueff ( 0, \mu, T, 2 ) =	\vdistance	\label{eq:U_0_mu_T_2}
			\\
			={}& \tfrac{\dimDirac}{4 \uppi} \, T^3 \big[ \Li_3 \big( - \exp \big( - \tfrac{\mu}{T} \big) \big) + ( \mu \to - \mu ) \big]	\vdistance	\nonumber
		\end{align}
	by taking the $\barsigma \to 0$ limit of \cref{eq:U_sigma_mu_T_2}.

\paragraph{\texorpdfstring{$T \neq 0, \barsigma = 0, \mu = 0$}{T != 0, sigma = 0, mu = 0}}

	It is straight forward to also set $\mu = 0$ in the previous formulae.
	For general $1 \leq \sdim < 3$ we find
		\begin{align}
			&	\Ueff ( 0, 0, T, \sdim ) =	\Vdistance	\label{eq:U_0_0_T_d}
			\\
			={}& - \tfrac{\dimDirac}{2} \, \tfrac{\Sd}{( 2 \uppi )^\sdim} \, 2 \int_{0}^{\infty} \dd p \, p^{\sdim - 1} \, T \ln \big[ 1 + \exp \big( - \tfrac{p}{T} \big) \big] \, ,	\Vdistance	\nonumber
		\end{align}
	while the special case $\sdim = 1$ evaluates to
		\begin{align}
			\Ueff ( 0, 0, T, 1 ) ={}& - \tfrac{\dimDirac}{2 \uppi} \, \tfrac{\uppi^2}{6} \, T^2 \, .	\label{eq:U_0_0_T_1}
		\end{align}
	For $\sdim = 2$ we have
		\begin{align}
			\Ueff ( 0, 0, T, 2 ) ={}& - \tfrac{\dimDirac}{4 \uppi} \, \tfrac{3}{2} \, \zeta ( 3 ) \, T^3 \, .	\label{eq:U_0_0_T_2}
		\end{align}

\subsection{\texorpdfstring{$T = 0$}{T = 0}}

	Next, we turn to the special cases, where $T = 0$.

\subsubsection{\texorpdfstring{$T = 0, \barsigma \neq 0$}{T != 0, sigma != 0}}
		
	We start off with nonzero background field $\barsigma$.

\paragraph{\texorpdfstring{$T = 0, \barsigma \neq 0, \mu \neq 0$}{T = 0, sigma != 0, mu != 0}}

	Using the explicit regularized expressions \cref{eq:l_0_sigma_mu_0_d_Lambda,eq:l_1_sigma_mu_0_d_Lambda} and inserting these in \cref{eq:effective_potential_regularized_main} we find
		\begin{align}
			& \Ueff^\Lambda ( \barsigma, \mu, 0, \sdim ) =	\Vdistance	\label{eq:U_sigma_mu_0_d_Lambda}
			\\
			={}& \tfrac{\dimDirac}{2} \, \big[ \barsigma^2 \, l_1^\Lambda ( \sigmaminvac, 0, 0, \sdim ) - l_0^\Lambda ( \barsigma, \mu, 0, \sdim ) \big] =	\Vdistance	\nonumber	
			\\
			={}& \tfrac{\dimDirac}{2} \, \tfrac{\Sd}{( 2 \uppi )^\sdim} \, \Big[ \barsigma^2 \, \tfrac{1}{2} \, \tfrac{| \sigmaminvac |^{\sdim - 1}}{\sdim} \, \big| \tfrac{\Lambda}{\sigmaminvac} \big|^\sdim \, \pFq{2}{1} \big( \tfrac{1}{2}, \tfrac{\sdim}{2}; \tfrac{\sdim + 2}{2}; - \tfrac{\Lambda^2}{\sigmaminvac^2} \big) +	\Vdistance	\nonumber
			\\
			& \quad - \tfrac{| \barsigma |^{\sdim + 1}}{\sdim} \, \Big[ \big| \tfrac{\Lambda}{\barsigma} \big|^\sdim \, \pFq{2}{1} \big( - \tfrac{1}{2}, \tfrac{\sdim}{2}; \tfrac{\sdim + 2}{2}; - \tfrac{\Lambda^2}{\barsigma^2} \big) +	\Vdistance	\nonumber
			\\
			& \qquad - \Theta \big( \tfrac{\barmu^2}{\barsigma^2} \big) \, \big| \tfrac{\barmu}{\barsigma} \big|^\sdim \, \big( \pFq{2}{1} \big( - \tfrac{1}{2}, \tfrac{\sdim}{2}; \tfrac{\sdim + 2}{2}; - \tfrac{\barmu^2}{\barsigma^2} \big) - \big| \tfrac{\mu}{\barsigma} \big| \big) \Big] \Big] =	\Vdistance	\nonumber
			\\
			={}& \tfrac{\dimDirac}{2} \, \tfrac{\Sd}{( 2 \uppi )^\sdim} \, \Big[ \tfrac{| \barsigma |}{\sdim} \, \Lambda^\sdim \, \Big[ \tfrac{1}{2} \, \tfrac{| \barsigma |}{\sigmaminvac} \, \pFq{2}{1} \big( \tfrac{1}{2}, \tfrac{\sdim}{2}; \tfrac{\sdim + 2}{2}; - \tfrac{\Lambda^2}{\sigmaminvac^2} \big) +	\Vdistance	\nonumber
			\\
			& \qquad - \pFq{2}{1} \big( - \tfrac{1}{2}, \tfrac{\sdim}{2}; \tfrac{\sdim + 2}{2}; - \tfrac{\Lambda^2}{\barsigma^2} \big) \Big] +	\Vdistance	\nonumber
			\\
			& \quad + \Theta \big( \tfrac{\barmu^2}{\barsigma^2} \big) \, \tfrac{| \barsigma |}{\sdim} \, | \barmu |^\sdim \, \big( \pFq{2}{1} \big( - \tfrac{1}{2}, \tfrac{\sdim}{2}; \tfrac{\sdim + 2}{2}; - \tfrac{\barmu^2}{\barsigma^2} \big) - \big| \tfrac{\mu}{\barsigma} \big| \big) \Big] \, .\Vdistance	\nonumber
		\end{align}
	Here, by sending $\Lambda \to \infty$ we remove the cutoff and find
		\begin{align}
			& \Ueff ( \barsigma, \mu, 0, \sdim ) =	\Vdistance	\label{eq:U_sigma_mu_0_d}
			\\
			={}& \tfrac{\dimDirac}{2} \, \tfrac{\Sd}{( 2 \uppi )^\sdim} \, \Big[ \tfrac{\Gamma ( \frac{\sdim}{2} ) \Gamma ( - \frac{\sdim + 1}{2} ) ( \sdim + 1)}{4 \sqrt{\uppi}} \, \big( \tfrac{1}{\sdim + 1} \, | \barsigma |^{\sdim + 1} - \tfrac{1}{2} \, \sigmaminvac^{\sdim - 1} \, \barsigma^2 \big) +	\Vdistance	\nonumber
			\\
			& \quad + \Theta \big( \tfrac{\barmu^2}{\barsigma^2} \big) \, \tfrac{| \barsigma |}{\sdim} \, | \barmu |^\sdim \, \big( \pFq{2}{1} \big( - \tfrac{1}{2}, \tfrac{\sdim}{2}; \tfrac{\sdim + 2}{2}; - \tfrac{\barmu^2}{\barsigma^2} \big) - \big| \tfrac{\mu}{\barsigma} \big| \big) \Big] \, .	\Vdistance	\nonumber
		\end{align}
	The equivalent expressions can be found by taking the limit $T\to0$ of \cref{eq:U_sigma_mu_T_d}.
	For the special case $\sdim = 1$ we recover \cite{Stoll:2021ori}
		\begin{align}
			\Ueff ( \barsigma, \mu, 0, 1 ) ={}& \tfrac{\dimDirac}{2 \uppi} \, \Big[ \tfrac{\barsigma^2}{4} \, \Big( \ln \big( \tfrac{\barsigma^2}{\sigmaminvac^2} \big) - 1 \Big) +	\vdistance	\label{eq:U_sigma_mu_0_1}
			\\
			& \quad + \Theta \big( \tfrac{\barmu^2}{\barsigma^2} \big) \, \big( \tfrac{\barsigma^2}{2} \arsinh \big( \tfrac{\barmu}{\barsigma} \big) - \tfrac{1}{2} \, \barmu \, | \mu | \big) \Big] \Big] \, ,	\vdistance	\nonumber
		\end{align}
	and for $\sdim = 2$ \cite[Eq.~4.38]{Lenz:2018thesis}
		\begin{align}
			\Ueff ( \barsigma, \mu, 0, 2 ) ={}& \tfrac{\dimDirac}{4 \uppi} \, \Big[ \barsigma^2 \, \Big( \tfrac{| \barsigma |}{3} - \tfrac{ \sigmaminvac }{2} \Big) +	\vdistance	\label{eq:U_sigma_mu_0_2}
			\\
			& \quad + \Theta \big( \tfrac{\barmu^2}{\barsigma^2} \big) \, \big( - \tfrac{\barsigma^3}{3} - \tfrac{| \mu |^3}{6} + \tfrac{\barsigma^2 | \mu |}{2} \big) \Big] \, .	\vdistance	\nonumber
		\end{align}

\paragraph{\texorpdfstring{$T = 0, \barsigma \neq 0, \mu = 0$}{T = 0, sigma != 0, mu = 0}}

	The results for $\mu = 0$ are a direct consequence of the previous results.
	In general, we find
		\begin{align}
			& \Ueff ( \barsigma, 0, 0, \sdim ) =	\Vdistance	\label{eq:U_sigma_0_0_d}
			\\
			={}& \tfrac{\dimDirac}{2} \, \tfrac{\Sd}{( 2 \uppi )^\sdim} \, \tfrac{\Gamma ( \frac{\sdim}{2} ) \Gamma ( - \frac{\sdim + 1}{2} ) ( \sdim + 1)}{4 \sqrt{\uppi}} \, \big( \tfrac{1}{\sdim + 1} \, | \barsigma |^{\sdim + 1} - \tfrac{1}{2} \, \sigmaminvac^{\sdim - 1} \, \barsigma^2 \big) \, ,	\Vdistance	\nonumber
		\end{align}
	which reduces for $\sdim = 1$ to
		\begin{align}
			\Ueff ( \barsigma, 0, 0, 1 ) ={}& \tfrac{\dimDirac}{2 \uppi} \, \tfrac{\barsigma^2}{4} \, \Big( \ln \big( \tfrac{\barsigma^2}{\sigmaminvac^2} \big) - 1 \Big) \, 	\label{eq:U_sigma_0_0_1}
		\end{align}
	and for $\sdim = 2$ to
		\begin{align}
			\Ueff ( \barsigma, 0, 0, 2 ) ={}& \tfrac{\dimDirac}{4 \uppi} \, \barsigma^2 \, \Big( \tfrac{| \barsigma |}{3} - \tfrac{| \sigmaminvac |}{2} \Big) \, .	\label{eq:U_sigma_0_0_2}
		\end{align}

\subsubsection{\texorpdfstring{$T = 0, \barsigma = 0$}{T != 0, sigma = 0}}

	Last, we turn to the case, where we study the potential again for $\barsigma = 0$.

\paragraph{\texorpdfstring{$T = 0, \barsigma = 0, \mu \neq 0$}{T = 0, sigma = 0, mu != 0}}
		
	Here, one finds from \cref{eq:U_sigma_mu_0_d} with \cref{eq:expansions_2f1}
		\begin{align}
			\Ueff ( 0, \mu, 0, \sdim ) ={}& \tfrac{\dimDirac}{2} \, \tfrac{\Sd}{( 2 \uppi )^\sdim} \, \Big[ - \tfrac{| \mu |^{\sdim +1}}{\sdim ( \sdim + 1 )} \Big]	\label{eq:U_0_mu_0_d}
		\end{align}
	For $\sdim = 1$ this is
		\begin{align}
			\Ueff ( 0, \mu, 0, 1 ) ={}& - \tfrac{\dimDirac}{2 \uppi} \, \tfrac{\mu^2}{2} \, ,	\label{eq:U_0_mu_0_1}
		\end{align}
	while for $\sdim = 2$ we have
		\begin{align}
			\Ueff ( 0, \mu, 0, 2 ) ={}& - \tfrac{\dimDirac}{4 \uppi} \, \tfrac{| \mu |^{3}}{6} \, .	\label{eq:U_0_mu_0_2}
		\end{align}
	The latter special cases can also be derived from the $T \neq 0$ formulas \cref{eq:U_0_mu_T_1,eq:U_0_mu_T_2} by sending $T \to 0$.
		
\paragraph{\texorpdfstring{$T = 0, \barsigma = 0, \mu = 0$}{T = 0, sigma = 0, mu = 0}}
	
	The trivial and last case is
		\begin{align}
			\Ueff ( 0, 0, 0, \sdim ) ={}& 0 \, .	\label{eq:U_0_0_0_d}
		\end{align}
	(Certainly, one could always add an arbitrary constant to the potential without changing the physical observables.)

\section{The bosonic wave-function renormalization}
\label{app:wave_function_renormalization}

	This appendix is dedicated to calculations as well as the presentation of detailed 	expressions and limiting cases for the bosonic wave-function renormalization \labelcref{eq:wave-function_renormalization_main}.
	We calculate the renormalized limits, such that the final results do not contain any \gls{uv} cutoff.
	Step by step, we provide expressions for the cases where $\barsigma$, $\mu$, and $T$ are zero or nonzero.
	Additionally, we evaluate the wave-function renormalization for the special cases $\sdim = 1$ and $\sdim = 2$ and demonstrate that we reproduce known literature results.
	All cases are collected in \cref{tab:z_explicit_expressions}, which links to the explicit formulae.

	However, we start by providing some useful intermediate steps for the derivation of the general formula for the bosonic wave-function renormalization \cref{eq:wave-function_renormalization_main}.
	The starting point is
		\begin{align}
			z ( \barsigma, \mu, T, \sdim ) = \tfrac{1}{2} \, \tfrac{\dd^2}{\dd q^2} \, \Gamma^{(2)} ( \barsigma, \mu, T, q, \sdim ) \Big|_{q = 0} \, ,
		\end{align}
	We note that the bosonic two-point function solely depends on the absolute/square of the spatial external momentum, we can use
		\begin{align}
			&	u = q^2 \, ,	&&	\Rightarrow	&&	\tfrac{1}{2} \, \tfrac{\dd^2}{\dd q^2} = \tfrac{\dd}{\dd u} + 2 u \, \tfrac{\dd^2}{\dd u^2}
		\end{align}
	to evaluate the derivative,
		\begin{align}
			z ( \barsigma, \mu, T, \sdim ) =  \big( \tfrac{\dd}{\dd u} + 2 u \, \tfrac{\dd^2}{\dd u^2} \big) \, \Gamma^{(2)} ( \barsigma, \mu, T, u, \sdim ) \Big|_{u = 0} \, .
		\end{align}
	Inserting the general expression \labelcref{eq:gamma2_main} for the bosonic two-point function, we obtain,
		\begin{align}
			& z ( \barsigma, \mu, T, \sdim ) =	\vdistance
			\\
			={}& \big( \tfrac{\dd}{\dd u} + 2 u \, \tfrac{\dd^2}{\dd u^2} \big) \, \big( \tfrac{\dimDirac}{2} \, ( u + 4 \, \barsigma^2 ) \, l_2 ( \barsigma, \mu, T, u, \sdim ) \big) \Big|_{u = 0} =	\vdistance	\nonumber
			\\
			={}& \tfrac{\dimDirac}{2} \, \big[ l_2 ( \barsigma, \mu, T, u, \sdim ) + 4 \barsigma^2 \, \tfrac{\dd}{\dd u} \, l_2 ( \barsigma, \mu, T, u, \sdim ) \big] \Big|_{u = 0} =	\vdistance	\nonumber
			\\
			={}& \tfrac{\dimDirac}{2} \, \big[ l_2 ( \barsigma, \mu, T, 0, \sdim ) - \tfrac{8}{6} \, \barsigma^2 \, l_3 ( \barsigma, \mu, T, \sdim ) \big]	\vdistance	\nonumber
		\end{align}
	where we defined the Matsubara sum and integral formula \labelcref{eq:l_3_sigma_mu_T_d}.
		\setlength{\extrarowheight}{3pt}
		\begin{table}
			\caption{\label{tab:z_explicit_expressions}%
			Direct links to the formulae for the bosonic wave-function renormalization $z ( \barsigma, \mu, T, \sdim )$.
			The formulae are simplified in terms of known functions as far as possible.
			}
			\begin{ruledtabular}
				\begin{tabular}{c c c c c c}
					$T$							&	$\barsigma$					&	$\mu$	&	$1 \leq \sdim < 3$				&	$\sdim = 1$	&	$\sdim =2$
					\\
					\hline
					\multirow{4}{*}{$\neq 0$}	&	\multirow{2}{*}{$\neq 0$}	&	$\neq 0$	&	\cref{eq:z_sigma_mu_T_d}	& 	\cref{eq:z_sigma_mu_T_1}	&	\cref{eq:z_sigma_mu_T_2}
					\\
					\cline{3-6}
												&								&	$= 0$		&	\cref{eq:z_sigma_0_T_d}		&	\cref{eq:z_sigma_0_T_1} 	&	\cref{eq:z_sigma_0_T_2}
					\\
					\cline{2-6}
												&	\multirow{2}{*}{$= 0$}		&	$\neq 0$	&	\cref{eq:z_0_mu_T_d}		&	\cref{eq:z_0_mu_T_1} 		&	\cref{eq:z_0_mu_T_2}
					\\
					\cline{3-6}
												&								&	$= 0$		&	\cref{eq:z_0_0_T_d}			&	\cref{eq:z_0_0_T_1} 		&	\cref{eq:z_0_0_T_2}
					\\
					\hline
					\multirow{4}{*}{$= 0$}		&	\multirow{2}{*}{$\neq 0$}	&	$\neq 0$	&	\cref{eq:z_sigma_mu_0_d}	& 	\cref{eq:z_sigma_mu_0_1}	&	\cref{eq:z_sigma_mu_0_2}
					\\
					\cline{3-6}
												&								&	$= 0$		&	\cref{eq:z_sigma_0_0_d}		& 	\cref{eq:z_sigma_0_0_1}		&	\cref{eq:z_sigma_0_0_2}
					\\
					\cline{2-6}
												&	\multirow{2}{*}{$= 0$}		&	$\neq 0$	&	\cref{eq:z_0_mu_0_d}		& 	\cref{eq:z_0_mu_0_1}		&	\cref{eq:z_0_mu_0_2}
					\\
					\cline{3-6}
												&								&	$= 0$		&	\cref{eq:z_0_0_0_d}			& 	\cref{eq:z_0_0_0_d}			&	\cref{eq:z_0_0_0_d}
				\end{tabular}
			\end{ruledtabular}
		\end{table}

\subsection{\texorpdfstring{$T \neq 0$}{T != 0}}

	We start with the wave-function renormalization in the heat bath with $T \neq 0$.
		
\subsubsection{\texorpdfstring{$T \neq 0, \barsigma \neq 0$}{T != 0, sigma != 0}}

	First, we study the wave-function renormalization for nontrivial background field configurations $\barsigma \neq 0$, \eg, in the phase of symmetry breaking.
		
\paragraph{\texorpdfstring{$T \neq 0, \barsigma \neq 0, \mu \neq 0$}{T != 0, sigma != 0, mu != 0}}

	For general $\mu \neq 0$ we simply insert \cref{eq:l_2_sigma_mu_T_q_d_final} for $q = 0$ and \cref{eq:l_3_sigma_mu_T_d} in \cref{eq:wave-function_renormalization_main}.
	For continuous $\sdim$ we find
		\begin{align}
			& z ( \barsigma, \mu, T, \sdim ) =	\Vdistance	\label{eq:z_sigma_mu_T_d}
			\\
			={}& \tfrac{\dimDirac}{2} \, \big[ l_2 ( \barsigma, \mu, T, 0, \sdim ) - \tfrac{8}{6} \, \barsigma^2 \, l_3 ( \barsigma, \mu, T, \sdim ) \big] =	\Vdistance	\nonumber
			\\
			={}& \tfrac{\dimDirac}{8} \, \tfrac{\Sd}{( 2 \uppi )^\sdim} \bigg( | \barsigma |^{\sdim - 3} \, \tfrac{\Gamma ( \frac{3 - \sdim}{2} ) \, \Gamma ( \frac{\sdim}{2} )}{\sqrt{\uppi}} \left[1- \tfrac{2}{3} \left(\tfrac{3-d}{2}\right)\right] +	\Vdistance	\nonumber
			\\
			& \quad - \int_{0}^{\infty} \dd p \, p^{\sdim - 1} \, \Big( \tfrac{1}{E^3} \, \big[ \nf \big( \tfrac{E + \mu}{T} \big) +	\Vdistance	\nonumber
			\\
			& \qquad - \tfrac{E}{T} \, \big[ \nf^2 \big( \tfrac{E + \mu}{T} \big) - \nf \big( \tfrac{E + \mu}{T} \big) \big] \big] +	\Vdistance	\nonumber
			\\
			& \quad - \barsigma^2 \, \tfrac{1}{E^5} \, \big[ \nf \big( \tfrac{E + \mu}{T} \big) - \tfrac{E}{T} \, \big[ \nf^2 \big( \tfrac{E + \mu}{T} \big) - \nf \big( \tfrac{E + \mu}{T} \big) \big] +	\Vdistance	\nonumber
			\\
			& \qquad + \big( \tfrac{E}{T} \big)^2 \, \big[ \tfrac{2}{3} \, \nf^3 \big( \tfrac{E + \mu}{T} \big) - \nf^2 \big( \tfrac{E + \mu}{T} \big) + \tfrac{1}{3} \, \nf \big( \tfrac{E + \mu}{T} \big) \big] +	\Vdistance	\nonumber
			\\
			& \quad + ( \mu \to - \mu ) \Big) \, .	\Vdistance	\nonumber
		\end{align}
	Setting $\sdim = 1$ leads to the known result \cite{Koenigstein:2021llr}
		\begin{align}
			& z ( \barsigma, \mu, T, 1 ) =	\Vdistance	\label{eq:z_sigma_mu_T_1}
			\\
			={}& \tfrac{\dimDirac}{8 \uppi} \, \bigg[ \tfrac{1}{3} \tfrac{1}{\barsigma^2} - \int_{0}^{\infty} \dd p \, \Big( \tfrac{1}{E^3} \, \big[ \nf \big( \tfrac{E + \mu}{T} \big) +	\Vdistance	\nonumber
			\\
			& \qquad - \tfrac{E}{T} \, \big[ \nf^2 \big( \tfrac{E + \mu}{T} \big) - \nf \big( \tfrac{E + \mu}{T} \big) \big] \big] +	\Vdistance	\nonumber
			\\
			& \quad - \barsigma^2 \, \tfrac{1}{E^5} \, \big[ \nf \big( \tfrac{E + \mu}{T} \big) - \tfrac{E}{T} \, \big[ \nf^2 \big( \tfrac{E + \mu}{T} \big) - \nf \big( \tfrac{E + \mu}{T} \big) \big] +	\Vdistance	\nonumber
			\\
			& \qquad + \big( \tfrac{E}{T} \big)^2 \, \big[ \tfrac{2}{3} \, \nf^3 \big( \tfrac{E + \mu}{T} \big) - \nf^2 \big( \tfrac{E + \mu}{T} \big) + \tfrac{1}{3} \, \nf \big( \tfrac{E + \mu}{T} \big) \big] +	\Vdistance	\nonumber
			\\
			& \quad + ( \mu \to - \mu ) \Big]	\Vdistance	\nonumber
		\end{align}
	On the other hand, for $\sdim = 2$, all integrals can be evaluated analytically,
		\begin{align}
			& z ( \barsigma, \mu, T, 2 ) =	\Vdistance	\label{eq:z_sigma_mu_T_2}
			\\
			={}& \tfrac{\dimDirac}{24 \uppi} \, \tfrac{1}{| \barsigma |} \, \Big( 1 - \nf \big( \tfrac{| \barsigma | + \mu}{T} \big) +	\Vdistance	\nonumber
			\\
			& \quad - \tfrac{1}{2} \, \tfrac{| \barsigma |}{T} \, \big[ \nf^2 \big( \tfrac{| \barsigma | + \mu}{T} \big) - \nf \big( \tfrac{| \barsigma | + \mu}{T} \big) \big] + ( \mu \to - \mu ) \Big) \, .	\Vdistance	\nonumber
		\end{align}

\paragraph{\texorpdfstring{$T \neq 0, \barsigma \neq 0, \mu = 0$}{T != 0, sigma != 0, mu = 0}}

	The cases for $\mu = 0$ can be inferred from the previous results.
	Hence,
		\begin{align}
			& z ( \barsigma, 0, T, \sdim ) =	\Vdistance	\label{eq:z_sigma_0_T_d}
			\\
			={}& \tfrac{\dimDirac}{8} \, \tfrac{\Sd}{( 2 \uppi )^\sdim} \bigg[ | \barsigma |^{\sdim - 3} \, \tfrac{\Gamma ( \frac{3 - \sdim}{2} ) \, \Gamma ( \frac{\sdim}{2} )}{\sqrt{\uppi}} \left[1- \tfrac{2}{3} \left(\tfrac{3-d}{2}\right)\right] +	\Vdistance	\nonumber
			\\
			& \quad - 2 \int_{0}^{\infty} \dd p \, p^{\sdim - 1} \, \Big( \tfrac{1}{E^3} \, \big[ \nf \big( \tfrac{E}{T} \big) +	\Vdistance	\nonumber
			\\
			& \qquad - \tfrac{E}{T} \, \big[ \nf^2 \big( \tfrac{E}{T} \big) - \nf \big( \tfrac{E}{T} \big) \big] \big] +	\Vdistance	\nonumber
			\\
			& \quad - \barsigma^2 \, \tfrac{1}{E^5} \, \big[ \nf \big( \tfrac{E}{T} \big) - \tfrac{E}{T} \, \big[ \nf^2 \big( \tfrac{E}{T} \big) - \nf \big( \tfrac{E}{T} \big) \big] +	\Vdistance	\nonumber
			\\
			& \qquad + \big( \tfrac{E}{T} \big)^2 \, \big[ \tfrac{2}{3} \, \nf^3 \big( \tfrac{E}{T} \big) - \nf^2 \big( \tfrac{E}{T} \big) + \tfrac{1}{3} \, \nf \big( \tfrac{E}{T} \big) \big] \Big) \bigg] \, .	\Vdistance	\nonumber
		\end{align}
	Furthermore, in the limit $\sdim = 1$ we have
		\begin{align}
			& z ( \barsigma, 0, T, 1 ) =	\Vdistance	\label{eq:z_sigma_0_T_1}
			\\
			={}& \tfrac{\dimDirac}{8 \uppi} \, \bigg[ \tfrac{1}{3} \tfrac{1}{\barsigma^2} - 2 \int_{0}^{\infty} \dd p \, \Big( \tfrac{1}{E^3} \, \big[ \nf \big( \tfrac{E}{T} \big) +	\Vdistance	\nonumber
			\\
			& \qquad - \tfrac{E}{T} \, \big[ \nf^2 \big( \tfrac{E}{T} \big) - \nf \big( \tfrac{E}{T} \big) \big] \big] +	\Vdistance	\nonumber
			\\
			& \quad - \barsigma^2 \, \tfrac{1}{E^5} \, \big[ \nf \big( \tfrac{E}{T} \big) - \tfrac{E}{T} \, \big[ \nf^2 \big( \tfrac{E}{T} \big) - \nf \big( \tfrac{E}{T} \big) \big] +	\Vdistance	\nonumber
			\\
			& \qquad + \big( \tfrac{E}{T} \big)^2 \, \big[ \tfrac{2}{3} \, \nf^3 \big( \tfrac{E}{T} \big) - \nf^2 \big( \tfrac{E}{T} \big) + \tfrac{1}{3} \, \nf \big( \tfrac{E}{T} \big) \big] \Big) \bigg] \, ,	\Vdistance	\nonumber
		\end{align}
	and
		\begin{align}
			& z ( \barsigma, 0, T, 2 ) =	\Vdistance	\label{eq:z_sigma_0_T_2}
			\\
			={}& \tfrac{\dimDirac}{24 \uppi} \, \tfrac{1}{| \barsigma |} \, \Big( 1 - 2 \, \nf \big( \tfrac{| \barsigma |}{T} \big) - \tfrac{| \barsigma |}{T} \, \big[ \nf^2 \big( \tfrac{| \barsigma |}{T} \big) - \nf \big( \tfrac{| \barsigma |}{T} \big) \big] \Big)	\Vdistance	\nonumber
		\end{align}
	for $\sdim = 2$.

\subsubsection{\texorpdfstring{$T \neq 0, \barsigma = 0$}{T != 0, sigma = 0}}

	Next, we turn to the wave-function renormalization in the symmetric phase for $\barsigma = 0$.
		
\paragraph{\texorpdfstring{$T \neq 0, \barsigma = 0, \mu \neq 0$}{T != 0, sigma = 0, mu != 0}}
	
	We start at nonzero chemical potential.
	Both the vacuum and the medium contribution separately exhibit an \gls{ir} divergence, which cancel each other.
	To account for this, we consider the vacuum part in its integral form together with the medium part and write
		\begin{align}
			& z ( 0, \mu, T, \sdim ) =	\Vdistance	\label{eq:z_0_mu_T_d}
			\\
			={}& \tfrac{\dimDirac}{8} \, \tfrac{\Sd}{( 2 \uppi )^\sdim} \int_{0}^{\infty} \dd p \, p^{\sdim - 4} \, \big[ 1 - \nf \big( \tfrac{p + \mu}{T} \big) +	\Vdistance	\nonumber
			\\
			& \quad + \tfrac{p}{T} \, \big[ \nf^2 \big( \tfrac{p + \mu}{T} \big) - \nf \big( \tfrac{p + \mu}{T} \big) \big] + ( \mu \to - \mu ) \big] \, .	\Vdistance	\nonumber
		\end{align}
	The tricky evaluation of this expression for $\sdim = 1$ is presented in \RciteSingle[Eq.~F.65]{Koenigstein:2023wso} and one finds
		\begin{align}
			z ( 0, \mu, T, 1 ) ={}& - \tfrac{\dimDirac}{2 \uppi} \, \tfrac{1}{8 T^2} \, \mathrm{DLi}_{-2} \big( \tfrac{\mu}{T} \big) \, ,	\label{eq:z_0_mu_T_1}
		\end{align}
	where we used the definition \labelcref{eq:DLi_definition}.
	While for $\sdim = 2$ integration by parts leads to
		\begin{align}
			z ( 0, \mu, T, 2 ) ={}& \tfrac{\dimDirac}{8 \uppi} \tfrac{1}{4 T \cosh^2 \left( \tfrac{\mu}{2 T} \right)} \, .	\label{eq:z_0_mu_T_2}
		\end{align}

\paragraph{\texorpdfstring{$T \neq 0, \barsigma = 0, \mu = 0$}{T != 0, sigma = 0, mu = 0}}

	At vanishing chemical potential \cref{eq:z_0_mu_T_d} simplifies to
		\begin{align}
			z ( 0, 0, T, \sdim ) ={}& \tfrac{\dimDirac}{8} \, \tfrac{\Sd}{( 2 \uppi )^\sdim} \int_{0}^{\infty} \dd p \, p^{\sdim - 4} \, \big[ 1 - 2 \, \nf \big( \tfrac{p}{T} \big) +	\Vdistance	\nonumber
			\\
			& \quad + 2 \, \tfrac{p}{T} \, \big[ \nf^2 \big( \tfrac{p}{T} \big) - \nf \big( \tfrac{p}{T} \big) \big] \big] \, .	\Vdistance	\label{eq:z_0_0_T_d}
		\end{align}
	It is possible to show that the integral approaches a zeta function for $\sdim \to 1$,
		\begin{align}
			z ( 0, 0, T, 1 ) ={}& \tfrac{\dimDirac}{2 \uppi} \, \tfrac{7}{16 \uppi^2} \, \zeta ( 3 ) \, \tfrac{1}{T^2} \, .	\label{eq:z_0_0_T_1}
		\end{align}
	For $\sdim = 2$ we can simply use \cref{eq:z_0_mu_T_2} and set $\mu = 0$,
		\begin{align}
			z ( 0, 0, T, 2 ) ={}& \tfrac{\dimDirac}{16 \uppi} \tfrac{1}{2 T} \, .	\label{eq:z_0_0_T_2}
		\end{align}

\subsection{\texorpdfstring{$T = 0$}{T = 0}}

	Having discussed all cases with a heat bath, we can next turn to $T = 0$.
		
\subsubsection{\texorpdfstring{$T = 0, \barsigma \neq 0$}{T = 0, sigma != 0}}

	Again, we start in the phase with a nontrivial expectation value of the bosonic field and therefore evaluate the wave-function renormalization at $\barsigma \neq 0$.

\paragraph{\texorpdfstring{$T = 0, \barsigma \neq 0, \mu \neq 0$}{T = 0, sigma != 0, mu != 0}}

	Keeping $\mu \neq 0$ the general expression for continuous $\sdim$ reads
		\begin{align}
			& z ( \barsigma, \mu, 0, \sdim ) =	\Vdistance	\label{eq:z_sigma_mu_0_d}
			\\
			={}& \tfrac{\dimDirac}{2} \, \big[ l_2 ( \barsigma, \mu, 0, 0, \sdim ) - \tfrac{8}{6} \, \barsigma^2 \, l_3 ( \barsigma, \mu, 0, \sdim ) \big] =	\Vdistance	\nonumber
			\\
			={}& \tfrac{\dimDirac}{2} \, \tfrac{\Sd}{( 2 \uppi )^\sdim} \, \tfrac{1}{4} \Big[
				| \barsigma |^{\sdim - 3} \, \tfrac{\Gamma ( \frac{3 - \sdim}{2} ) \, \Gamma ( \frac{\sdim}{2} )}{\sqrt{\uppi}} \left[1- \tfrac{2}{3} \left(\tfrac{3-d}{2}\right)\right] +	\Vdistance	\nonumber
				\\
				& \quad - \Theta \big( \tfrac{\barmu^2}{\barsigma^2} \big) \Big( \tfrac{\barmu^\sdim}{\barsigma^3} \, \tfrac{1}{\sdim} \, \pFq{2}{1} \big( \tfrac{3}{2}, \tfrac{\sdim}{2}; \tfrac{\sdim + 2}{2}; - \tfrac{\barmu^2}{\barsigma^2} \big) +	\Vdistance	\nonumber
				\\
				& \qquad - \tfrac{\barmu^\sdim}{| \barsigma |^3} \, \tfrac{1}{\sdim} \, \pFq{2}{1} \big( \tfrac{5}{2}, \tfrac{\sdim}{2}; \tfrac{\sdim + 2}{2}; - \tfrac{\barmu^2}{\barsigma^2} \big) + \tfrac{\barmu^{\sdim - 2}}{| \mu |} +	\Vdistance	\nonumber
				\\
				& \qquad - \tfrac{\sdim - 2}{3} \, \tfrac{\barsigma^2 \, \barmu^{\sdim - 4}}{| \mu |} - \tfrac{\barsigma^2}{3} \, \tfrac{\barmu^{\sdim - 2}}{| \mu |^3} \Big) \Big]	\Vdistance	\nonumber
		\end{align}
	where we used \cref{eq:l_2_sigma_mu_0_q_d} for $q = 0$ and \cref{eq:l_3_sigma_mu_0_d}.
	For $\sdim = 1$ this simplifies drastically, see also \RciteSingle[Eq.~F.68]{Koenigstein:2023wso} 
		\begin{align}
			z ( \barsigma, \mu, 0, 1 ) ={}& \tfrac{\dimDirac}{2 \uppi} \, \tfrac{1}{12} \, \tfrac{1}{\barsigma^2} \Big[ 1 - \Theta \big( \tfrac{\barmu^2}{\barsigma^2} \big) \, \big| \tfrac{\mu}{\barmu} \big|^3 \Big] \, 	\label{eq:z_sigma_mu_0_1}
		\end{align}
	and for $\sdim = 2$
		\begin{align}
			z ( \barsigma, \mu, 0, 2 ) ={}& \tfrac{\dimDirac}{4 \uppi} \, \tfrac{1}{6} \, \tfrac{1}{| \barsigma |} \Big[ 1 - \Theta \big( \tfrac{\barmu^2}{\barsigma^2} \big) \Big] \, .	\label{eq:z_sigma_mu_0_2}
		\end{align}

\paragraph{\texorpdfstring{$T = 0, \barsigma \neq 0, \mu = 0$}{T = 0, sigma != 0, mu = 0}}

	Having zero chemical potential, the previous expressions are even simpler.
	For continuous $\sdim$ only the vacuum contribution remains
		\begin{align}
			& z ( \barsigma, 0, 0, \sdim ) =	\Vdistance	\label{eq:z_sigma_0_0_d}
			\\
			={}& \tfrac{\dimDirac}{2} \, \tfrac{\Sd}{( 2 \uppi )^\sdim} \, \tfrac{1}{4}\, | \barsigma |^{\sdim - 3} \, \tfrac{\Gamma ( \frac{3 - \sdim}{2} ) \, \Gamma ( \frac{\sdim}{2} )}{\sqrt{\uppi}} \left[1- \tfrac{2}{3} \left(\tfrac{3-d}{2}\right)\right] \, ,	\Vdistance	\nonumber
		\end{align}
	which again simplifies for $\sdim = 1$,
		\begin{align}
			z ( \barsigma, 0, 0, 1 ) ={}& \tfrac{\dimDirac}{2\uppi} \, \tfrac{1}{12} \, \tfrac{1}{\barsigma^2} \, ,	\label{eq:z_sigma_0_0_1}
		\end{align}
	and for $\sdim = 2$,
		\begin{align}
			z ( \barsigma, 0, 0, 2 ) ={}& \tfrac{\dimDirac}{4 \uppi} \, \tfrac{1}{6} \, \tfrac{1}{| \barsigma |} \, .	\label{eq:z_sigma_0_0_2}
		\end{align}

\subsubsection{\texorpdfstring{$T = 0, \barsigma = 0$}{T = 0, sigma = 0}}
	
	Next, we turn to the symmetric phase at $T = 0$, hence $\barsigma = 0$.

\paragraph{\texorpdfstring{$T = 0, \barsigma = 0, \mu \neq 0$}{T = 0, sigma = 0, mu != 0}}

	Using the expansion \cref{eq:expansions_2f1} for \cref{eq:z_sigma_mu_0_d} one finds
		\begin{align}
			z ( 0, \mu, 0, \sdim ) ={}& - \tfrac{\dimDirac}{2} \, \tfrac{\Sd}{( 2 \uppi )^\sdim} \, \tfrac{1}{4} \, \tfrac{\sdim - 2}{\sdim - 3} \, | \mu |^{\sdim - 3} \, .	\label{eq:z_0_mu_0_d}
		\end{align}
	This is easily evaluated for $\sdim = 1$,
		\begin{align}
			z ( 0, \mu, 0, 1 ) ={}& - \tfrac{\dimDirac}{2 \uppi} \, \tfrac{1}{8} \tfrac{1}{\mu^2} \, ,	\label{eq:z_0_mu_0_1}
		\end{align}
	and $\sdim = 2$,
		\begin{align}
			z ( 0, \mu, 0, 2 ) ={}& 0 \, .	\label{eq:z_0_mu_0_2}
		\end{align}

\paragraph{\texorpdfstring{$T = 0, \barsigma = 0, \mu = 0$}{T = 0, sigma = 0, mu = 0}}

	Lastly, if one evaluates the wave-function renormalization in vacuum at the trivial evaluation point it is ill conditioned,
		\begin{align}
			z ( 0, 0, 0, d ) \in \{ \, & \pm \infty, 0 \} \, ,	\label{eq:z_0_0_0_d}
		\end{align}
	when taking the respective limits from the previous results.

\section{The bosonic two-point function}
\label{app:bosonic_two-point_function}

	In this appendix we calculate the bosonic two-point function \labelcref{eq:gamma2_main}.
	We use the regularized integrals \labelcref{eq:l_1_sigma_mu_T_d_Lambda,eq:l_2_sigma_mu_T_q_d_final} as well as \labelcref{eq:l_3_sigma_mu_T_d} and calculate the renormalized limits.
	Here, we present results for $q \neq 0$ and the limit $q = 0$.
	For the sake of clearness, we prepared \cref{tab:gamma2_explicit_expressions} which links to the different cases with (non-)vanishing $\barsigma$, $\mu$, $T$, as well as the special cases with $\sdim = 1$ and $\sdim = 2$.
	The limiting cases for $\sdim = 1$ are discussed in detail in \Rcite{Koenigstein:2021llr,Koenigstein:2023wso}, whereas the $\sdim = 2$ formulae are briefly discussed in \Rcite{Buballa:2020nsi,Pannullo:2023one}.
		\setlength{\extrarowheight}{3pt}
		\begin{table}
			\caption{\label{tab:gamma2_explicit_expressions}%
			Quick links to the equations for explicit evaluation of the bosonic two-point function $\Gamma^{(2)} ( \bar{\sigma}, \mu, T, q, \sdim )$.
			The formulae are simplified in terms of known functions as far as possible.
			}
			\begin{ruledtabular}
				\begin{tabular}{c c c c c c c}
					$T$	&	$\sigma$	&	$\mu$	&	$q$	&	$1 \leq \sdim < 3$	&	$\sdim = 1$	&	$\sdim = 2$
					\\
					\hline
					\multirow{8}{*}{$\neq 0$}	&	\multirow{4}{*}{$\neq 0$}	&	\multirow{2}{*}{$\neq 0$}	&	$\neq 0$	&	\cref{eq:gamma2q_sigma_mu_T_d}	&	\cref{eq:gamma2q_sigma_mu_T_1}	&	\cref{eq:gamma2q_sigma_mu_T_2}
					\\
					\cline{4-7}
						&	&	&	$= 0$	&	\cref{eq:gamma20_sigma_mu_T_d}	&	\cref{eq:gamma20_sigma_mu_T_1}	&	\cref{eq:gamma20_sigma_mu_T_2}
					\\
					\cline{3-7}
						&	&	\multirow{2}{*}{$= 0$}	&	$\neq 0$	&	\cref{eq:gamma2q_sigma_0_T_d}	&	\cref{eq:gamma2q_sigma_0_T_1}	&	\cref{eq:gamma2q_sigma_0_T_2}
					\\
					\cline{4-7}
						&	&	&	$= 0$	&	\cref{eq:gamma20_sigma_0_T_d}	&	\cref{eq:gamma20_sigma_0_T_1}	&	\cref{eq:gamma20_sigma_0_T_2}
					\\
					\cline{2-7}
						&	\multirow{4}{*}{$= 0$}	&	\multirow{2}{*}{$\neq 0$}	&	$\neq 0$	&	\cref{eq:gamma2q_0_mu_T_d}	&	\cref{eq:gamma2q_0_mu_T_1}	&	\cref{eq:gamma2q_0_mu_T_2}
					\\
					\cline{4-7}
						&	&	&	$= 0$	&	\cref{eq:gamma20_0_mu_T_d}	&	\cref{eq:gamma20_0_mu_T_1}	&	\cref{eq:gamma20_0_mu_T_2}
					\\
					\cline{3-7}
						&	&	\multirow{2}{*}{$= 0$}	&	$\neq 0$	&	\cref{eq:gamma2q_0_0_T_d}	&	\cref{eq:gamma2q_0_0_T_1}	&	\cref{eq:gamma2q_0_0_T_2}
					\\
					\cline{4-7}
						&	&	&	$= 0$	&	\cref{eq:gamma20_0_0_T_d}	&	\cref{eq:gamma20_0_0_T_1}	&	\cref{eq:gamma20_0_0_T_2}
					\\
					\hline
					\multirow{8}{*}{$= 0$}	&	\multirow{4}{*}{$\neq 0$}	&	\multirow{2}{*}{$\neq 0$}	&	$\neq 0$	&	\cref{eq:gamma2q_sigma_mu_0_d}	&	\cref{eq:gamma2q_sigma_mu_0_1}	&	\cref{eq:gamma2q_sigma_mu_0_2}
					\\
					\cline{4-7}
						&	&	&	$= 0$	&	\cref{eq:gamma20_sigma_mu_0_d}	&	\cref{eq:gamma20_sigma_mu_0_1}	&	\cref{eq:gamma20_sigma_mu_0_2}
					\\
					\cline{3-7}
						&	&	\multirow{2}{*}{$= 0$}	&	$\neq 0$	&	\cref{eq:gamma2q_sigma_0_0_d}	&	\cref{eq:gamma2q_sigma_0_0_1}	&	\cref{eq:gamma2q_sigma_0_0_2}
					\\
					\cline{4-7}
						&	&	&	$= 0$	&	\cref{eq:gamma20_sigma_0_0_d}	&	\cref{eq:gamma20_sigma_0_0_1}	&	\cref{eq:gamma20_sigma_0_0_2}
					\\
					\cline{2-7}
						&	\multirow{4}{*}{$= 0$}	&	\multirow{2}{*}{$\neq 0$}	&	$\neq 0$	&	\cref{eq:gamma2q_0_mu_0_d}	&	\cref{eq:gamma2q_0_mu_0_1}	&	\cref{eq:gamma2q_0_mu_0_2}
					\\
					\cline{4-7}
						&	&	&	$= 0$	&	\cref{eq:gamma20_0_mu_0_d}	&	\cref{eq:gamma20_0_mu_0_1}	&	\cref{eq:gamma20_0_mu_0_2}
					\\
					\cline{3-7}
						&	&	\multirow{2}{*}{$= 0$}	&	$\neq 0$	&	\cref{eq:gamma2q_0_0_0_d}	&	\cref{eq:gamma2q_0_0_0_1}	&	\cref{eq:gamma2q_0_0_0_2}
					\\
					\cline{4-7}
						&	&	&	$= 0$	&	\cref{eq:gamma20_0_0_0_d}	&	\cref{eq:gamma20_0_0_0_1}	&	\cref{eq:gamma20_0_0_0_2}
				\end{tabular}
			\end{ruledtabular}
		\end{table}

\subsection{\texorpdfstring{$T \neq 0$}{T != 0}}

	We start at nonzero temperature.
		
\subsubsection{\texorpdfstring{$T \neq 0, \barsigma \neq 0$}{T != 0, sigma != 0}}

	Furthermore, we first consider points in the regime, where $\barsigma \neq 0$.

\paragraph{\texorpdfstring{$T \neq 0, \barsigma \neq 0, \mu \neq 0$}{T != 0, sigma != 0, mu != 0}}

	For $\mu \neq 0$ and general $\sdim$, we simply insert \cref{eq:l_1_sigma_mu_T_d_Lambda,eq:l_2_sigma_mu_T_q_d_final,eq:l_3_sigma_mu_T_d} in the regularized expression \labelcref{eq:gamma2_regularized_main} and send $\Lambda \to \infty$.
	We use \cref{eq:expansions_2f1} and obtain,
		\begin{align}
			& \Gamma^{(2)} ( \barsigma, \mu, T, q, \sdim ) =	\Vdistance	\label{eq:gamma2q_sigma_mu_T_d}
			\\
			={}& \tfrac{1}{\ffcoupling} - \dimDirac \big[ l_1 ( \sigma, \mu, T, \sdim ) - \tfrac{1}{2} \, ( q^2 + 4 \, \barsigma^2 ) \, l_2 ( \barsigma, \mu, T, q, \sdim ) \big] =	\Vdistance	\nonumber
			\\
			={}& \dimDirac \, \big[ l_1 ( \sigmaminvac, 0, 0, \sdim ) - l_1 ( \sigma, \mu, T, \sdim ) +	\Vdistance	\nonumber
			\\
			& \quad + \tfrac{1}{2} \, ( q^2 + 4 \, \barsigma^2 ) \, l_2 ( \barsigma, \mu, T, q, \sdim ) \big] =	\Vdistance	\nonumber
			\\
			={}& \lim_{\Lambda \to \infty} \dimDirac \, \bigg[ \tfrac{\Sd}{( 2 \uppi )^\sdim} \, \tfrac{1}{2} \, \tfrac{| \sigmaminvac |^{\sdim - 1}}{\sdim} \, \big| \tfrac{\Lambda}{\sigmaminvac} \big|^\sdim \, \pFq{2}{1} \big( \tfrac{1}{2}, \tfrac{\sdim}{2}; \tfrac{\sdim + 2}{2}; - \tfrac{\Lambda^2}{\sigmaminvac^2} \big) +	\Vdistance	\nonumber
			\\
			& \quad - \tfrac{\Sd}{( 2 \uppi )^\sdim} \tfrac{1}{2} \bigg( \tfrac{1}{\sdim | \barsigma |} \, \Lambda^\sdim \, \pFq{2}{1} \big( \tfrac{1}{2}, \tfrac{\sdim}{2}; \tfrac{\sdim + 2}{2}; - \tfrac{\Lambda^2}{\barsigma^2} \big) +	\Vdistance	\nonumber
			\\
			& \qquad - \int_{0}^{\infty} \dd p \, p^{\sdim - 1} \, \tfrac{1}{E} \big[ \nf \big( \tfrac{E + \mu}{T} \big) + \nf \big( \tfrac{E - \mu}{T} \big) \big] \bigg) +	\Vdistance	\nonumber
			\\
			& \quad + \tfrac{1}{2} \, ( q^2 + 4 \, \barsigma^2 ) \, \tfrac{\Sd}{( 2 \uppi )^\sdim} \int_{0}^{1} \dd x \int_{0}^{\infty} \dd p \, p^{\sdim - 1} \, \tfrac{1}{4 \tilde{E}^3} \times	\Vdistance	\nonumber
			\\
			& \qquad \times \big[ 1 - \nf \big( \tfrac{\tilde{E} + \mu}{T} \big) + \tfrac{\tilde{E}}{T} \, \big[ \nf^2 \big( \tfrac{\tilde{E} + \mu}{T} \big) - \nf \big( \tfrac{\tilde{E} + \mu}{T} \big) \big] +	\Vdistance	\nonumber
			\\
			& \qquad \quad + ( \mu \to - \mu ) \big] \bigg] =	\Vdistance	\nonumber
			\\
			={}& \tfrac{\dimDirac}{2} \, \tfrac{\Sd}{( 2 \uppi )^\sdim} \, \bigg[ \big( | \sigmaminvac |^{\sdim - 1} - | \barsigma |^{\sdim - 1} \big) \, \tfrac{\Gamma ( \frac{1 - \sdim}{2} ) \, \Gamma ( \frac{\sdim}{2} )}{2 \sqrt{\uppi}} +	\Vdistance	\nonumber
			\\
			& \quad + \int_{0}^{\infty} \dd p \, p^{\sdim - 1} \, \tfrac{1}{E} \big[ \nf \big( \tfrac{E + \mu}{T} \big) + \nf \big( \tfrac{E - \mu}{T} \big) \big] +	\Vdistance	\nonumber
			\\
			& \quad + \big( \tfrac{q^2}{4} + \barsigma^2 ) \int_{0}^{1} \dd x \int_{0}^{\infty} \dd p \, p^{\sdim - 1} \, \tfrac{1}{\tilde{E}^3} \times	\Vdistance	\nonumber
			\\
			& \qquad \times \big[ 1 - \nf \big( \tfrac{\tilde{E} + \mu}{T} \big) + \tfrac{\tilde{E}}{T} \, \big[ \nf^2 \big( \tfrac{\tilde{E} + \mu}{T} \big) - \nf \big( \tfrac{\tilde{E} + \mu}{T} \big) \big] +	\Vdistance	\nonumber
			\\
			& \qquad \quad + ( \mu \to - \mu ) \big] \bigg] \, .	\Vdistance	\nonumber
		\end{align}
	For $q \to 0$ the $x$-integral is trivial and we find by evaluating another vacuum contribution,
		\begin{align}
			& \Gamma^{(2)} ( \barsigma, \mu, T, 0, \sdim ) =	\Vdistance	\label{eq:gamma20_sigma_mu_T_d}
			\\
			={}& \tfrac{\dimDirac}{2} \, \tfrac{\Sd}{( 2 \uppi )^\sdim} \, \bigg[ \big( | \sigmaminvac |^{\sdim - 1} - \sdim \, | \barsigma |^{\sdim - 1} \big) \, \tfrac{\Gamma ( \frac{1 - \sdim}{2} ) \, \Gamma ( \frac{\sdim}{2} )}{2 \sqrt{\uppi}} +	\Vdistance	\nonumber
			\\
			& \quad + \int_{0}^{\infty} \dd p \, p^{\sdim - 1} \, \tfrac{1}{E} \big[ \nf \big( \tfrac{E + \mu}{T} \big) + \nf \big( \tfrac{E - \mu}{T} \big) \big] +	\Vdistance	\nonumber
			\\
			& \quad - \barsigma^2 \int_{0}^{\infty} \dd p \, p^{\sdim - 1} \, \tfrac{1}{E^3} \, \big[ \nf \big( \tfrac{E + \mu}{T} \big) +	\Vdistance	\nonumber
			\\
			& \qquad - \tfrac{E}{T} \, \big[ \nf^2 \big( \tfrac{E + \mu}{T} \big) - \nf \big( \tfrac{E + \mu}{T} \big) \big]  + ( \mu \to - \mu ) \big] \bigg] \, .	\Vdistance	\nonumber
		\end{align}
	For $\sdim = 1$ we obtain
		\begin{align}
			& \Gamma^{(2)} ( \barsigma, \mu, T, q, 1 ) =	\Vdistance	\label{eq:gamma2q_sigma_mu_T_1}
			\\
			={}& \tfrac{\dimDirac}{2 \uppi} \, \Big[ \tfrac{1}{2} \, \ln \big( \tfrac{\barsigma^2}{\sigmaminvac^2} \big) + \int_{0}^{\infty} \dd p \, \tfrac{1}{E} \big[ \nf \big( \tfrac{E + \mu}{T} \big) + \nf \big( \tfrac{E - \mu}{T} \big) \big] +	\Vdistance	\nonumber
			\\
			& \quad + \big( \tfrac{q^2}{4} + \barsigma^2 \big) \int_{0}^{1} \dd x \int_{0}^{\infty} \dd p \, \tfrac{1}{\tilde{E}^3} \, \big[ 1 - \nf \big( \tfrac{\tilde{E} + \mu}{T} \big) +	\Vdistance	\nonumber
			\\
			& \qquad + \tfrac{\tilde{E}}{T} \, \big[ \nf^2 \big( \tfrac{\tilde{E} + \mu}{T} \big)  - \nf \big( \tfrac{\tilde{E} + \mu}{T} \big) \big] + ( \mu \to - \mu ) \big] \, .	\Vdistance	\nonumber
		\end{align}
	Here, in the limit $q \to 0$,
		\begin{align}
			& \Gamma^{(2)} ( \barsigma, \mu, T, 0, 1 ) =	\Vdistance	\label{eq:gamma20_sigma_mu_T_1}
			\\
			={}& \tfrac{\dimDirac}{2 \uppi} \, \Big[ \tfrac{1}{2} \, \ln \big( \tfrac{\barsigma^2}{\sigmaminvac^2} \big) + 1 + \int_{0}^{\infty} \dd p \, \tfrac{1}{E} \big[ \nf \big( \tfrac{E + \mu}{T} \big) + \nf \big( \tfrac{E - \mu}{T} \big) \big] +	\Vdistance	\nonumber
			\\
			& \quad - \barsigma^2 \int_{0}^{\infty} \dd p \, \tfrac{1}{E^3} \, \big[ \nf \big( \tfrac{E + \mu}{T} \big) +	\Vdistance	\nonumber
			\\
			& \qquad - \tfrac{E}{T} \, \big[ \nf^2 \big( \tfrac{E + \mu}{T} \big)  - \nf \big( \tfrac{E + \mu}{T} \big) \big] + ( \mu \to - \mu ) \big] \, .	\Vdistance	\nonumber
		\end{align}
	For the special case $\sdim = 2$ the momentum integrals can be evaluated analytically,
		\begin{align}
			& \Gamma^{(2)} ( \barsigma, \mu, T, q, 2 ) =	\Vdistance	\label{eq:gamma2q_sigma_mu_T_2}
			\\
			={}& \tfrac{\dimDirac}{4 \uppi} \, \bigg[ | \barsigma | - | \sigmaminvac | + T \ln \big[ 1 + \exp \big( - \tfrac{| \barsigma | + \mu}{T} \big) \big] +	\Vdistance	\nonumber
			\\
			& \quad + ( \tfrac{q^2}{4} + \barsigma^2 ) \, \int_{0}^{1} \dd x \, \tfrac{1}{\tilde{\Delta}} \, \big[ 1 - \nf \big( \tfrac{| \tilde{\Delta} | + \mu}{T} \big) \big] +	\Vdistance	\nonumber
			\\
			& \quad + ( \mu \to - \mu ) \bigg]	\Vdistance	\nonumber
		\end{align}
	In the limit $q \to 0$, we find
		\begin{align}
			& \Gamma^{(2)} ( \barsigma, \mu, T, 0, 2 ) =	\Vdistance	\label{eq:gamma20_sigma_mu_T_2}
			\\
			={}& \tfrac{\dimDirac}{4 \uppi} \, \bigg[ 2 \, | \barsigma | - | \sigmaminvac | + T \ln \big[ 1 + \exp \big( - \tfrac{| \barsigma | + \mu}{T} \big) \big] +	\Vdistance	\nonumber
			\\
			& \quad - | \barsigma | \, \nf \big( \tfrac{| \barsigma | + \mu}{T} \big) + ( \mu \to - \mu ) \bigg]	\Vdistance	\nonumber
		\end{align}

\paragraph{\texorpdfstring{$T \neq 0, \barsigma \neq 0, \mu = 0$}{T != 0, sigma != 0, mu = 0}}

	For studying the cases with vanishing chemical potential, we simply have to insert $\mu = 0$ in the previous expressions,
		\begin{align}
			& \Gamma^{(2)} ( \barsigma, 0, T, q, \sdim ) =	\Vdistance	\label{eq:gamma2q_sigma_0_T_d}
			\\
			={}& \tfrac{\dimDirac}{2} \, \tfrac{\Sd}{( 2 \uppi )^\sdim} \, \bigg[ \big( | \sigmaminvac |^{\sdim - 1} - | \barsigma |^{\sdim - 1} \big) \, \tfrac{\Gamma ( \frac{1 - \sdim}{2} ) \, \Gamma ( \frac{\sdim}{2} )}{2 \sqrt{\uppi}} +	\Vdistance	\nonumber
			\\
			& \quad + 2 \int_{0}^{\infty} \dd p \, p^{\sdim - 1} \, \tfrac{1}{E} \, \nf \big( \tfrac{E}{T} \big) +	\Vdistance	\nonumber
			\\
			& \quad + \big( \tfrac{q^2}{4} + \barsigma^2 ) \int_{0}^{1} \dd x \int_{0}^{\infty} \dd p \, p^{\sdim - 1} \, \tfrac{1}{\tilde{E}^3} \times	\Vdistance	\nonumber
			\\
			& \qquad \times \big[ 1 - 2 \, \nf \big( \tfrac{\tilde{E}}{T} \big) + 2 \, \tfrac{\tilde{E}}{T} \, \big[ \nf^2 \big( \tfrac{\tilde{E}}{T} \big) - \nf \big( \tfrac{\tilde{E}}{T} \big) \big] \big] \bigg] \, .	\Vdistance	\nonumber
		\end{align}
	In the limit of vanishing external momentum, this reduces to
		\begin{align}
			& \Gamma^{(2)} ( \barsigma, 0, T, 0, \sdim ) =	\Vdistance	\label{eq:gamma20_sigma_0_T_d}
			\\
			={}& \tfrac{\dimDirac}{2} \, \tfrac{\Sd}{( 2 \uppi )^\sdim} \, \bigg[ \big( | \sigmaminvac |^{\sdim - 1} - \sdim \, | \barsigma |^{\sdim - 1} \big) \, \tfrac{\Gamma ( \frac{1 - \sdim}{2} ) \, \Gamma ( \frac{\sdim}{2} )}{2 \sqrt{\uppi}} +	\Vdistance	\nonumber
			\\
			& \quad + 2 \int_{0}^{\infty} \dd p \, p^{\sdim - 1} \, \tfrac{1}{E} \, \nf \big( \tfrac{E}{T} \big) +	\Vdistance	\nonumber
			\\
			& \quad - 2 \, \barsigma^2 \int_{0}^{\infty} \dd p \, p^{\sdim - 1} \, \tfrac{1}{E^3} \times	\Vdistance	\nonumber
			\\
			& \qquad \times \big[ \nf \big( \tfrac{E}{T} \big) - \tfrac{E}{T} \, \big[ \nf^2 \big( \tfrac{E}{T} \big) - \nf \big( \tfrac{E}{T} \big) \big] \big] \bigg] \, .	\Vdistance	\nonumber
		\end{align}
	For $\sdim = 1$ we find
		\begin{align}
			& \Gamma^{(2)} ( \barsigma, 0, T, q, 1 ) =	\Vdistance	\label{eq:gamma2q_sigma_0_T_1}
			\\
			={}& \tfrac{\dimDirac}{2 \uppi} \, \bigg[ \tfrac{1}{2} \, \ln \big( \tfrac{\barsigma^2}{\sigmaminvac^2} \big) + 2 \int_{0}^{\infty} \dd p \, \tfrac{1}{E} \, \nf \big( \tfrac{E}{T} \big) +	\Vdistance	\nonumber
			\\
			& \quad + \big( \tfrac{q^2}{4} + \barsigma^2 ) \int_{0}^{1} \dd x \int_{0}^{\infty} \dd p \, \tfrac{1}{\tilde{E}^3} \times	\Vdistance	\nonumber
			\\
			& \qquad \times \big[ 1 - 2 \, \nf \big( \tfrac{\tilde{E}}{T} \big)  + 2 \, \tfrac{\tilde{E}}{T} \, \big[ \nf^2 \big( \tfrac{\tilde{E}}{T} \big) - \nf \big( \tfrac{\tilde{E}}{T} \big) \big] \big] \bigg] \, ,	\Vdistance	\nonumber
		\end{align}
	which has the $q \to 0$ limit
		\begin{align}
			& \Gamma^{(2)} ( \barsigma, 0, T, 0, 1 ) =	\Vdistance	\label{eq:gamma20_sigma_0_T_1}
			\\
			={}& \tfrac{\dimDirac}{2 \uppi} \, \bigg[ \tfrac{1}{2} \, \ln \big( \tfrac{\barsigma^2}{\sigmaminvac^2} \big) + 1 + 2 \int_{0}^{\infty} \dd p \, \tfrac{1}{E} \, \nf \big( \tfrac{E}{T} \big) +	\Vdistance	\nonumber
			\\
			& \quad - 2 \, \barsigma^2 \int_{0}^{\infty} \dd p \, \tfrac{1}{E^3} \times	\Vdistance	\nonumber
			\\
			& \qquad \times \big[ \nf \big( \tfrac{E}{T} \big) + \tfrac{E}{T} \, \big[ \nf^2 \big( \tfrac{E}{T} \big) - \nf \big( \tfrac{E}{T} \big) \big] \big] \bigg] \, ,	\Vdistance	\nonumber
		\end{align}
	For $\sdim = 2$ the explicit expression reads
		\begin{align}
			& \Gamma^{(2)} ( \barsigma, 0, T, q, 2 ) =	\Vdistance	\label{eq:gamma2q_sigma_0_T_2}
			\\
			={}& \tfrac{\dimDirac}{4 \uppi} \, \bigg[ | \barsigma | - | \sigmaminvac | + 2 \, T \ln \big[ 1 + \exp \big( - \tfrac{| \barsigma |}{T} \big) \big] \big] +	\Vdistance	\nonumber
			\\
			& \quad + \big( \tfrac{q^2}{4} + \barsigma^2 \big) \int_{0}^{1} \dd x \, \tfrac{1}{| \tilde{\Delta} |} \, \big[ 1 - 2 \, \nf \big( \tfrac{| \tilde{\Delta} |}{T} \big) \big] \bigg] \, ,	\Vdistance	\nonumber
		\end{align}
	which is
		\begin{align}
			& \Gamma^{(2)} ( \barsigma, 0, T, 0, 2 ) =	\Vdistance	\label{eq:gamma20_sigma_0_T_2}
			\\
			={}& \tfrac{\dimDirac}{4 \uppi} \, \bigg[ 2 \, | \barsigma | - | \sigmaminvac | + 2 \, T \ln \big[ 1 + \exp \big( - \tfrac{| \barsigma |}{T} \big) \big] \big] +	\Vdistance	\nonumber
			\\
			& \quad - 2 \, | \barsigma | \, \nf \big( \tfrac{| \barsigma |}{T} \big) \bigg] \, ,	\Vdistance	\nonumber
		\end{align}
	in the $q \to 0$ limit.

\subsubsection{\texorpdfstring{$T \neq 0, \barsigma = 0$}{T != 0, sigma = 0}}
	
	Next, we turn to the symmetric regime, $\barsigma = 0$, at nonzero temperature.

\paragraph{\texorpdfstring{$T \neq 0, \barsigma = 0, \mu \neq 0$}{T != 0, sigma = 0, mu != 0}}

	Here, we start with the cases with $\mu \neq 0$.
	For general $\sdim$ we find,
		\begin{align}
			& \Gamma^{(2)} ( 0, \mu, T, q, \sdim ) =	\Vdistance	\label{eq:gamma2q_0_mu_T_d}
			\\
			={}& \tfrac{\dimDirac}{2} \, \tfrac{\Sd}{( 2 \uppi )^\sdim} \, \bigg[  | \sigmaminvac |^{\sdim - 1} \, \tfrac{\Gamma ( \frac{1 - \sdim}{2} ) \, \Gamma ( \frac{\sdim}{2} )}{2 \sqrt{\uppi}} +	\Vdistance	\nonumber
			\\
			& \quad + \int_{0}^{\infty} \dd p \, p^{\sdim - 2} \, \big[ \nf \big( \tfrac{p + \mu}{T} \big) + \nf \big( \tfrac{p - \mu}{T} \big) \big] +	\Vdistance	\nonumber
			\\
			& \quad + \tfrac{q^2}{4} \int_{0}^{1} \dd x \int_{0}^{\infty} \dd p \, p^{\sdim - 1} \, \tfrac{1}{\tilde{p}^3} \, \big[ 1 - \nf \big( \tfrac{\tilde{p} + \mu}{T} \big) +	\Vdistance	\nonumber
			\\
			& \qquad + \tfrac{\tilde{p}}{T} \, \big[ \nf^2 \big( \tfrac{\tilde{p} + \mu}{T} \big) - \nf \big( \tfrac{\tilde{p} + \mu}{T} \big) \big] + ( \mu \to - \mu ) \big] \bigg] \, .	\Vdistance	\nonumber
			\\
			={}& \tfrac{\dimDirac}{2} \, \tfrac{\Sd}{( 2 \uppi )^\sdim} \, \bigg[  | \sigmaminvac |^{\sdim - 1} \, \tfrac{\Gamma ( \frac{1 - \sdim}{2} ) \, \Gamma ( \frac{\sdim}{2} )}{2 \sqrt{\uppi}} +	\Vdistance	\nonumber
			\\
			& \quad - T^{\sdim - 1} \, \Gamma ( \sdim - 1 ) \big[ \Li_{\sdim - 1} \big( - \ee^{\frac{\mu}{T}} \big) + \Li_{\sdim - 1} \big( - \ee^{- \frac{\mu}{T}} \big) \big]  +	\Vdistance	\nonumber
			\\
			& \quad + \tfrac{q^2}{4} \int_{0}^{1} \dd x \int_{0}^{\infty} \dd p \, p^{\sdim - 1} \, \tfrac{1}{\tilde{p}^3} \, \big[ 1 - \nf \big( \tfrac{\tilde{p} + \mu}{T} \big) +	\Vdistance	\nonumber
			\\
			& \qquad + \tfrac{\tilde{p}}{T} \, \big[ \nf^2 \big( \tfrac{\tilde{p} + \mu}{T} \big) - \nf \big( \tfrac{\tilde{p} + \mu}{T} \big) \big] + ( \mu \to - \mu ) \big] \bigg] \, .	\Vdistance	\nonumber
		\end{align}
	At $q = 0$ the remaining integral (the last term) vanishes and we find
		\begin{align}
			& \Gamma^{(2)} ( 0, \mu, T, 0, \sdim ) =	\Vdistance	\label{eq:gamma20_0_mu_T_d}
			\\
			={}& \tfrac{\dimDirac}{2} \, \tfrac{\Sd}{( 2 \uppi )^\sdim} \, \bigg[  | \sigmaminvac |^{\sdim - 1} \, \tfrac{\Gamma ( \frac{1 - \sdim}{2} ) \, \Gamma ( \frac{\sdim}{2} )}{2 \sqrt{\uppi}} +	\Vdistance	\nonumber
			\\
			& \quad - T^{\sdim - 1} \, \Gamma ( \sdim - 1 ) \big[ \Li_{\sdim - 1} \big( - \ee^{\frac{\mu}{T}} \big) + \Li_{\sdim - 1} \big( - \ee^{- \frac{\mu}{T}} \big) \big] \bigg] \, .	\Vdistance	\nonumber
		\end{align}
	The limit $\sdim = 1$ is special, because of a tricky cancellation of \gls{ir} divergences.
	We obtain
		\begin{align}
			& \Gamma^{(2)} ( 0, \mu, T, q, 1 ) =	\Vdistance	\label{eq:gamma2q_0_mu_T_1}
			\\
			={}& \tfrac{\dimDirac}{2 \uppi} \, \bigg[ \tfrac{1}{2} \ln \Big( \tfrac{( 2 T )^2}{\sigmaminvac^2} \Big) - \upgamma - \mathrm{DLi}_0 \big( \tfrac{\mu}{T} \big) +	\Vdistance	\nonumber
			\\
			& \quad + \tfrac{q^2}{4} \int_{0}^{1} \dd x \int_{0}^{\infty} \dd p \, \tfrac{1}{\tilde{p}^3} \, \big[ 1 - \nf \big( \tfrac{\tilde{p} + \mu}{T} \big) +	\Vdistance	\nonumber
			\\
			& \qquad + \tfrac{\tilde{p}}{T} \, \big[ \nf^2 \big( \tfrac{\tilde{p} + \mu}{T} \big) - \nf \big( \tfrac{\tilde{p} + \mu}{T} \big) \big] + ( \mu \to - \mu ) \big] \bigg] \, ,	\Vdistance	\nonumber
		\end{align}
	while the $q \to 0$ limit is easily obtained by discarding the integral,
		\begin{align}
			& \Gamma^{(2)} ( 0, \mu, T, 0, 1 ) =	\Vdistance	\label{eq:gamma20_0_mu_T_1}
			\\
			={}& \tfrac{\dimDirac}{2 \uppi} \, \Big[ \tfrac{1}{2} \ln \Big( \tfrac{( 2 T )^2}{\sigmaminvac^2} \Big) - \upgamma - \mathrm{DLi}_0 \big( \tfrac{\mu}{T} \big) \Big] \, ,	\Vdistance	\nonumber
		\end{align}
	On the other hand, for $\sdim = 2$,
		\begin{align}
			& \Gamma^{(2)} ( 0, \mu, T, q, 2 ) =	\Vdistance	\label{eq:gamma2q_0_mu_T_2}
			\\
			={}& \tfrac{\dimDirac}{4 \uppi} \, \bigg[ - | \sigmaminvac | + T \ln \big[ 1 + \exp \big( \tfrac{\mu}{T} \big) \big] + ( \mu \to - \mu ) +	\Vdistance	\nonumber
			\\
			& \quad + \tfrac{q^2}{4} \int_{0}^{1} \dd x \, \tfrac{1}{\sqrt{ q^2 \, x \, ( 1 - x ) }} \, \Big[ 1 - \nf \Big( \tfrac{\sqrt{ q^2 \, x \, ( 1 - x ) } + \mu}{T} \Big) +	\Vdistance	\nonumber
			\\
			& \qquad - \nf \Big( \tfrac{\sqrt{ q^2 \, x \, ( 1 - x ) } - \mu}{T} \Big) \Big] \bigg] \, .	\Vdistance	\nonumber
		\end{align}
	Also here, we can simply drop the remaining integral to study $q = 0$,
		\begin{align}
			& \Gamma^{(2)} ( 0, \mu, T, 0, 2 ) =	\Vdistance	\label{eq:gamma20_0_mu_T_2}
			\\
			={}& \tfrac{\dimDirac}{4 \uppi} \, \big[ - | \sigmaminvac | + T \ln \big[ 1 + \exp \big( \tfrac{\mu}{T} \big) \big] + ( \mu \to - \mu ) \big] \, .	\Vdistance	\nonumber
		\end{align}

\paragraph{\texorpdfstring{$T \neq 0, \barsigma = 0, \mu = 0$}{T != 0, sigma = 0, mu != 0}}
	Setting also $\mu = 0$ in the previous formulae, we find for general $\sdim$,
		\begin{align}
			& \Gamma^{(2)} ( 0, 0, T, q, \sdim ) =	\Vdistance	\label{eq:gamma2q_0_0_T_d}
			\\
			={}& \tfrac{\dimDirac}{2} \, \tfrac{\Sd}{( 2 \uppi )^\sdim} \, \bigg[ | \sigmaminvac |^{\sdim - 1} \, \tfrac{\Gamma ( \frac{1 - \sdim}{2} ) \, \Gamma ( \frac{\sdim}{2} )}{2 \sqrt{\uppi}} +	\Vdistance	\nonumber
			\\
			& \quad + 2 \, T^{\sdim - 1} \, \Gamma ( \sdim - 1 ) \, \eta ( \sdim - 1 ) +	\Vdistance	\nonumber
			\\
			& \quad + \tfrac{q^2}{4} \int_{0}^{1} \dd x \int_{0}^{\infty} \dd p \, p^{\sdim - 1} \, \tfrac{1}{\tilde{p}^3} \, \big[ 1 - 2 \, \nf \big( \tfrac{\tilde{p}}{T} \big) +	\Vdistance	\nonumber
			\\
			& \qquad + 2 \, \tfrac{\tilde{p}}{T} \, \big[ \nf^2 \big( \tfrac{\tilde{p}}{T} \big) - \nf \big( \tfrac{\tilde{p}}{T} \big) \big] \big] \bigg] \, .	\Vdistance	\nonumber
		\end{align}
	Here, $\eta ( s )$ is the Dirichlet eta function \labelcref{eq:dirichlet_eta_function}. 
	In the $q \to 0$ limit, the expression reduces to
		\begin{align}
			& \Gamma^{(2)} ( 0, 0, T, 0, \sdim ) =	\Vdistance	\label{eq:gamma20_0_0_T_d}
			\\
			={}& \tfrac{\dimDirac}{2} \, \tfrac{\Sd}{( 2 \uppi )^\sdim} \, \Big[ | \sigmaminvac |^{\sdim - 1} \, \tfrac{\Gamma ( \frac{1 - \sdim}{2} ) \, \Gamma ( \frac{\sdim}{2} )}{2 \sqrt{\uppi}} +	\Vdistance	\nonumber
			\\
			& \quad + 2 \, T^{\sdim - 1} \, \Gamma ( \sdim - 1 ) \, \eta ( \sdim - 1 ) \Big] \, .	\Vdistance	\nonumber
		\end{align}
	Taking the limit of $\sdim = 1$ carefully, we find
		\begin{align}
			& \Gamma^{(2)} ( 0, 0, T, q, 1 ) =	\Vdistance	\label{eq:gamma2q_0_0_T_1}
			\\
			={}& \tfrac{\dimDirac}{2 \uppi} \, \bigg[ \tfrac{1}{2} \ln \Big( \tfrac{( \uppi T )^2}{\sigmaminvac^2} \Big) - \upgamma + q^2 \int_{0}^{1} \dd x \int_{0}^{\infty} \dd p \, \tfrac{1}{4 \tilde{p}^3} \times	\Vdistance	\nonumber
			\\
			& \qquad \times \big[ 1 - 2 \, \nf \big( \tfrac{\tilde{p}}{T} \big) + 2 \, \tfrac{\tilde{p}}{T} \, \big[ \nf^2 \big( \tfrac{\tilde{p}}{T} \big) - \nf \big( \tfrac{\tilde{p}}{T} \big) \big] \big] \bigg] \, ,	\Vdistance	\nonumber
		\end{align}
	and
		\begin{align}
			\Gamma^{(2)} ( 0, 0, T, 0, 1 ) ={}& \tfrac{\dimDirac}{2 \uppi} \, \Big[ \tfrac{1}{2} \ln \Big( \tfrac{( \uppi T )^2}{\sigmaminvac^2} \Big) - \upgamma \Big] \, ,	\Vdistance	\label{eq:gamma20_0_0_T_1}
		\end{align}	
	while for $\sdim = 2$ we have
		\begin{align}
			& \Gamma^{(2)} ( 0, 0, T, q, 2 ) =	\Vdistance	\label{eq:gamma2q_0_0_T_2}
			\\
			={}& \tfrac{\dimDirac}{4 \uppi} \, \bigg[ - | \sigmaminvac | + T \ln ( 4 ) + q^2 \, \tfrac{1}{4} \int_{0}^{1} \dd x \, \tfrac{1}{\sqrt{ q^2 \, x \, ( 1 - x ) }} \times	\Vdistance	\nonumber
			\\
			& \qquad \times \Big[ 1 - 2 \, \nf \Big( \tfrac{\sqrt{ q^2 \, x \, ( 1 - x ) }}{T} \Big) \Big] \bigg] \, .	\Vdistance	\nonumber
		\end{align}
	For $q = 0$, this reduces to
		\begin{align}
			\Gamma^{(2)} ( 0, 0, T, 0, 2 ) ={}& \tfrac{\dimDirac}{4 \uppi} \, \big[ - | \sigmaminvac | + T \ln ( 4 ) \big] \, .	\label{eq:gamma20_0_0_T_2}
		\end{align}

\subsection{\texorpdfstring{$T = 0$}{T = 0}}

	Having completed the $T \neq 0$ cases, we turn to the zero-temperature limit.
		
\subsubsection{\texorpdfstring{$T = 0, \barsigma \neq 0$}{T = 0, sigma != 0}}

	We start at nonzero background field $\barsigma \neq 0$.
		
\paragraph{\texorpdfstring{$T = 0, \barsigma \neq 0, \mu \neq 0$}{T = 0, sigma != 0, mu != 0}}

	For $\mu \neq 0$ we can insert the zero-temperature integrals \labelcref{eq:l_1_sigma_mu_0_d_Lambda,eq:l_2_sigma_mu_0_q_d} in \cref{eq:gamma2_regularized_main} and take the limit $\Lambda \to \infty$ by using \cref{eq:expansions_2f1}.
	This results in
		\begin{align}
			& \Gamma^{(2)} ( \barsigma, \mu, 0, q, \sdim ) =	\Vdistance	\label{eq:gamma2q_sigma_mu_0_d}
			\\
			={}& \tfrac{1}{\ffcoupling} - \dimDirac \big[ l_1 ( \sigma, \mu, 0, \sdim ) - \tfrac{1}{2} \, ( q^2 + 4 \, \barsigma^2 ) \, l_2 ( \barsigma, \mu, 0, q, \sdim ) \big] =	\Vdistance	\nonumber
			\\
			={}& \lim_{\Lambda \to \infty} \dimDirac \, \big[ l_1^\Lambda ( \sigmaminvac, 0, 0, \sdim ) - l_1^\Lambda ( \sigma, \mu, 0, \sdim ) +	\Vdistance	\nonumber
			\\
			& \quad + \tfrac{1}{2} \, ( q^2 + 4 \, \barsigma^2 ) \, l_2 ( \barsigma, \mu, 0, q, \sdim ) \big] =	\Vdistance	\nonumber
			\\
			={}& \dimDirac \, \tfrac{\Sd}{( 2 \uppi )^\sdim} \, \tfrac{1}{2} \, \bigg[ \big( | \sigmaminvac |^{\sdim - 1} - | \barsigma |^{\sdim - 1} \big) \, \tfrac{\Gamma ( \frac{1 - \sdim}{2} ) \, \Gamma ( \frac{\sdim}{2} )}{2 \sqrt{\uppi}} +	\Vdistance	\nonumber
			\\
			& \quad + \Theta \big( \tfrac{\barmu^2}{\barsigma^2} \big) \,  \tfrac{| \barsigma |^{\sdim - 1}}{\sdim} \, \big| \tfrac{\barmu}{\barsigma} \big|^\sdim \, \pFq{2}{1} \big( \tfrac{1}{2}, \tfrac{\sdim}{2}; \tfrac{\sdim + 2}{2}; - \tfrac{\barmu^2}{\barsigma^2} \big) +	\Vdistance	\nonumber
			\\
			& \quad + \big( \tfrac{q^2}{4} + \barsigma^2 ) \int_{0}^{1} \dd x \, \Big[ \tilde{\Delta}^{\sdim - 3} \, \tfrac{\Gamma ( \frac{3 - \sdim}{2} ) \, \Gamma ( \frac{\sdim}{2} )}{\sqrt{\uppi}} +	\Vdistance	\nonumber
			\\
			& \qquad - \Theta \Big( \tfrac{\tilde{\mu}^2}{\tilde{\Delta}^2} \Big) \Big( \tfrac{\tilde{\mu}^\sdim}{\tilde{\Delta}^3} \, \tfrac{1}{\sdim} \, \pFq{2}{1} \big( \tfrac{3}{2}, \tfrac{\sdim}{2}; \tfrac{\sdim + 2}{2}; - \tfrac{\tilde{\mu}^2}{\tilde{\Delta}^2} \big) + \tfrac{\tilde{\mu}^{\sdim - 2}}{| \mu |} \Big) \Big] \bigg] \, .	\Vdistance	\nonumber
		\end{align}
	In the limit of vanishing external momentum $q$ the result reduces to
		\begin{align}
			& \Gamma^{(2)} ( \barsigma, \mu, 0, 0, \sdim ) =	\Vdistance	\label{eq:gamma20_sigma_mu_0_d}
			\\
			={}& \dimDirac \, \tfrac{\Sd}{( 2 \uppi )^\sdim} \, \tfrac{1}{2} \, \bigg[ \big( | \sigmaminvac |^{\sdim - 1} - | \barsigma |^{\sdim - 1} \big) \, \tfrac{\Gamma ( \frac{1 - \sdim}{2} ) \, \Gamma ( \frac{\sdim}{2} )}{2 \sqrt{\uppi}} +	\Vdistance	\nonumber
			\\
			& \quad + \Theta \big( \tfrac{\barmu^2}{\barsigma^2} \big) \,  \tfrac{| \barsigma |^{\sdim - 1}}{\sdim} \, \big| \tfrac{\barmu}{\barsigma} \big|^\sdim \, \pFq{2}{1} \big( \tfrac{1}{2}, \tfrac{\sdim}{2}; \tfrac{\sdim + 2}{2}; - \tfrac{\barmu^2}{\barsigma^2} \big) +	\Vdistance	\nonumber
			\\
			& \quad + \barsigma^2 \Big[ | \barsigma |^{\sdim - 3} \, \tfrac{\Gamma ( \frac{3 - \sdim}{2} ) \, \Gamma ( \frac{\sdim}{2} )}{\sqrt{\uppi}} +	\Vdistance	\nonumber
			\\
			& \qquad - \Theta \Big( \tfrac{\barmu^2}{\barsigma^2} \Big) \Big( \tfrac{\barmu^\sdim}{\barsigma^3} \, \tfrac{1}{\sdim} \, \pFq{2}{1} \big( \tfrac{3}{2}, \tfrac{\sdim}{2}; \tfrac{\sdim + 2}{2}; - \tfrac{\barmu^2}{\barsigma^2} \big) + \tfrac{\barmu^{\sdim - 2}}{| \mu |} \Big) \Big] \bigg] \, .	\Vdistance	\nonumber
		\end{align}
	For $\sdim = 1$ the two-point function at vanishing temperature reads
		\begin{align}
			& \Gamma^{(2)} ( \barsigma, \mu, 0, q, 1 ) =	\Vdistance	\label{eq:gamma2q_sigma_mu_0_1}
			\\
			={}& \tfrac{\dimDirac}{2 \uppi} \, \Bigg[ \tfrac{1}{2} \ln \big( \tfrac{\barsigma^2}{\sigmaminvac^2} \big) + \sqrt{1 + \tfrac{4 \barsigma^2}{q^2}} \, \arcoth \Big( \sqrt{1 + \tfrac{4 \barsigma^2}{q^2}} \Big) +	\Vdistance	\nonumber
			\\
			& \quad + \Theta \Big( \tfrac{\barmu^2}{\barsigma^2} \Big) \Bigg( \artanh \big( \big| \tfrac{\barmu}{\mu} \big| \big) - \tfrac{1}{2} \, \sqrt{1 + \tfrac{4 \barsigma^2}{q^2}} \times \Vdistance	\nonumber
			\\
			& \qquad \times \Bigg[ \artanh \Bigg( \frac{\frac{2 \barsigma^2}{\mu q} + | \frac{\barmu}{\mu} |}{\sqrt{1 + \frac{4 \barsigma^2}{q^2}}} \Bigg) + ( \mu \to - \mu ) \Bigg] \Bigg) \Bigg] \, , \Vdistance	\nonumber
		\end{align}
	and for $q = 0$,
		\begin{align}
			& \Gamma^{(2)} ( \barsigma, \mu, 0, 0, 1 ) =	\Vdistance	\label{eq:gamma20_sigma_mu_0_1}
			\\
			={}& \tfrac{\dimDirac}{2 \uppi} \, \Big[ \tfrac{1}{2} \ln \big( \tfrac{\barsigma^2}{\sigmaminvac^2} \big) + 1 + \Theta \Big( \tfrac{\barmu^2}{\barsigma^2}  \Big) \Big( \artanh \big( \big| \tfrac{\barmu}{\mu} \big| \big) - \big| \tfrac{\mu}{\barmu} \big| \Big) \Big] \, . \Vdistance	\nonumber
		\end{align}
	For $\sdim = 2$, 
		\begin{align}
			& \Gamma^{(2)} ( \barsigma, \mu, 0, q, 2 ) =	\Vdistance	\label{eq:gamma2q_sigma_mu_0_2}
			\\
			={}& \tfrac{\dimDirac}{4 \uppi} \, \Bigg[ | \barsigma | - | \sigmaminvac | + \Theta \left( \tfrac{\barmu^2}{\barsigma^2} \right) \, ( | \mu | - | \barsigma | ) +	\Vdistance	\nonumber
			\\
			&  + \big( \tfrac{q^2}{4} + \barsigma^2 \big) \Theta \left(\tfrac{ \tfrac{q^2}{4}-\barmu^2}{\barsigma^2 }\right)  \tfrac{1}{2|q|} \arctan\left(\sqrt{\tfrac{\tfrac{q^2}{4}-   \Theta \big( \tfrac{\barmu^2}{\barsigma^2} \big) \barmu^2 }{ \barsigma^2 +\Theta \big( \tfrac{\barmu^2}{\barsigma^2} \big) \barmu^2 }}\right)  \Bigg] \, ,	\Vdistance	\nonumber
		\end{align}
	which simplifies to
		\begin{align}
			& \Gamma^{(2)} ( \barsigma, \mu, 0, 0, 2 ) =	\Vdistance	\label{eq:gamma20_sigma_mu_0_2}
			\\
			={}& \tfrac{\dimDirac}{4 \uppi} \, \big[ 2 | \barsigma | - | \sigmaminvac | + \Theta \big( \tfrac{\barmu^2}{\barsigma^2} \big) \, ( | \mu | - 2 | \barsigma | ) \big]	\Vdistance	\nonumber
		\end{align}
	for $q = 0$.

\paragraph{\texorpdfstring{$T = 0, \barsigma \neq 0, \mu = 0$}{T = 0, sigma != 0, mu = 0}}

	Next, we use the previous expressions and set $\mu = 0$.
	First, we obtain
		\begin{align}
			& \Gamma^{(2)} ( \barsigma, 0, 0, q, \sdim ) =	\Vdistance	\label{eq:gamma2q_sigma_0_0_d}
			\\
			={}& \dimDirac \, \tfrac{\Sd}{( 2 \uppi )^\sdim} \, \tfrac{1}{2} \, \bigg[ \big( | \sigmaminvac |^{\sdim - 1} - | \barsigma |^{\sdim - 1} \big) \, \tfrac{\Gamma ( \frac{1 - \sdim}{2} ) \, \Gamma ( \frac{\sdim}{2} )}{2 \sqrt{\uppi}} +	\Vdistance	\nonumber
			\\
			& \quad + \big( \tfrac{q^2}{4} + \barsigma^2 \big) \int_{0}^{1} \dd x \, \tilde{\Delta}^{\sdim - 3} \, \tfrac{\Gamma ( \frac{3 - \sdim}{2} ) \, \Gamma ( \frac{\sdim}{2} )}{\sqrt{\uppi}} \bigg] \, ,	\Vdistance	\nonumber
		\end{align}
	and
		\begin{align}
			& \Gamma^{(2)} ( \barsigma, 0, 0, 0, \sdim ) =	\Vdistance	\label{eq:gamma20_sigma_0_0_d}
			\\
			={}& \dimDirac \, \tfrac{\Sd}{( 2 \uppi )^\sdim} \, \tfrac{1}{2} \, \bigg[ \big( | \sigmaminvac |^{\sdim - 1} - \sdim \, | \barsigma |^{\sdim - 1} \big) \, \tfrac{\Gamma ( \frac{1 - \sdim}{2} ) \, \Gamma ( \frac{\sdim}{2} )}{2 \sqrt{\uppi}} \Big] \, .	\Vdistance	\nonumber
		\end{align}
	For spatial dimension $\sdim = 1$ the $\mu = 0$ case reads
		\begin{align}
			& \Gamma^{(2)} ( \barsigma, 0, 0, q, 1 ) =	\Vdistance	\label{eq:gamma2q_sigma_0_0_1}
			\\
			={}& \tfrac{\dimDirac}{2 \uppi} \, \Big[ \tfrac{1}{2} \ln \big( \tfrac{\barsigma^2}{\sigmaminvac^2} \big) + \sqrt{1 + \tfrac{4 \barsigma^2}{q^2}} \, \arcoth \Big( \sqrt{1 + \tfrac{4 \barsigma^2}{q^2}} \Big) \Big]	\Vdistance	\nonumber
		\end{align}
	and the corresponding $q \to 0$ limit is
		\begin{align}
			\Gamma^{(2)} ( \barsigma, 0, 0, 0, 1 ) ={}& \tfrac{\dimDirac}{2 \uppi} \, \Big[ \tfrac{1}{2} \ln \big( \tfrac{\barsigma^2}{\sigmaminvac^2} \big) + 1 \Big] \, .	\label{eq:gamma20_sigma_0_0_1}
		\end{align}
	The other special case, $\sdim = 2$, simplifies to
		\begin{align}
			& \Gamma^{(2)} ( \barsigma, 0, 0, q, 2 ) =	\Vdistance	\label{eq:gamma2q_sigma_0_0_2}
			\\
			={}& \tfrac{\dimDirac}{4 \uppi} \, \Big[ | \barsigma | - | \sigmaminvac | + | q | \, \Big( 1 + \tfrac{4 \barsigma^2}{q^2} \Big) \tfrac{1}{2} \arctan \Big( \sqrt{\tfrac{q^2}{4 \barsigma^2}} \Big) \Big] \, ,	\Vdistance	\nonumber
		\end{align}
	and has the $q = 0$ limit
		\begin{align}
			\Gamma^{(2)} ( \barsigma, 0, 0, 0, 2 ) ={}& \tfrac{\dimDirac}{4 \uppi} \, \big( 2 \, | \barsigma | - | \sigmaminvac | \big) \, .	\label{eq:gamma20_sigma_0_0_2}
		\end{align}

\subsubsection{\texorpdfstring{$T = 0, \barsigma = 0$}{T = 0, sigma = 0}}

	Finally, we turn to the expressions in the symmetric phase, where the evaluation point is $\barsigma = 0$.

\paragraph{\texorpdfstring{$T = 0, \barsigma = 0, \mu \neq 0$}{T = 0, sigma = 0, mu != 0}}

	In a first step, we again study the cases with $\mu \neq 0$ and start with general $\sdim$,
		\begin{align}
			& \Gamma^{(2)} ( 0, \mu, 0, q, \sdim ) =	\Vdistance	\label{eq:gamma2q_0_mu_0_d}
			\\
			={}& \dimDirac \, \tfrac{\Sd}{( 2 \uppi )^\sdim} \, \tfrac{1}{2} \, \bigg[ | \sigmaminvac |^{\sdim - 1} \, \tfrac{\Gamma ( \frac{1 - \sdim}{2} ) \, \Gamma ( \frac{\sdim}{2} )}{2 \sqrt{\uppi}} + \tfrac{| \mu |^{\sdim - 1}}{\sdim - 1} +	\Vdistance	\nonumber
			\\
			& \quad + \tfrac{q^2}{4} \int_{0}^{1} \dd x \, \Big[ [ q^2 \, x \, ( 1 - x ) ]^{\frac{\sdim - 3}{2}} \, \tfrac{\Gamma ( \frac{3 - \sdim}{2} ) \, \Gamma ( \frac{\sdim}{2} )}{\sqrt{\uppi}} +	\Vdistance	\nonumber
			\\
			& \qquad - \Theta \Big( \tfrac{\mu^2 - q^2 \, x \, ( 1 - x )}{q^2 \, x \, ( 1 - x )} \Big) \Big( \tfrac{[ \mu^2 - q^2 \, x \, ( 1 - x ) ]^\frac{\sdim}{2}}{[ q^2 \, x \, ( 1 - x ) ]^\frac{3}{2}} \, \tfrac{1}{\sdim} \times	\Vdistance	\nonumber
			\\
			& \qquad \quad \times \pFq{2}{1} \big( \tfrac{3}{2}, \tfrac{\sdim}{2}; \tfrac{\sdim + 2}{2}; - \tfrac{\mu^2 - q^2 \, x \, ( 1 - x )}{q^2 \, x \, ( 1 - x )} \big) +	\Vdistance	\nonumber
			\\
			& \qquad \quad + \tfrac{[ \mu^2 - q^2 \, x \, ( 1 - x ) ]^{\frac{\sdim - 2}{2}}}{| \mu |} \Big) \Big] \bigg] \, .	\Vdistance	\nonumber
		\end{align}
	The limit $q \to 0$ of this expression is,
		\begin{align}
			& \Gamma^{(2)} ( 0, \mu, 0, 0, \sdim ) =	\Vdistance	\label{eq:gamma20_0_mu_0_d}
			\\
			={}& \dimDirac \, \tfrac{\Sd}{( 2 \uppi )^\sdim} \, \tfrac{1}{2} \, \Big[ | \sigmaminvac |^{\sdim - 1} \, \tfrac{\Gamma ( \frac{1 - \sdim}{2} ) \, \Gamma ( \frac{\sdim}{2} )}{2 \sqrt{\uppi}} + \tfrac{| \mu |^{\sdim - 1}}{\sdim - 1}	\Big] \, .	\Vdistance	\nonumber
		\end{align}
	However, if we study $\sdim = 1$ we can use
		\begin{align}
			\Gamma^{(2)} ( 0, \mu, 0, q, 1 ) ={}& \tfrac{\dimDirac}{4 \uppi} \, \ln \big( \tfrac{| 4 \mu^2 - q^2 |}{\sigmaminvac^2} \big) \, ,	\label{eq:gamma2q_0_mu_0_1}
		\end{align}
	and
		\begin{align}
			\Gamma^{(2)} ( 0, \mu, 0, 0, 1 ) ={}& \tfrac{\dimDirac}{4 \uppi} \, \ln \big( \tfrac{4 \mu^2}{\sigmaminvac^2} \big) \, .	\label{eq:gamma20_0_mu_0_1}
		\end{align}
	For the $\sdim = 2$ limiting case,
		\begin{align}
			& \Gamma^{(2)} ( 0, \mu, 0, q, 2 ) =	\Vdistance	\label{eq:gamma2q_0_mu_0_2}
			\\
			={}& \tfrac{\dimDirac}{4 \uppi} \, \bigg[ | \mu | - | \sigmaminvac | +	\Vdistance	\nonumber
			\\
			& \quad + \Theta \Big( \tfrac{ q^2 -4 \mu^2}{4 q^2} \Big) |q| \bigg[\tfrac{\uppi}{4}- \tfrac{1}{2} \arctan\left(\sqrt{\tfrac{4\mu^2}{q^2-4\mu^2}}\right) \bigg] \bigg]	\Vdistance	\nonumber
		\end{align}
	with
		\begin{align}
			\Gamma^{(2)} ( 0, \mu, 0, 0, 2 ) ={}& \tfrac{\dimDirac}{4 \uppi} \, \big( | \mu | - | \sigmaminvac | \big) \, .	\label{eq:gamma20_0_mu_0_2}
		\end{align}

\paragraph{\texorpdfstring{$T = 0, \barsigma = 0, \mu = 0$}{T = 0, sigma = 0, mu = 0}}

	Lastly, we consider $\mu = T = \barsigma = 0$.
	For continuous $\sdim$, we find
		\begin{align}
			& \Gamma^{(2)} ( 0, 0, 0, q, \sdim ) =	\Vdistance	\label{eq:gamma2q_0_0_0_d}
			\\
			={}& \dimDirac \, \tfrac{\Sd}{( 2 \uppi )^\sdim} \, \tfrac{1}{2} \, \bigg[ | \sigmaminvac |^{\sdim - 1} \, \tfrac{\Gamma ( \frac{1 - \sdim}{2} ) \, \Gamma ( \frac{\sdim}{2} )}{2 \sqrt{\uppi}} +	\Vdistance	\nonumber
			\\
			& \quad + \big( \tfrac{q^2}{4} \big)^{\frac{\sdim - 1}{2}} \tfrac{\Gamma ( \frac{3 - \sdim}{2} ) \, \Gamma ( \frac{\sdim - 1}{2} )}{2} \bigg] \, .	\Vdistance	\nonumber
		\end{align}
	In the limit of $q = 0$, this reduces to
		\begin{align}
			\Gamma^{(2)} ( 0, 0, 0, 0, \sdim ) ={}& \dimDirac \, \tfrac{\Sd}{( 2 \uppi )^\sdim} \, \tfrac{1}{2} \, | \sigmaminvac |^{\sdim - 1} \, \tfrac{\Gamma ( \frac{1 - \sdim}{2} ) \, \Gamma ( \frac{\sdim}{2} )}{2 \sqrt{\uppi}} \, .	\label{eq:gamma20_0_0_0_d}
		\end{align}
	For $\sdim = 1$ we find
		\begin{align}
			\Gamma^{(2)} ( 0, 0, 0, 0, 1 ) ={}& \tfrac{\dimDirac}{4 \uppi} \, \ln \Big( \tfrac{q^2}{\sigmaminvac^2} \Big) \, ,	\label{eq:gamma20_0_0_0_1}
		\end{align}
	while the $q \to 0$ limit is manifestly \gls{ir} divergent in one spatial dimension,
		\begin{align}
			\Gamma^{(2)} ( 0, 0, 0, q, 1 ) ={}& - \infty \, .	\label{eq:gamma2q_0_0_0_1}
		\end{align}
	On the other hand, for $\sdim = 2$, we find
		\begin{align}
			\Gamma^{(2)} ( 0, 0, 0, q, 2 ) ={}& \tfrac{\dimDirac}{4 \uppi} \, \big( - | \sigmaminvac | + \tfrac{\uppi}{4} \, | q | \big) \, ,	\label{eq:gamma2q_0_0_0_2}
		\end{align}
	which simplifies to
		\begin{align}
			\Gamma^{(2)} ( 0, 0, 0, 0, 2 ) ={}& - \tfrac{\dimDirac}{4 \uppi} \, | \sigmaminvac | \, .	\label{eq:gamma20_0_0_0_2}
		\end{align}

\newpage

\FloatBarrier

\bibliography{bib/general.bib,bib/gn.bib,bib/inhomo.bib,bib/instanton.bib,bib/lattice.bib,bib/math.bib,bib/numerics.bib,bib/qcd.bib,bib/rg.bib,bib/software.bib,bib/symmetries.bib,bib/thermal_qft.bib,bib/thies.bib,bib/virasoro_algebra.bib,bib/zero-dim-qft.bib} 


\end{document}